\documentclass[fleqn,usenatbib]{mnras}

\usepackage[T1]{fontenc}

\DeclareRobustCommand{\VAN}[3]{#2}
\let\VANthebibliography\thebibliography
\def\thebibliography{\DeclareRobustCommand{\VAN}[3]{##3}\VANthebibliography}

\usepackage{amssymb}	

\usepackage{graphicx}
\usepackage{amsmath}
\usepackage{threeparttable}
\usepackage{caption}
\usepackage{bibunits}
\usepackage{color}
\usepackage{ragged2e}
\usepackage{adjustbox}
\usepackage{url}
\usepackage{soul}
\newcommand{\cii}{[C\,{\sc ii}]}
\newcommand{\ci}{[C\,{\sc i}]}

\newcommand{\hii}{H\,{\sc ii}}

\newcommand{\oiii}{[O\,{\sc iii}]}

\newcommand{\oiiitocii}{$L_\mathrm{[O\,\textsc{iii}]}/L_\mathrm{[C\,\textsc{ii}]}$}

\newcommand{\oiiil}{[O\,{\sc iii}] 88\,$\mu{\rm m}$}
\newcommand{\ciil}{[C\,{\sc ii}] 158\,$\mu{\rm m}$}

\definecolor{referee}{RGB}{0,0,0}
\definecolor{referee2}{RGB}{0,0,0}

\title[\oiii{} at $z =$~6--7.6]{
Gas conditions of a star-formation selected sample in the first billion years}

\author[Tom Bakx \& Hiddo Algera et al.]{Tom J. L. C. Bakx$^{1}$\thanks{E-mail: tom.bakx@chalmers.se},
Hiddo S. B. Algera$^{2,3}$,
Bram Venemans$^{4}$,
Laura Sommovigo$^{5}$,
Seiji Fujimoto$^{6}$,\newauthor
Stefano Carniani$^{7}$,
Masato Hagimoto$^{8}$,
Takuya Hashimoto $^{9,10}$,
Akio K. Inoue$^{11,12}$,
Dragan Salak$^{13,14}$,\newauthor
Stephen Serjeant$^{15}$,
Livia Vallini$^{16}$, 
Stephen Eales$^{17}$,
Andrea Ferrara$^{7}$, 
Yoshinobu Fudamoto$^{18}$,\newauthor
Chihiro Imamura$^{8}$,
Shigeki Inoue$^{13}$, 
Kirsten K. Knudsen$^{1}$,
Hiroshi Matsuo$^{3,19}$, 
Yuma Sugahara$^{11,3}$, \newauthor
Yoichi Tamura$^{8}$,
Akio Taniguchi$^{8}$, and
Satoshi Yamanaka$^{20}$.
\\
$^{1}$ Department of Space, Earth, \& Environment, Chalmers University of Technology, Chalmersplatsen 4 412 96 Gothenburg, Sweden \\
$^{2}$ Hiroshima Astrophysical Science Center, Hiroshima University, 1-3-1 Kagamiyama, Higashi-Hiroshima, Hiroshima 739-8526, Japan \\
$^{3}$ National Astronomical Observatory of Japan, 2-21-1, Osawa, Mitaka, Tokyo, Japan \\
$^{4}$ Leiden Observatory, Leiden University, Niels Bohrweg 2, NL-2333 CA Leiden, Netherlands \\
$^{5}$ Center for Computational Astrophysics, Flatiron Institute, 162 5th Avenue, New York, NY 10010, USA \\
$^{6}$ Department of Astronomy, The University of Texas at Austin, Austin, TX 78712, USA \\
$^{7}$ Scuola Normale Superiore, Piazza dei Cavalieri 7, I-56126 Pisa, Italy \\
$^{8}$ Department of Physics, Graduate School of Science, Nagoya University, Nagoya, Aichi 464-8602, Japan \\
$^{9}$ Division of Physics, Faculty of Pure and Applied Sciences, University of Tsukuba, Tsukuba, Ibaraki 305-8571, Japan \\
$^{10}$ Tomonaga Center for the History of the Universe (TCHoU), Faculty of Pure and Applied Sciences, University of Tsukuba, Tsukuba, Ibaraki 305-8571, Japan \\
$^{11}$ Department of Physics, School of Advanced Science and Engineering, Faculty of Science and Engineering,\\\, Waseda University, 3-4-1, Okubo, Shinjuku, Tokyo 169-8555, Japan  \\
$^{12}$ Waseda Research Institute for Science and Engineering, Faculty of Science and Engineering, Waseda University, 3-4-1 Okubo, Shinjuku, Tokyo 169-8555, Japan  \\
$^{13}$ Institute for the Advancement of Higher Education, Hokkaido University, Kita 17 Nishi 8, Kita-ku, Sapporo, Hokkaido 060-0817, Japan \\
$^{14}$ Department of Cosmosciences, Graduate School of Science, Hokkaido University, Kita 10 Nishi 8, Kita-ku, Sapporo, Hokkaido 060-0810, Japan \\
$^{15}$ School of Physical Sciences, The Open University, Milton Keynes MK7 6AA, UK \\
$^{16}$ INAF-Osservatorio di Astrofisica e Scienza dello Spazio, via Gobetti 93/3, I-40129, Bologna, Italy \\
$^{17}$ School of Physics and Astronomy, Cardiff University, The Parade, Cardiff CF24 3AA, UK \\
$^{18}$ Center for Frontier Science, Chiba University, 1-33 Yayoi-cho, Inage-ku, Chiba 263-8522, Japan \\
$^{19}$ Graduate University for Advanced Studies (SOKENDAI), 2-21-1 Osawa, Mitaka, Tokyo 181-8588, Japan \\
$^{20}$ General Education Department, National Institute of Technology, Toba College, 1-1, Ikegami-cho, Toba, Mie 517-8501, Japan 
}

\date{Accepted 2024 June 27. Received 2024 June 27; in original form 2024 March 18}

\pubyear{2024}
\usepackage{newtxtext,newtxmath}
\begin{document}
\label{firstpage}
\pagerange{\pageref{firstpage}--\pageref{lastpage}}
\maketitle

\begin{abstract}
We present Atacama Large Millimetre/submillimetre Array (ALMA) observations of the \oiiil{} emission of a sample of thirteen galaxies at $z = 6$ to 7.6 selected as \cii{}-emitting companion sources of quasars. {\color{referee2}
To disentangle the origins of the luminous Oxygen line in the $z > 6$ Universe, we looked at emission-line galaxies that are selected through an excellent star-formation tracer \cii{} with star-formation rates between 9 and 162~$\rm M_{\odot}/yr$.}
Direct observations reveal \oiii{} emission in just a single galaxy (\oiiitocii{}$\,= 2.3$), and a stacked image shows no \oiii{} detection, providing deep upper limits on the \oiiitocii{} ratios in the $z > 6$ Universe (\oiiitocii{}$\,< 1.2$ at $3 \sigma$). {\color{referee2}
While the fidelity of this sample is high, no obvious optical/near-infrared counterpart is seen in the JWST imaging available for four galaxies. Additionally accounting for low-$z$ CO emitters, line stacking shows that our sample-wide result remains robust: The enhanced \oiiitocii{} reported} in the first billion years of the Universe is likely due to the selection towards bright, blue Lyman-break galaxies with high surface star-formation rates or young stellar populations. 
{\color{referee2} The deep upper limit on the rest-frame 90~\micron{} continuum emission ($< 141 \mu$Jy at $3 \sigma$), implies} a low average dust temperature ($T_\mathrm{dust} \lesssim 30\,$K) and high dust mass ($M_\mathrm{dust} \sim 10^8\,\mathrm{M}_\odot$). As more normal galaxies are explored in the early Universe, synergy between JWST and ALMA is fundamental to further investigate the ISM properties of the a broad range of samples of high-$z$ galaxies.
\end{abstract}




\section{Introduction}
Observations of distant galaxies by the \textit{James Webb Space Telescope (JWST)} have continued to find a surprisingly large population of $z > 10$ galaxies \citep{Adams2022a,Boylan-Kolchin2022,Castellano2022,Donnan2022,Finkelstein2022,morishita22,Naidu2022,Atek2022,Harikane2023,Yan2022}, building upon earlier work by \textit{Hubble} 
(e.g., \citealt{Bouwens2015,Oesch2018}). 
These rest-frame UV and optical observations imply a strong evolution of the cosmic star-formation rate density (CSFRD) by an order of magnitude within a very short time-scale ($\sim 200$~Myr) at $z > 6$, suggesting galaxies efficiently grow from primordial gas through first and second-generation stars \citep[e.g.,][]{Harikane2022}. 
Meanwhile, sub-mm observations with institutes such as the Atacama Large Millimetre/submillimetre Array (ALMA) have revealed dust in even lower-mass ($M_{\star} \approx 10^9 M_{\odot}$) systems at $z> 6$ \citep{Watson2015,Tamura2019,Tamura2023,Akins2022,Inami2022}, indicating that up to half of all star-formation may remain obscured from \textit{JWST} and \textit{HST} observations \citep{Zavala2021,Algera2023,Barrufet2023,Fujimoto2023}. {\color{referee} Formation of this dust requires environments where metals are rapidly formed and nucleated in supernovae and evolved stars. Subsequently, these small dust seeds can efficiently grow in dense regions of the inter-stellar medium (ISM).} In fact, these high dust masses actually provide a relatively strict constraint on the gas and stellar conditions in the early Universe, as {\color{referee} the observed stellar masses require efficient dust growth, and in particular little dust destruction, to explain the dust masses based on local dust (re)processing mechanisms \citep{Lesniewska2019,Dayal2022, DiCesare2023,Witstok2023,Sommovigo2022, Schneider2024}. }
As a consequence, recent studies have aimed to study the gas, dust and star-forming properties of distant galaxies to provide better constraints on galaxy evolution processes in the early Universe.

The majority of ALMA explorations of the ISM of normal star-forming galaxies beyond $z > 4.5$ have centred around the \ciil{} emission line, including several large programs (e.g., ALPINE; \citealt{LeFevre2020}, REBELS; \citealt{Bouwens2021Rebels}, CRISTAL, and ASPIRE). \cii{} is one of the dominant cooling lines of galaxies, and it is therefore often used to infer global star-formation rates, as confirmed out to $z \approx 6$ \citep{Schaerer2020,Faisst2020}. In the local Universe, \cii{} primarily originates from photo-dissociation regions (PDRs; roughly 70\%; \citealt{Stacey2010,Wolfire2022}), with the remainder from X-ray and cosmic ray dominated regions, ionised regions \citep{Meijerink2007,Wolfire2022}, low density warm gas, shocks \citep{Appleton2013} and/or diffuse \textsc{H\,i} clouds \citep[$< 30$~\%][]{Madden1997,Cormier2019}.
Its luminosity, even out to high redshifts, enables \cii{} to trace fainter features, such as outflows \citep{Ginolfi2019} and extended haloes surrounding galaxies in the early Universe \citep{Fujimoto2019,Fujimoto2021,Fudamoto2022}, although the origins of these haloes are still debated \citep[][and references therein]{Pizzati2020,Pizzati2023}. 
Attempts to observe \cii{} out to the highest redshifts ($z > 8$) have proven difficult \citep[e.g.,][]{Carniani2020}. Instead, emission from doubly-ionized Oxygen (\oiiil{}) has more often been detected as a bright emission line in the distant Universe \citep[e.g.,][]{Inoue2016,Hashimoto2018,Harikane2019,Tamura2019,Tamura2023}. This line is emitted from ionized (i.e., \hii{}) regions, and has been reported to be a dominant emission line also in low-metallicity dwarf galaxies in the local Universe \citep{Cormier2015,Cormier2019,Madden2018,Ura2023}. Subsequent studies of early galaxies have revealed Oxygen emission in around twenty sources beyond $z > 6$ \citep[][and references therein]{Witstok2022,Algera2023b}. Meanwhile, deep observations to detect \cii{} provide high luminosity ratios of Oxygen to Carbon emission (\oiiitocii{}~$> 3$; e.g., \citealt{carniani:2017oiii,Carniani2020,Harikane2019}), roughly one order of magnitude higher than observed in even metal-poor galaxies in the local Universe (\oiiitocii{}~$\approx 1$; e.g., \citealt{Madden2013,delooze14})\footnote{With the notable exception of POX 186 that has an \oiiitocii{}$\,\approx 10$; \citet{Cormier2015}} with a handful of low-mass systems reported up to \oiiitocii{}~$\approx 3$, indicating unique ISM conditions in the early Universe, although recent follow-up of more massive, evolved systems in the $z > 6$ Universe have indicated relatively lower ratios \citep[\oiiitocii{}\,$= 0.8 - 1.5$][]{Algera2023b}.

The origins of a high \oiiitocii{} at $z > 6$ are manifold. 
Galaxies at $z > 6$ are within one billion years from the birth of the Universe, and this short time means that the processes that govern galaxy evolution in the local Universe might be different. 
Similar to the argument that young Universe could restrict the galaxy evolutionary processes, the \oiiitocii{} line ratio could indicate that galaxies in the $z > 6$ Universe are intrinsically different from the populations found in the local Universe. 
Finally, observational constraints play an important role in high-$z$ galaxy studies, and could also influence the observed ISM conditions in the early Universe. 

These broad categorizations between the origins of an elevated \oiiitocii{} ratio are mirrored by models, finding a detailed set of phenomena that could underlie this high ratio.
Indeed, extensive photo-ionization studies indicate that the high \oiiitocii{} ratio can be reproduced through a combination of interlinked causes attributed to galaxy build-up arguments (low metallicity, {\color{referee} faster build-up of Oxygen relative to Carbon}, dust build-up, CMB attenuation effects), different galaxy populations (strong ionizing radiation fields, low PDR covering fractions, high-density gas, high Lyman-continuum escape fractions, top-heavy IMF), or observational constraints (spatially-extended \cii{} haloes, inclination effects, selection biases) \citep{Harikane2019,Carniani2020,Sugahara2021,Katz2022,Fujimoto2023}.

It is not surprising that galaxies emerging from primordial gas have a low metallicity (c.f., the observations of a surprisingly-enriched galaxy at $z = 10.6$, GN-z11; \citealt{Bunker2023}). Both models (e.g., \citealt{Vallini2015} and \citealt{Katz2019}) and observations (e.g., \citealt{Cormier2019}) find that the \ciil{} luminosity decreases with decreasing gas metallicity and, therefore, the lack of \cii{} detections in the $z>6$ Universe may indicate that these galaxies are very metal-poor systems.\footnote{Although the CLOUDY modeling reported in \cite{Harikane2019} does not indicate any effect of metallicity, their models assume high extinction ($\rm A_V$) within their clouds, allowing most 11.2~eV$ > E > $13.6~eV photons to interact with Carbon to form C$^+$.} The additional heating from the CMB could further affect the contrast against which the \cii{} and \oiii{} lines are observed, and provide an extra background emission for the excitation of Carbon and Oxygen \citep{Lagache2018,Laporte2019,Harikane2019}.
While Oxygen is typically produced by core-collapse supernovae, Carbon is primarily produced by later-stage AGB stars. As a consequence, the bulk of Oxygen is produced in the first 50 Myr, while it takes a billion years before Carbon reaches its expected chemical abundance \citep{Maiolino2019}. This causes a star-formation history dependence on the \oiiitocii{} ratio, which could be noticeable particularly at $z > 6$.
Meanwhile, both supernovae and AGB stars produce dust, which is an efficient drain for elements out of the gas phase \citep[e.g.,][]{DeCia2013} and thus reduces the line luminosities. Carbon is efficiently removed from the ISM through dust formation \citep{Konstantopoulou2022}, although the analysis of dust in distant Universe is in line with more silicate-rich dust in the high-$z$ Universe instead \citep[][; c.f., \citealt{Witstok2023b}]{Behrens2018,Ferrara2022}.

Changes to the ISM conditions in these galaxies could also drive a high \oiiitocii{} ratio.
For instance, strong interstellar radiation fields can deplete the C$^+$ ion abundance by turning Carbon into higher ionization states (e.g., C$^{++}$;  \citealt{Ferrara2019,Vallini2021}), which also creates an abundance of \oiii{} (O$^{++}$) in the ionized layer \citep{Arata2020,Harikane2019}. In the $z = 0 - 5$ Universe, the brightest dusty star-forming galaxies appear to be relatively sub-luminous in \cii{}, i.e., the so-called \cii{} deficit, which has been explained by thermal saturation \citep{Rybak2019}, although the effects of optical depth, dust extinction, high kinetic temperatures and a deficit of neutral gas cannot be ruled out \citep{Casey2014}. While the warm neutral medium appears to be the dominant region for \cii{} emission for nearby dwarf galaxies, star-bursting galaxy phases producing young \hii{} regions can dominate the emission, where the high ionization fields produce a similar \cii{} deficit \citep{Bisbas2022}. A denser ISM would also inhibit \cii{} emission. However, this effect does not appear to cause an elevated \oiiitocii{} ratio in direct models of galaxies \citep{Katz2022}, perhaps due to a similar decrease in \oiii{} emission because of its low electron critical density ($n_e \sim 500$~cm$^{-3}$). 
Since the \cii{} emitting regions mostly consist of neutral or atomic gas, a low \cii{} emission could originate from low gas masses \citep{Zanella2018,Aravena2024} or low gas surface densities ($\Sigma_{\rm gas}$; \citealt{Ferrara2017,Vallini2021,Vallini2024}).
Similarly, the low \cii{} emission could also be associated to a low (0~to~10 per cent) PDR covering fraction due to the compact size of high-$z$ galaxies or galactic outflows driving gas away \citep{Cormier2019,Harikane2019}. The latter scenario seems to be also supported by recent observational evidence revealing outflowing gas in star-forming galaxies at $z = 4 - 6$ \citep{Gallerani2018,Fujimoto2019,Fujimoto2021,Ginolfi2019,Sugahara2019,Spilker2020,Carniani2023,Xu2023}. The lower dark-matter halos of galaxies in the early Universe, prior to the bulk of their dark matter accretion, could struggle to prevent the outflows of such gas \citep{Arata2019}. 
Consequently, high \oiiitocii{} could be linked to Lyman Continuum emitters, which was shown through correlations in observations \citep{Ura2023} and in direct hydrodynamical simulations \citep{Katz2022}.

\begin{table*}
    \centering
    \caption{ \color{referee} Star-formation selected sample of quasar companion sources}
    \label{tab:sample}
    \begin{tabular}{lccccccc}
    \hline 
Source & RA & DEC & $z_\mathrm{[C\,II]}$ & SFR$_\mathrm{[C\,II]}$ & $f$ & $\Delta V_{\rm QSO}$ & EW \\
 & [hms] & [dms] & & [M$_{\odot}$/yr] & [0 - 1] & [km/s] & [km/s] \\ \hline
J0100+2802C1 & 01 00 14.04 & +28 02 17.4 & 6.324 & 75.8   $\pm$ 12.3 & 0.91 & 110 & $>$\textit{1160} \\
J0842+1218C1 & 08 42 28.97 & +12 18 55.0 & 6.065 & 161.4  $\pm$ 11.6 & 1 & 410 & 4300 \\
J0842+1218C2 & 08 42 29.67 & +12 18 46.3 & 6.064 & 38.3   $\pm$ 8.0 & 1 & 450 & $>$\textit{2050} \\
J1306+0356C1 & 13 06 08.33 & +03 56 26.2 & 6.034 & 114.2  $\pm$ 13.3 & 1 & 40 & 3220 \\
J1319+0950C1 & 13 19 11.65 & +09 50 38.2 & 6.050 & 80.8   $\pm$ 19.5 & 0.95 & 3300 & $>$\textit{1150} \\
J1342+0928C1 & 13 42 08.24 & +09 28 43.4 & 7.533 & 12.2   $\pm$ 3.7 & 1 & 240 & $>$\textit{1000} \\
J23183113C1 & 23 18 18.72 & -31 13 49.9 & 6.410 & 8.7    $\pm$ 1.9 & 1 & 1300 & $>$\textit{500} \\
J23183113C2 & 23 18 19.36 & -31 13 48.9 & 6.457 & 16.6   $\pm$ 4.9 & 0.99 & 580 & $>$\textit{460} \\
J23183029C1 & 23 18 33.36 & -30 29 44.5 & 6.122 & 84.1   $\pm$ 19.0 & 0.91 & 930 & $>$\textit{1900} \\
P036+03C1  & 02 26 00.80 & +03 03 02.4 & 6.460 & 36.2   $\pm$ 6.8 & 1 & 3200 & $>$\textit{620} \\
P183+05C1  & 12 12 26.32 & +05 05 29.6 & 6.435 & 39.9   $\pm$ 11.7 & 0.92 & 130 & $>$\textit{1320} \\
P183+05C2  & 12 12 28.07 & +05 05 34.6 & 6.894 & 29.0   $\pm$ 6.4 & 1 & 17000 & $>$\textit{430} \\
P23120C2 & 15 26 37.62 & -20 49 58.6 & 7.086 & 13.4   $\pm$ 2.2 & 1    & 19000 & $>$\textit{750} \\ \hline
\end{tabular}
    \raggedright \justify \vspace{-0.2cm}
\textbf{Notes:} 
Col. 1: Name of the companion source.
Col. 2 and 3: RA and DEC positions of the companion source.
Col. 4: The \cii{}-based redshift. 
Col. 5: The \cii{}-derived star-formation rate using the scaling relation from \citet{Schaerer2020}. There exist an additional 0.2 to 0.3~dex uncertainty on the absolute SFR estimates, which are not included in the reported uncertainties as our study is mostly interested in the relative SFRs between the sources.
Col. 6: The fidelity value (i.e., \textit{true positive} probability) derived by \citet{Venemans2020} by comparing the \textsc{FindClump} extractions of both the positive and negative maps.
Col. 7: Redshift difference to the quasar expressed as a velocity.
Col. 8: The equivalent width (EW) of the velocity-integrated spectral line flux relative to the associated dust continuum expressed in [km/s].
\end{table*}

The lack of \cii{} emission could also be linked with observational limitations. While the cause for the extended emission is still unclear (e.g., \citealt{Fujimoto2019}), extended -- and therefore missed -- emission of \cii{} could artificially boost the \oiiitocii{} \citep{Carniani2020}. For example, \citet{carniani:2017oiii} show that about 70 per cent of the diffuse \cii{} emission of BDF-3299, a star-forming galaxy at $z = 7$, is missed in ALMA observations with angular resolution of 0.3~arcsec, while the total emission is recovered in the data sets with a beam of 0.6~arcsec.
Similarly, a change in angle of incidence could affect the detectability of \cii{} relative to \oiii{}. \cite{Kohandel2019} found that the disc inclination particularly changes the line width of the \cii{} line, while the \oiii{} line width remains similar, which may cause higher observed ratios. Different kinematic and spatial distributions of the two spectral lines would then allow narrower lines (typically \oiii{}) to more easily push above the detection limit.

Counter to the abundant evidence of high \oiiitocii{} observed in the early Universe, hydrodynamical models of galaxies in the $z \sim 8$ Universe struggle to achieve similar values to the ones observed. For example, the suite of zoom-in simulations ``SERRA'' \citep{Pallottini2017_Chemistry,Pallottini2022} and zoom-in models ``SPHINX'' \citep{Katz2019,Katz2021} include galaxies up to a value of \oiiitocii{} $\approx 1$. By increasing the ionized gas fractions, requiring a top-heavy initial mass function (IMF), and increasing the oxygen-to-carbon abundance, models by \cite{Katz2022} are able to match the line ratios in observed high-redshift galaxies. Post-hoc photo-ionization models using \textsc{CLOUDY} \citep{Ferland2017} on the hydrodynamical models of \cite{Yajima2017} by \cite{Arata2020} and physically-motivated gas models \citep[][]{Vallini2021,Vallini2024} are also able to explain these high ratios in sources with high star-burstiness (i.e., upwards deviation from the star-forming main-sequence) and high star-formation rate surface densities. 

The strong variation between models, and their ability to characterize the observed ISM conditions of distant galaxies, provides a unique opportunity to benchmark our understanding of the first epochs of galaxy formation in the early Universe.
However, all existing $z > 6$ galaxies with both lines detected have been selected through their strong dependence on a tracer of on-going star-formation, either through {\it only}-obscured star-formation (in the case of SMGs) or through {\it only}-unobscured star-formation from UV-selected sources (e.g., \citealt{Harikane2019,Witstok2022}).
Such a pre-selection towards bursty star-forming galaxies could offset the \oiiitocii{} ratio, since many galaxies are selected with very blue UV spectra (i.e., Lyman-break galaxies; LBGs, selected with \textit{Hubble} and \textit{JWST}; \citealt{Sun_2023}){\color{referee}, although even the extreme LBGs appear to have metallicities similar to the typical metallicity in the $z = 6 - 8$ epoch \citep{Nakajima2023,Curti2024}.}
While the current sample of sources with both \oiii{} and \cii{} emission is well-constrained in their star-formation rate, they do not necessarily represent the behaviour of typical star-forming galaxies at this epoch, as they clearly represent a specific phase in their galactic evolution (i.e., SMGs trace a dusty starbursting phase, while UV-selected galaxies trace bursts of {\color{referee} unobscured} young stars).

An alternative approach would be to select sources based on their total star-formation rates. The ALMA Large Program ALPINE studied $4 < z < 6$ star-forming galaxies, finding agreement between the \cii{} luminosity and star-formation rate out to high redshift \citep{Schaerer2020,Harikane2019,Carniani2020,Faisst2022,Pallottini2022}. This agreement, in accordance with the local relation found by \cite{delooze14}, indicates that a purely star-formation selected sample can be constructed based on the \ciil{} luminosity. The main difficulty in selecting such a sample lies in the lack of deep spectroscopic observations that can find typical (10 to 100 M$_{\rm \odot}$/yr) galaxies at $z > 6$.

In this paper, we discuss the \oiii{} observations of a sample of galaxies found as likely \cii{} emitting companions to bright $z > 6$ quasars \citep{Venemans2020}. In Section~\ref{sec:sample}, we describe the sample and the observations. In Section~\ref{sec:results}, we report the results of direct observations, as well as a stack. We perform our analysis of this study in Section \ref{sec:Analysis} including an evaluation of false positives and biases originating from selecting emission line galaxies in the same field as quasars. We discuss the effect on the \oiiitocii{} ratio in Section~\ref{sec:OiiiCiiDiscussion}. We conclude in Section~\ref{sec:conclusions}. Throughout this paper, we assume a flat $\Lambda$-CDM cosmology with the best-fit parameters derived from the \textit{Planck} results \citep{Planck2020,Planck2021}, which are $\Omega_\mathrm{m} = 0.315$, $\Omega_\mathrm{\Lambda} = 0.685$ and $h = 0.674$. We report dust temperatures as they would be in a CMB background at $z = 0$. 

\section{Sample and Observations}
\label{sec:sample}

\subsection{Quasar companion galaxy sample}
In an effort to create a SFR-selected sample, we use the catalogue reported in \cite{Venemans2020}. They revealed a population of line-emitting galaxies discovered in same field as the quasars they were targeting in the $6 < z < 7.6$ Universe. Because of the high fidelity of their ALMA observations, many of these sources are found to emit what is most-likely \ciil{}. 
Unlike any previous $z>4$ selection, these targets are selected solely by their emission line, and thus if they are true \cii{} emitters, represent a \ciil{}-selected sample. 
As a consequence, this selection addresses a problem revealed by studies of high-redshift galaxies, namely that their selection strongly influences the population of galaxies being traced; i.e., sub-mm selected sources are typically dust extincted, while rest-frame UV observations trace young {\color{referee} unobscured} stellar populations.
Instead of directly tracing unobscured starlight or obscured starlight by means of dust, the sample analyzed in this work directly traces the total SFR in a way that is little or un-affected by dust obscuration \citep{Schaerer2020,Harikane2019,Carniani2020,Faisst2022,Pallottini2022}. 
Adopting the \cii{}-SFR relation from \citet{Schaerer2020}, which was determined for the ALPINE sample at slightly lower redshift, our companion galaxies  (see Table~\ref{tab:sample}) span a range of $9 \lesssim \mathrm{SFR}_\text{\cii{}} / (M_\odot\,\mathrm{yr}^{-1}) \lesssim 162$, with a mean (median) SFR of $55\,M_\odot\,\mathrm{yr}^{-1}$ ($38\,M_\odot\,\mathrm{yr}^{-1}$). Although the \cite{Schaerer2020} relation is similarly-tight as the local relation \citep{delooze14} and follows predictions from phenomenological \citep{Lagache2018} and hydrodynamical models \citep{Pallottini2022}, there still exist uncertainties in the SFR estimates from the infrared and UV emission on the ALPINE sources themselves \citep{Sommovigo2022}, and the sources that remained undetected in \cii{} or dust-continuum in ALPINE could affect the scatter on the relation we adopt. Some studies further indicate that the \cii{}-to-SFR relation could deviate at higher redshift \citep[$z > 6$;][]{Harikane2019}. These caveats increase the systematic uncertainties on the star-formation rates with around 0.2 to 0.3~dex, although we note for the purposes of this study, our interest is mostly in the relative star-formation rates which are less affected by this systematic uncertainty. 
We evaluate the sample completion in Section~\ref{sec:sampleCompleteness} and potential biases as a result of selecting \cii{}-bright galaxies around known quasars in Section~\ref{sec:biases}.

\begin{figure}
    \centering
    \includegraphics[width=0.9\linewidth]{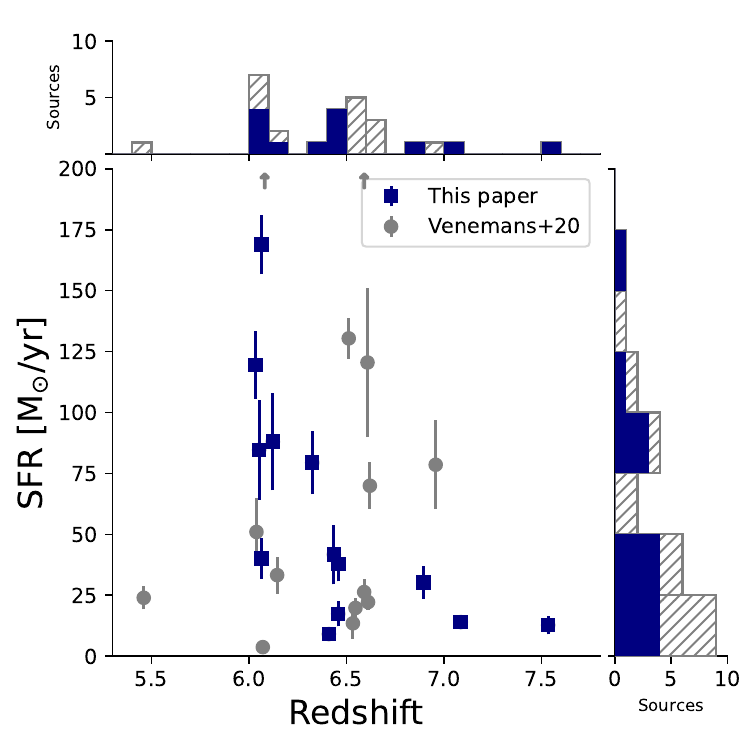}
    \caption{The star-formation rate selected sample of galaxies found in quasar fields, originally identified by \citet{Venemans2020}. We select thirteen sources with redshifts that allow for the efficient observation of the \oiii{} emission ({\it blue squares}). The two arrows indicate sources with \cii{}-based star-formation rates in excess of $> 200$~M$_{\odot}$/yr. Uniquely, the sample is representative of quasar companion galaxies in general, with no prior selection towards higher star-formation rates, and is based solely on the atmospheric windows where \oiii{} would be observable with ALMA. }
    \label{fig:sampleDescription}
\end{figure}
Out of the 27 companion sources reported in \citet{Venemans2020}, a total of 13 fall into a favorable redshift range that enables ground-based \oiii{} observations with ALMA if the emission line is \cii{}. As shown in Figure~\ref{fig:sampleDescription}, the star-formation rates of these 13 sources match the range of SFRs seen for quasar companion galaxies. The observations were designed to investigate their \oiiitocii{} luminosity ratios, and 
we aimed for a $5\sigma$ detection of a line ratio of 2.5 at the high star-formation rates which represents a deep survey to evaluate the line ratio where most of the $z > 6$ galaxies are detected. At lower SFR ($< 80$~M$_{\odot}$/yr), our sensitivity limit to the line ratio increases to \oiiitocii{}$ = 5$ at $5 \sigma$. The \oiiil{} line of four quasars also falls within the primary beam and bandwidth of our observations, and will be presented in a subsequent paper (Algera \& Bakx et al.\ in prep.).

\subsection{ALMA observations} 
Using ALMA Band\,8 observations, we target the \oiii{} line of our 13 targets between 390~to~490~GHz. In total, our observations comprise of 11 separate pointings near 10 different quasars. A summary of the observations is presented in Table~\ref{tab:observationdetails}. The companion sources remain mostly unresolved between 0.15 and 0.3 arcsec resolution in \cii{} emission, and for the best possible comparison, we hence matched the observed resolution of the \cii{} emission from the earlier \cite{Venemans2020} observations, equating to beam sizes between roughly 0.2 and 0.4~arcsec, slightly varying from source to source. The slightly larger range in angular resolutions compared to the \cii{} observations was chosen to facilitate scheduling constraints, especially since our sample spans a large range in RA across the sky.

The quasars J0423-0120, J2253+1608, J0854+2006, J1058+0133, J1256-0547, J1229+0203, J1924-2914, and J1517-2422 were used as bandpass and flux calibrators, and quasars J0112+2244, J0217+0144, J0840+1312, J0831+0429, J1229+0203, J1224+0330, J1332+0200, J1254+1141, J2258-2758, J1507-1652, and J1347+1217 were used as phase calibrators.

Data reduction was performed following the standard calibration procedure and using the ALMA pipeline. Then, we use \textsc{CASA} \citep{CASATEAM2022} for imaging the {\it uv}-visibilities using Briggs weighting with a robust parameter of 2.0 to maximize the depth of the observations at the expense of slightly increasing the final synthesized beam size. The resulting beam sizes range between  $0\farcs2$ to $0\farcs4$ at a depth between 0.38 to 8~mJy/beam in 35~km/s bins (see Table \ref{tab:observationdetails}).

\begin{table*}
    \centering
    \caption{Parameters of the ALMA observations}
    \label{tab:observationdetails}
    \begin{tabular}{lcccccccc} \hline
Source & UTC start time          & Baseline length & N$_{\rm ant}$ &  Frequency & T$_{\rm int}$ & PWV & $\sigma_{\rm 35 km/s}$ &  Beam\\
 & $[$YYYY-MM-DD hh:mm:ss$]$  & [m] 			 & 			 	 &  [GHz]     & [min] &  [mm] & [mJy]& [$" \times{} "$] \\ \hline
J0100+2802C1 & 2022-06-11 14:41:04 & 15 -- 1213 & 40 & 462.346 & 87.15 & 0.35 & 0.65 & 0.33 $\times{}$ 0.25 \\
& 2022-06-11 12:52:17 & 15 -- 1213 & 43 & 462.346 & 87.28 & 0.37 \\
& 2022-06-11 11:24:04 & 15 -- 1213 & 43 & 462.346 &	87.62 & 0.38 \\
J0842+1218C1+C2 & 2022-06-09 20:52:35 & 15 -- 783\enspace & 39 & 479.326	& 85.04	& 0.36 & 1.11 / 8.36 & 0.32 $\times{}$ 0.25 \\
& 2022-05-26 00:23:58 & 15 -- 783\enspace & 41 & 479.326 & 87.21	& 0.42 \\
& 2022-01-05 08:37:04 & 14 -- 783\enspace & 39 & 479.326 	& 84.85	& 0.45 \\
J1306+0356C1 & 2022-05-18 04:07:47 & 15 -- 740\enspace & 44 & 481.449 & 	35.06 & 0.61 & 3.00 & 0.38 $\times{}$ 0.31 \\
J1319+0950C1 & 2022-05-26 01:53:31 & 15 -- 783\enspace & 46 & 480.229	 & 74.55 & 0.50 & 0.73 & 0.3 $\times{}$ 0.29 \\
& 2022-05-25 04:39:00	& 15 -- 783\enspace & 46 & 480.229 &  74.20	 & 0.45 \\
J1342+0928C1 & 2022-06-11 04:35:13	& 15 -- 1213 & 42 & 396.667 & 103.97 & 0.41 & 0.41 & 0.38 $\times{}$ 0.29 \\
& 2022-06-11 02:51:11	& 15 -- 1213 & 41 & 396.667 & 104.48 & 0.46 \\
J23183113C1+C2 & 2022-05-26 12:24:01 & 15 -- 783\enspace &  46 &  457.450 	& 92.81	& 0.59 & 0.66 / 1.28 & 0.32 $\times{}$ 0.27 \\
& 2022-05-25 10:44:26 & 15 -- 783\enspace & 43 & 457.450 & 91.82 & 0.67 \\
J23183029C1 & 2022-05-15 13:16:47	& 15 -- 680\enspace & 40 & 477.303 & 79.32 & 0.32 \\
& 2022-05-15 11:40:24	& 15 -- 740\enspace &  41 & 477.303 & 77.88	& 0.35 \\
P036+03C1 & 2022-06-13 12:22:12	& 15 -- 1301 & 42 & 453.939 & 66.30 & 0.19  & 0.82 & 0.26 $\times{}$ 0.18 \\
P183+05C1 & 2022-06-09 22:30:25	& 15 -- 783\enspace & 43 & 455.422 &  83.77	& 0.47 & 0.53  & 0.31 $\times{}$ 0.27\\
& 2022-05-26 03:38:39 & 15 -- 783\enspace & 40 &	455.422 & 84.36	& 0.28 \\
P183+05C2 & 2022-05-23 02:41:49 & 15 -- 783\enspace & 47 &	428.891 & 85.45	& 0.64 & 0.45 & 0.34 $\times{}$ 0.28 \\
& 2022-01-01 11:53:25 & 14 -- 783\enspace & 40 & 428.891 & 	85.64 & 0.72 \\
P23120C2 & 2022-05-21 05:41:19 & 15 -- 783\enspace & 46 & 418.697	& 85.73 & 0.99 & 0.38 & 0.35 $\times{}$ 0.31 \\
& 2022-05-18 06:14:41 & 15 -- 740\enspace &	44 & 418.697 & 85.96 & 0.55 \\ \hline
    \end{tabular}
    \raggedright \justify \vspace{-0.2cm}
\textbf{Notes:} 
Col. 1: Source name.
Col. 2: UTC start time of the observations.
Col. 3: Length between the nearest and furthest antennae.
Col. 4: Number of antennae participating in the observations.
Col. 5: The Local Oscillator frequency.
Col. 6: The total observation time including overheads.
Col. 7: The precipitable water vapor during the observations.
Col. 8: The $1 \sigma$ standard deviation in 35~km/s bins of the observations centred on the \oiii{} frequency. Observations with multiple companions are denoted with a slash (/), where the first value corresponds to the first companion.
Col. 9: The beam size at the \oiii{} frequency, which does not vary appreciably for sources with multiple companions.
\end{table*}

\section{Results}
\label{sec:results}

\begin{table*}
    
    \caption{Emission line properties of our companion sources.}
    \label{tab:emission_line_parameters}
    
    
    \begin{tabular}{lccccccccc}
        \hline
        Source & $S_\mathrm{[C\,II]}$ & $\mathrm{FWHM}_\mathrm{[C\,II]}$ & $L_\mathrm{[C\,II]}$ & $S_\mathrm{[O\,III]}$ & $\mathrm{FWHM}_\mathrm{[O\,III]}$ & $L_\mathrm{[O\,III]}$ & $L_\mathrm{[O\,III]} / L_\mathrm{[C\,II]}$ & $S_{\rm 158 \mu m}$ & $S_{\rm 88 \mu m}$\\
    \hline 
    &  $\mathrm{Jy\,km\,s}^{-1}$ & $\mathrm{km\,s}^{-1}$ & $10^8 L_\odot$ & $\mathrm{Jy\,km\,s}^{-1}$ & $\mathrm{km\,s}^{-1}$ & $10^8 L_\odot$ & &  [mJy] & [mJy] \\
    \hline 
    
J0100+2802C1 & $0.80 \pm 0.13$ & $435 \pm 78\enspace$  & \enspace8.5 $\pm$ 1.4 & \textit{ $<$ 0.27} & \textit{435} & \textit{6.0} &  \textit{0.7} & \textit{0.69} & \textit{0.52} \\

J0842+1218C1 & $1.81 \pm 0.13$ & $331 \pm 26\enspace$  & 18.1 $\pm$ 1.3 & 1.77$_{-0.31}^{+0.35}$  & 153$_{-35}^{+48}$ & $40 \pm 10$ & $2.3 \pm 0.6$ & $0.42 \pm 0.05$ & $1.16 \pm 0.31$ \\

J0842+1218C2 & $0.43 \pm 0.09$ & $268 \pm 62\enspace$  & \enspace4.3 $\pm$ 0.9 &  \textit{ $<$ 5.87} & \textit{268} & \textit{58.5} & \textit{14.2} & \textit{0.21} & \textit{7.88} \\

J1306+0356C1 & $1.29 \pm 0.15$ & $200 \pm 26\enspace$  & 12.8 $\pm$ 1.5 & \textit{ $<$ 0.51} & \textit{200} & \textit{19.6} & \textit{1.6} & $0.40 \pm 0.07$ & \textit{1.21} \\

J1319+0950C1 & $0.91 \pm 0.22$ & $490 \pm 127$ & \enspace9.1 $\pm$ 2.2 & \textit{ $<$ 0.34} & \textit{490} & \textit{8.3} & \textit{1.0} & \textit{0.79} & \textit{0.61} \\

J1342+0928C1 & $0.10 \pm 0.03$ & $222 \pm 75\enspace$  & \enspace1.4 $\pm$ 0.4 & \textit{ $<$ 0.74} & \textit{222} & \textit{3.5} & \textit{2.7} & \textit{0.10} & \textit{0.28} \\

J23183113C1 & $0.09 \pm 0.02$ & $102\pm26\enspace$  & \enspace0.9 $\pm$ 0.2 & \textit{ $<$ 0.94} & \textit{102} & \textit{4.0} & \textit{4.3} & \textit{0.18} & \textit{0.66} \\

J23183113C2 & $0.17 \pm 0.05$ & $111 \pm 35\enspace$  & \enspace1.8 $\pm$ 0.5 &  \textit{ $<$ 0.31} &  \textit{111} & \textit{5.8}  & \textit{3.3} & \textit{0.37} & \textit{0.75} \\

J23183029C1 & $0.93 \pm 0.21$ & $427\pm104$   & 10.2 $\pm$ 2.3 & \textit{ $<$ 1.84} & \textit{427} & \textit{8.4} & \textit{0.9} & \textit{0.49} & \textit{0.38} \\

P036+03C1  & $0.37 \pm 0.07$ & $103 \pm 23\enspace$  & \enspace4.1 $\pm$ 0.8 & \textit{ $<$ 0.47} & \textit{103} & \textit{4.3} & \textit{1.1} & \textit{0.60} & \textit{0.58} \\ 

P183+05C1  & $0.41 \pm 0.12$ & $373\pm121$   & \enspace4.5 $\pm$ 1.3 & \textit{ $<$ 0.41} & \textit{373} & \textit{4.6} & \textit{1.1} & \textit{0.31} & \textit{0.49} \\ 

P183+05C2  & $0.27 \pm 0.06$ & $155 \pm 35\enspace$  & \enspace3.2 $\pm$ 0.7 & \textit{ $<$ 0.39} & \textit{155} & \textit{3.8} & \textit{1.2} & \textit{0.63} & \textit{0.32} \\

P23120C2 & $0.12 \pm 0.02$ & $209 \pm 45\enspace$  & \enspace1.5 $\pm$ 0.3 & \textit{ $<$ 0.78} & \textit{209} & \textit{3.9} & \textit{2.7} & \textit{0.16} & \textit{0.33} \\

    \hline 
    \end{tabular}
    \raggedright \justify \vspace{-0.2cm}
\textbf{Notes:} 
Col. 1: Name of the companion source.
Col. 2 \& 5: The velocity-integrated flux of the \cii{} and \oiii{} line, resp.
Col. 3 \& 6: The full-width at half-maximum of the \cii{} and \oiii{} line, resp.
Col. 4 \& 7: The line luminosity of the \cii{} line and \oiii{} line, resp.
Col. 8: The \oiiitocii{} luminosity ratio. 
Col. 9 \& 10: The continuum flux densities at rest-frame $158\,\mu$m and $88\,\mu$m, resp.
The values in columns 2 and 3 are taken from \citet{Venemans2020}. For sources where the \oiii{} line is not detected, italics indicate $3 \sigma$ upper limits on the flux and luminosity, and fiducial values for the velocities. Continuum flux densities in italics also correspond to $3\sigma$ upper limits.
\end{table*}

\begin{figure*}
    \centering
    \includegraphics[width=0.24\textwidth]{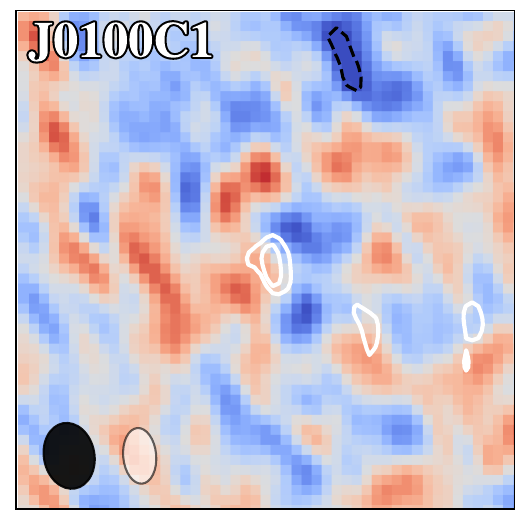}
    \includegraphics[width=0.24\textwidth]{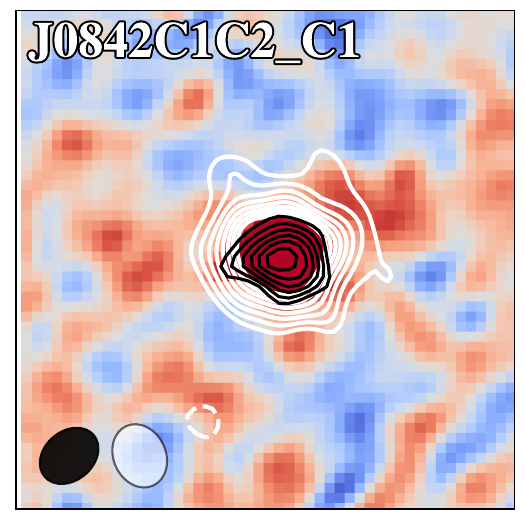}
    \includegraphics[width=0.24\textwidth]{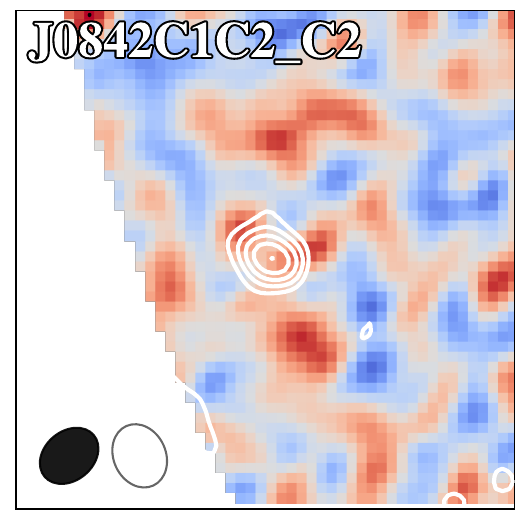}
    \includegraphics[width=0.24\textwidth]{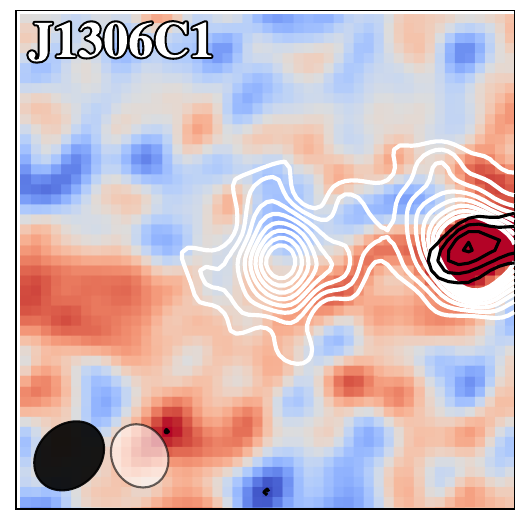}
    \includegraphics[width=0.24\textwidth]{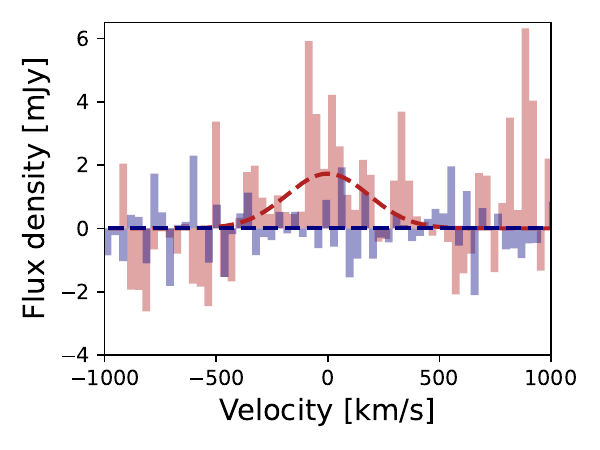}
    \includegraphics[width=0.24\textwidth]{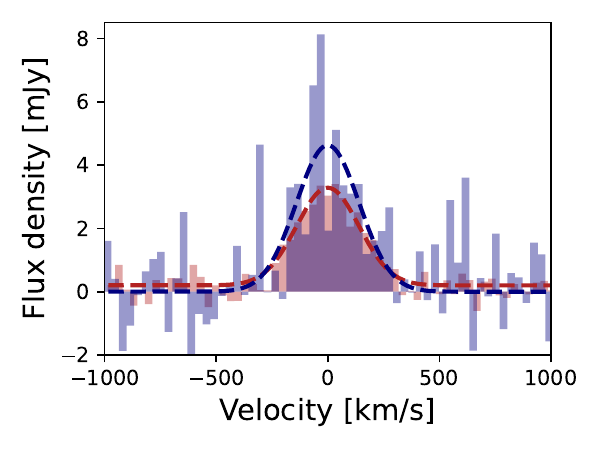}
    \includegraphics[width=0.24\textwidth]{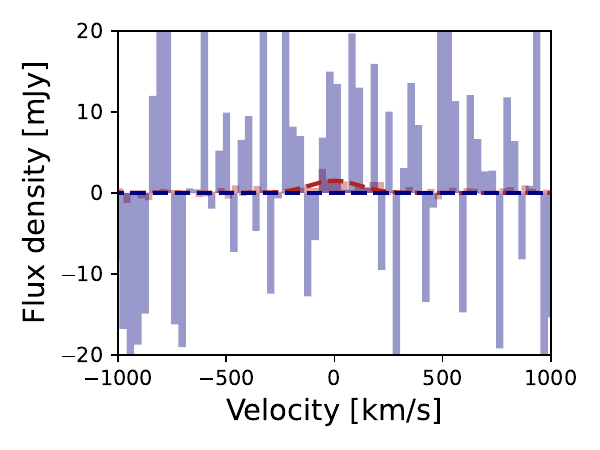}
    \includegraphics[width=0.24\textwidth]{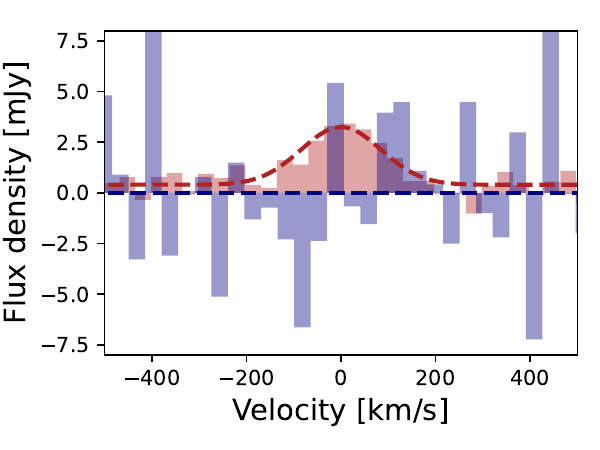}
    \includegraphics[width=0.24\textwidth]{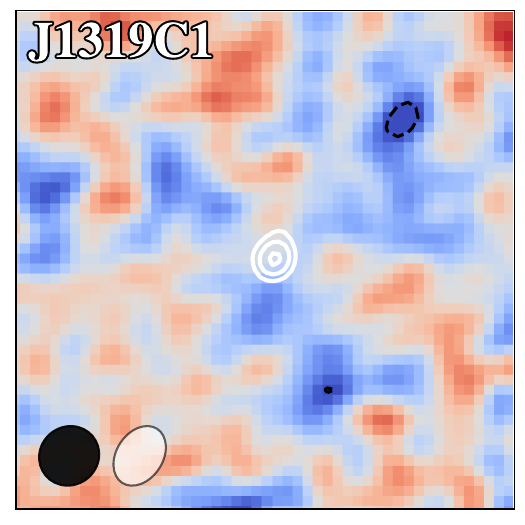}
    \includegraphics[width=0.24\textwidth]{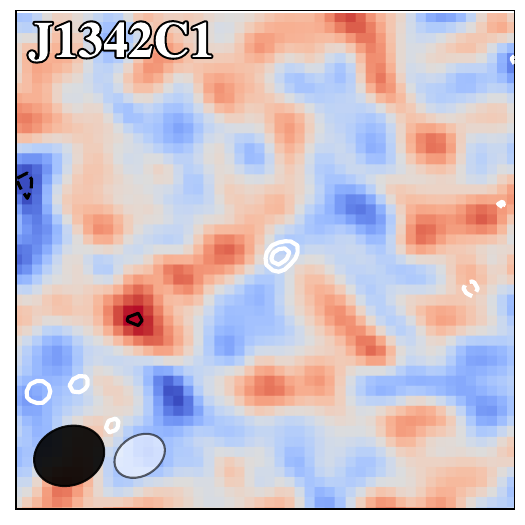}
    \includegraphics[width=0.24\textwidth]{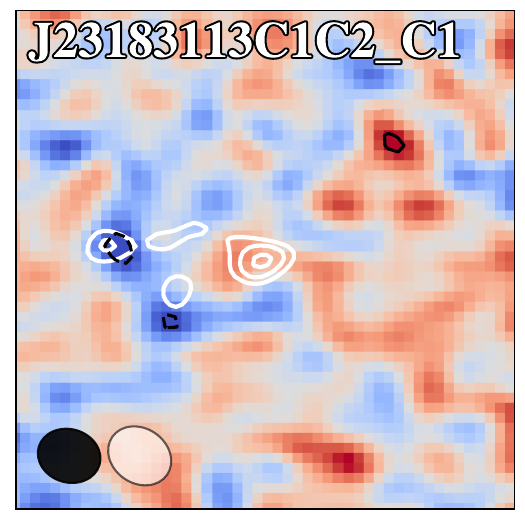}
    \includegraphics[width=0.24\textwidth]{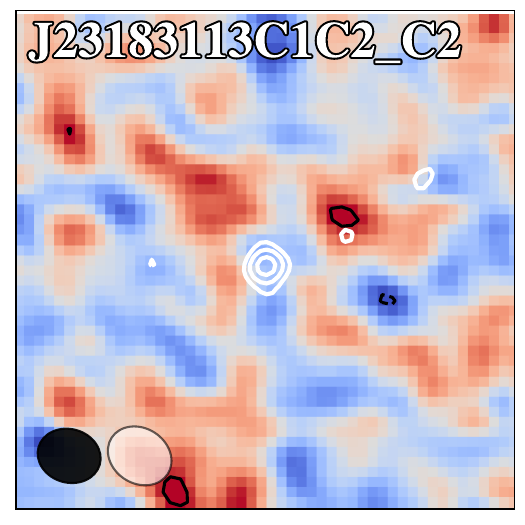}
    \includegraphics[width=0.24\textwidth]{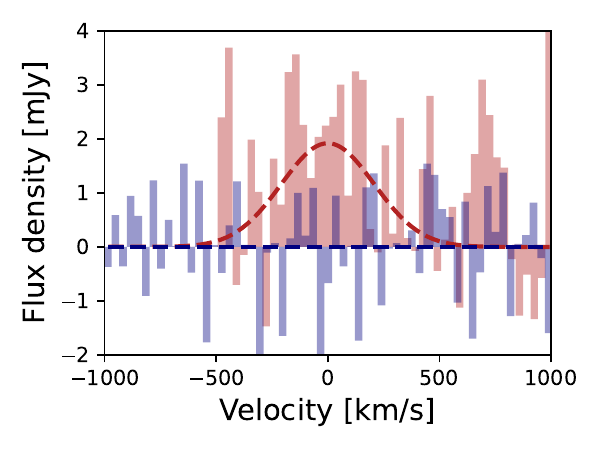}
    \includegraphics[width=0.24\textwidth]{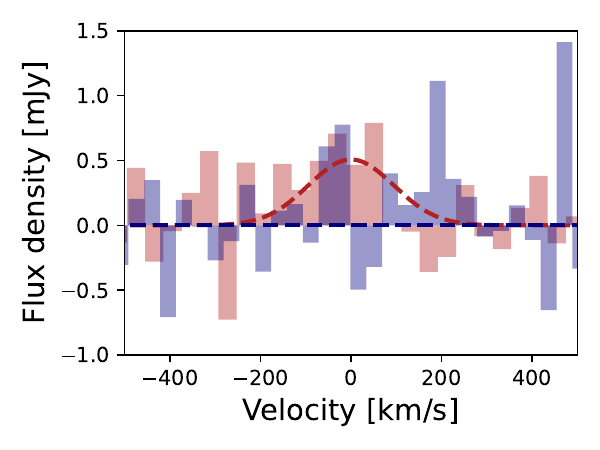} 
    \includegraphics[width=0.24\textwidth]{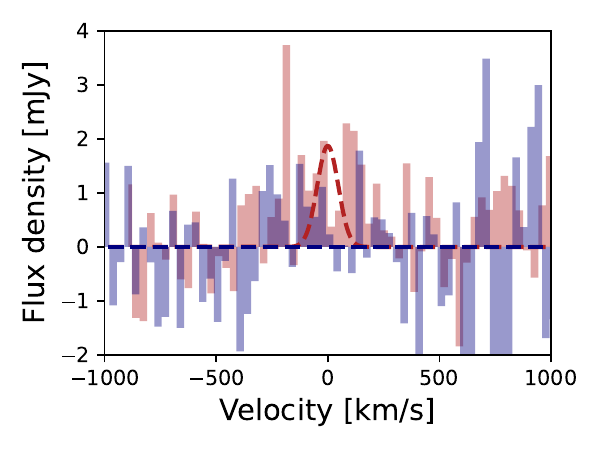} 
    \includegraphics[width=0.24\textwidth]{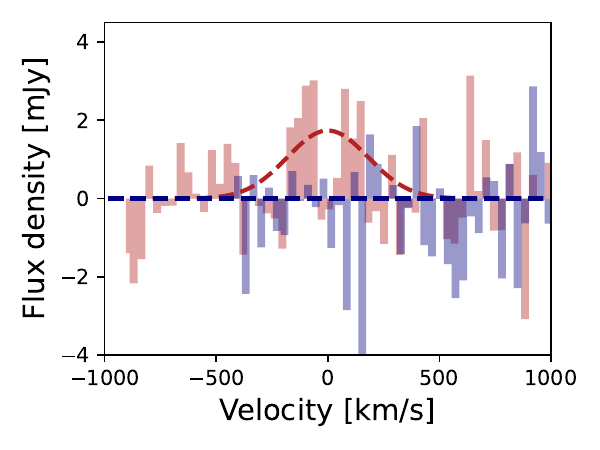}    
    \caption{{\it Top panel per source:} Moment-0 maps for the companion sources ($2\farcs5\times 2\farcs5$). Background: \oiii{}, contours: \cii{} (white) and \oiii{} (black), drawn at $3 - 10 \sigma$ in steps of $1\sigma$, where $\sigma$ is the RMS noise in the moment-0 map. Dashed contours indicate negative emission, and the colorscale runs from $-3\sigma$ to $+3\sigma$. {\it Bottom panel per source:} The aperture-extracted line emission shown for \cii{} in {\it red}, and the \oiii{} emission in {\it blue}. Line fits are shown in the same colour. Note the presence of the quasar in the panel labeled J1306C1, which will be discussed in Algera \& Bakx et al.\ in prep. }
    \label{fig:enter-label}
\end{figure*}\addtocounter{figure}{-1}
\begin{figure*}
    \centering
    \includegraphics[width=0.24\textwidth]{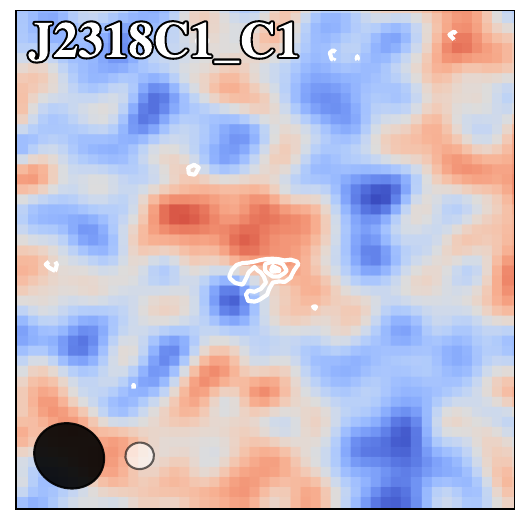}
    \includegraphics[width=0.24\textwidth]{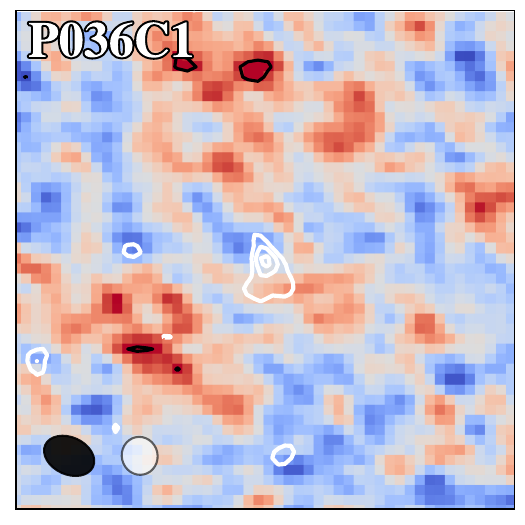}
    \includegraphics[width=0.24\textwidth]{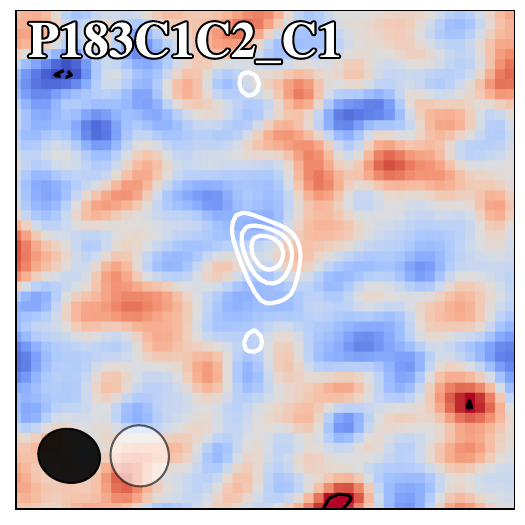}
    \includegraphics[width=0.24\textwidth]{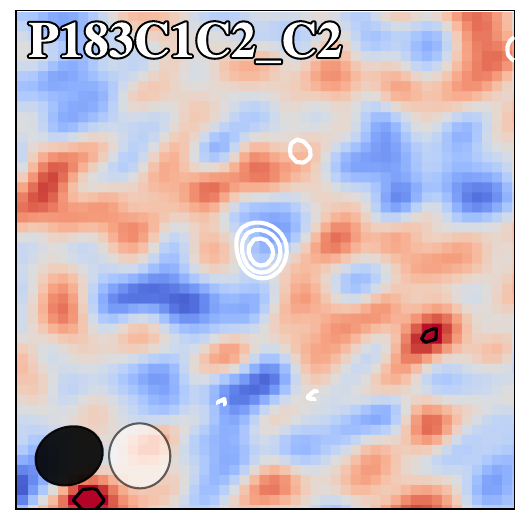}
    \includegraphics[width=0.24\textwidth]{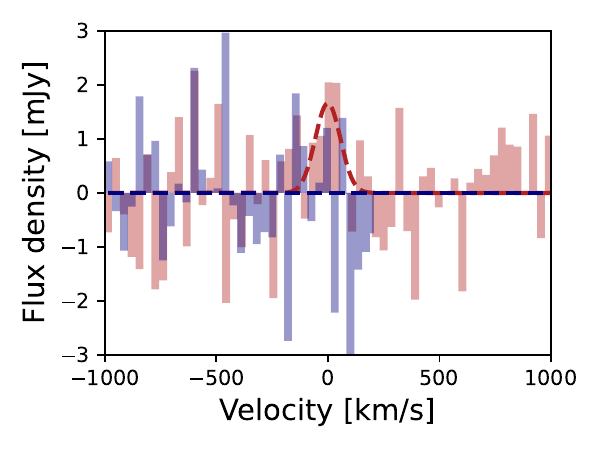}
    \includegraphics[width=0.24\textwidth]{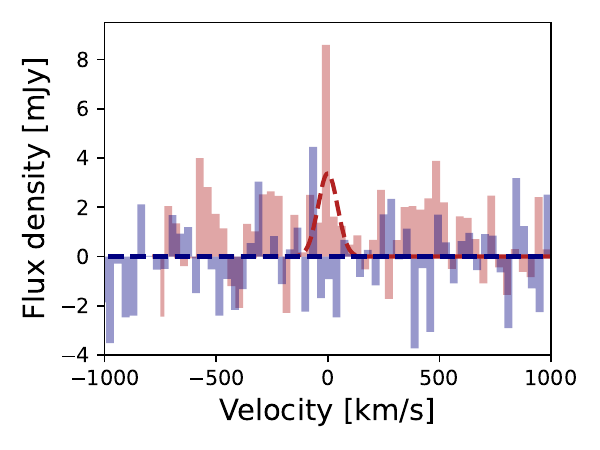}
    \includegraphics[width=0.24\textwidth]{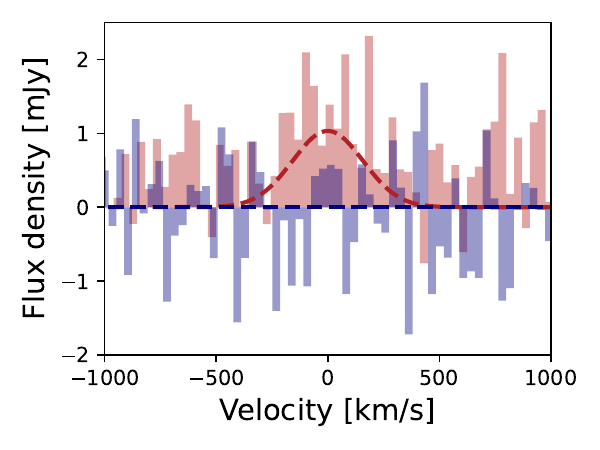}
    \includegraphics[width=0.24\textwidth]{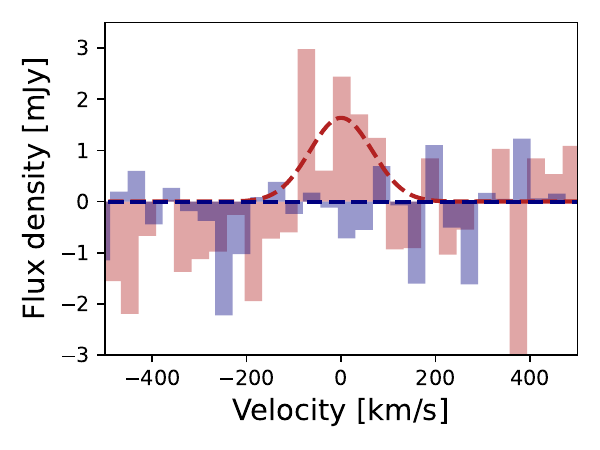}
    \includegraphics[width=0.24\textwidth]{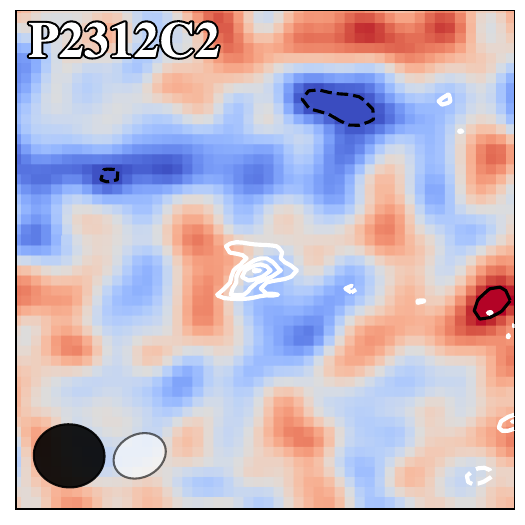}
    \includegraphics[width=0.24\textwidth]{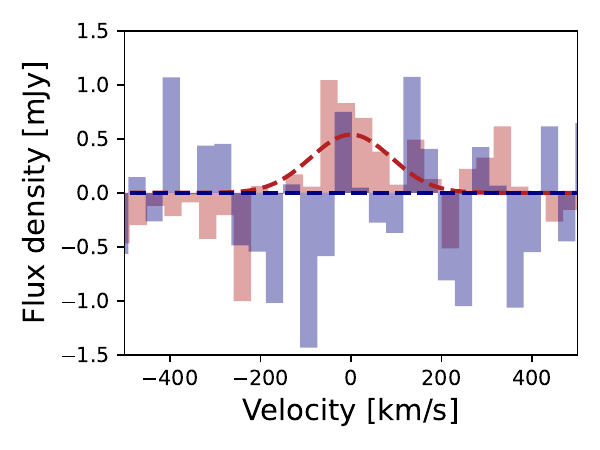}
    \caption{{\it Continued from the previous page.}} 
    \label{fig:moment0}
\end{figure*}

\subsection{Line emission}
Based on the known positions of the \cii{}-emitting companion galaxies, we extract their 1D spectra across the Band 8 datacubes in $0\farcs3$ apertures, after having explored a variety of aperture radii. In addition, we create moment-0 maps around the expected frequency center of \oiii{} across  $1.2\times$ the known \cii{} FWHM from \citet{Venemans2020}. This velocity range maximizes the S/N of the line, assuming a Gaussian profile (e.g., \citealt{Novak2019}). 

We find that a single source, J0842C1, is robustly detected in \oiii{} emission, while the remainder of our sample is not detected (Fig.\ \ref{fig:moment0} and Table~\ref{tab:emission_line_parameters}). We further test the existence of extended emission by evaluating maps tapered at $1"$, but find no additional sources with emission. We determine the line flux of the single detection through fitting its extracted 1D spectrum with a Gaussian, while we place upper limits on the \oiii{} fluxes for the rest of the sample. To this end, we adopt $3$ times the noise in the moment-0 maps as our $3\sigma$ upper limit on the line flux (in units of $\mathrm{Jy\,km/s}$), whereby we implicitly assume that the \oiii{} and \cii{} emission have similar intrinsic line widths, and that the \oiii{} emission is spatially unresolved. Conservatively, we further multiply the upper limits by a factor of $1 / 0.84$ to account for the fact that our moment-0 maps were collapsed across $1.2\times$ the expected line FWHM, which includes only $\approx84\%$ of the total flux assuming a Gaussian profile.

\subsection{Dust continuum emission}
\label{sec:dustContinuum}

We use the Band 8 continuum images to search for dust emission emanating from the targeted quasar companion galaxies. We show the rest-frame $\sim90\,\mu$m and $\sim160\,\mu$m continuum emission of our targets in Fig. \ref{fig:continuum} in the Appendix. Our single \oiii{}-detected galaxy, J0842C1, is also detected in dust continuum emission in both Bands 6 and 8, while the remainder of our sample does not show any rest-frame $90\,\mu$m continuum emission at the $> 3\sigma$ level. One further companion galaxy, J1306C1, is robustly continuum-detected at rest-frame $160\,\mu$m. Note that no previous selection towards dust continuum detections was made, and instead only sources with good atmospheric transmission at their \oiii{} frequencies were selected in this study.

We measure the continuum flux densities of our targets in both bands using a $0\farcs5$ aperture. For those that are not continuum-detected, we adopt $3\times$ the error on the aperture flux as the corresponding upper limits instead.\footnote{These upper limits are more conservative than adopting thrice the RMS in the continuum maps as our $3\sigma$ upper limits, as the latter implicitly assume our targets are spatially unresolved.}

\subsection{Stacked line and continuum emission}
\label{sec:stacking}

Given that most of our sample remains undetected in both \oiii{} and underlying continuum emission, we turn to stacking to assess the average level of line and continuum emission of SFR-selected galaxies in the epoch of reionization.

We explore several weighting schemes in our stacking experiments, including variance-weighted, SFR-weighted, and \cii{} observing depth-based. Although there is an argument to be made for star-formation rate based weighting, such a stack overly represents the brightest sources, while the main goal of our observations is to probe also the lower-SFR ($\sim 10$~M$_{\odot}\,\mathrm{yr}^{-1}$) regime. Instead, we opt for a simpler inverse-variance weighting, based solely on the noise in our \cii{} and \oiii{} data cubes. For completeness, we provide the SFR-weighted stack in Appendix \ref{app:stackingBasedOnSFR}.

We stack both the moment-0 maps of \cii{} and \oiii{} and the continuum maps in a similar way, in order to obtain a measurement of the typical \oiiitocii{} ratio and dust continuum ratio of SFR-selected galaxies, respectively. We include galaxies with detected emission lines and dust continua in the stacks to avoid biasing our results by treating detections and upper limits separately, although our results do not change if we leave these sources out. 

Given that the adopted sensitivity of our Band 8 observations was tuned based on the known $L_\text{\cii{}}$ of the quasar companion sources in \citet{Venemans2020}, the typical RMS noise in our \oiii{} datacubes is lower for the \cii{}-fainter sources. As such, performing a simple inverse-variance-weighted stack in both bands individually would bias our results towards up-weighting the deeper cubes likely containing \oiii{}-faint systems, and hence towards a lower average \oiiitocii{}. To circumvent this bias, we weight our $i^\mathrm{th}$ source by $w_i = 1 / \max(\sigma_{i,\mathrm{B6}}^2, \sigma_{i,\mathrm{B8}}^2)$, that is, by the maximum of the noise levels in the $i^\mathrm{th}$ Band 6 or Band 8 continua and moment-0 maps. This ensures that we weigh the same source equally in both bands, while still benefiting from downweighting the noisiest data compared to a simple, unweighted average. 

\begin{figure}
    \centering
    \includegraphics[width=0.45\textwidth]{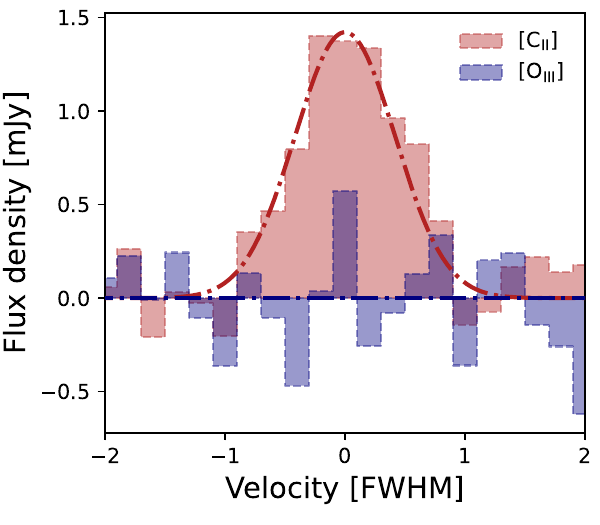}
    \includegraphics[width=0.23\textwidth]{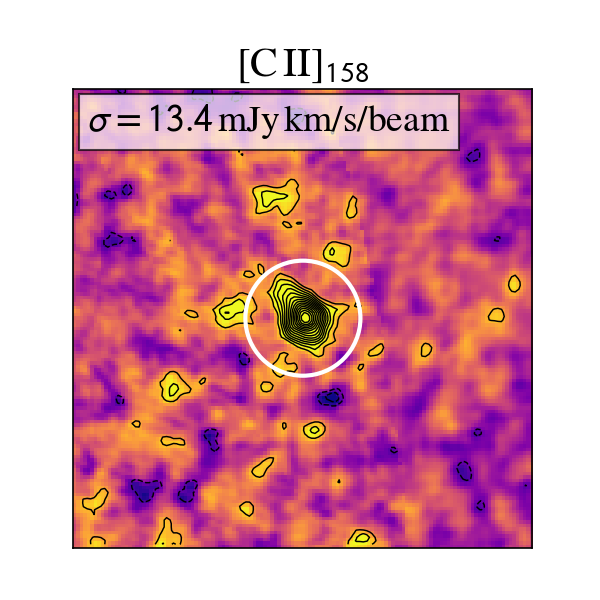} %
    \includegraphics[width=0.23\textwidth]{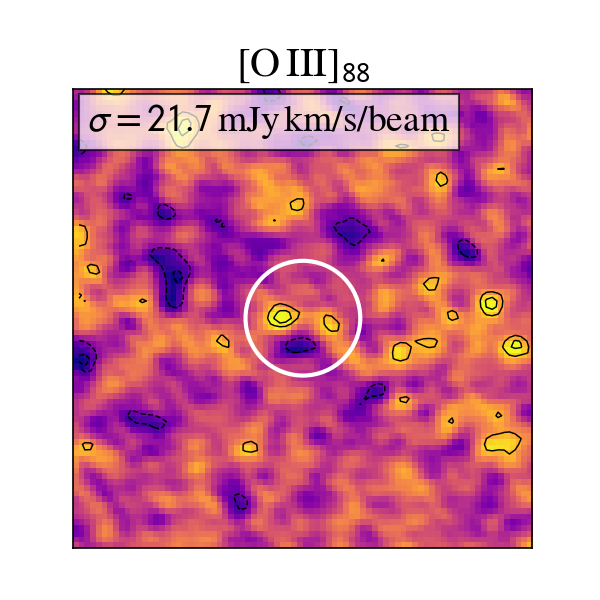}
    \includegraphics[width=0.23\textwidth]{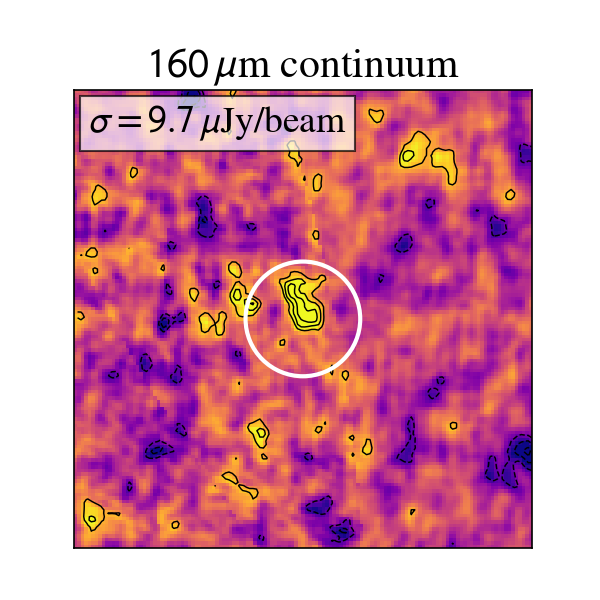} %
    \includegraphics[width=0.23\textwidth]{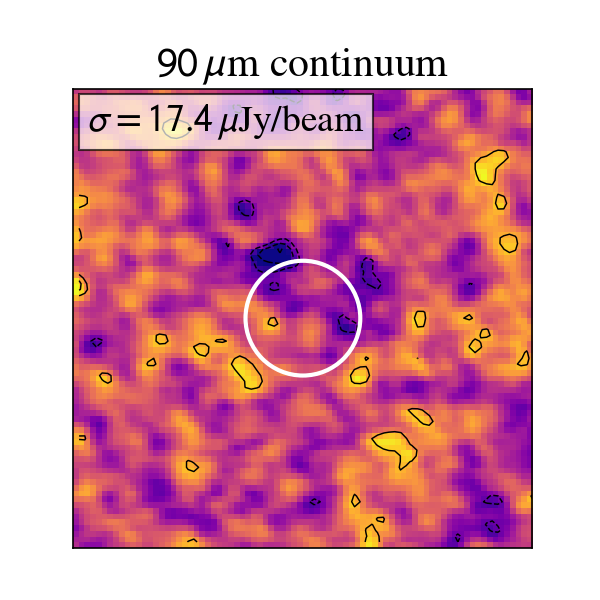}
    \caption{The stacked spectrum of \oiii{} ({\it blue}) and \cii{} ({\it red}) normalized to a full-width at half-maximum per source prior to stacking (\textit{top panel}). The \cii{} emission is detected, while no \oiii{} is seen in this very deep spectral stack.
    Image-plane stacks of the \cii{} and \oiii{} moment-0 maps ($4\farcs0\times4\farcs0$; \textit{middle left and right panels}, resp.), and rest-frame $160\,\mu$m and $90\,\mu$m continua (\textit{bottom left and right panels}, resp). Contours are $\pm2, 3, 4, \ldots \sigma$ and the colorscale runs from $-3\sigma$ to $+3\sigma$. A $0.5''$ radius aperture, in which the fluxes are extracted, is overplotted. The stacked \cii{} and $160\,\mu$m continuum emission are both clearly detected, while there is no discernible emission in the \oiii{} and rest-frame $90\,\mu$m stacks.}
    \label{fig:stacks}
\end{figure}

The stacked \cii{} and $160\,\mu$m continuum emission are both clearly detected, while there is no discernible emission in the \oiii{} and rest-frame $90\,\mu$m stacks (Figure \ref{fig:stacks}). We extract the line and continuum fluxes in a $0\farcs5$ aperture, finding $S_\text{\cii{}} = 410 \pm 45\,\mathrm{mJy\,km/s}$; $S_\text{\oiii{}} = 36 \pm 60\,\mathrm{mJy\,km/s}$; $S_{160\,\mu\mathrm{m}} = 126 \pm 34\,\mu$Jy; and $S_{90\,\mu\mathrm{m}} = -65 \pm 47\,\mu$Jy. Both the \oiii{} line flux and the underlying continuum are consistent with zero to within $< 1.5\sigma$. As such, we adopt $3\times$ the uncertainty on the aperture flux as an upper limit on the \oiii{} and $90\,\mu\mathrm{m}$ continuum fluxes. We emphasize that these are relatively conservative limits, given that the uncertainties on the aperture flux densities are larger than the noise in the stack by a typical factor of $\sim3$.

Similarly, we create a stacked spectrum using a relatively small aperture ({$0\farcs3$}) using the same stacking procedure, except in frequency space. The noise is based on the aperture size and varies as a function of frequency, while we normalize each spectrum to the fraction of the full-width at half-maximum (FWHM) for a fair comparison across all sources. The \cii{} emission is detected, while no \oiii{} is seen in this very deep spectral stack (Figure \ref{fig:stacks}). Finally, additional tests using curve-of-growth analysis and the stacking $1"$ $uv$-tapered maps also do not find any \oiii{} emission, while the \cii{} emission is recovered in each experiment.

\section{Analysis}
\label{sec:Analysis}

\subsection{The Oxygen to Carbon Ratios of SFR-selected Galaxies}
\label{sec:oiiitociiGraphSection}
\begin{figure*}
    \centering
    \includegraphics[width=0.75\textwidth]{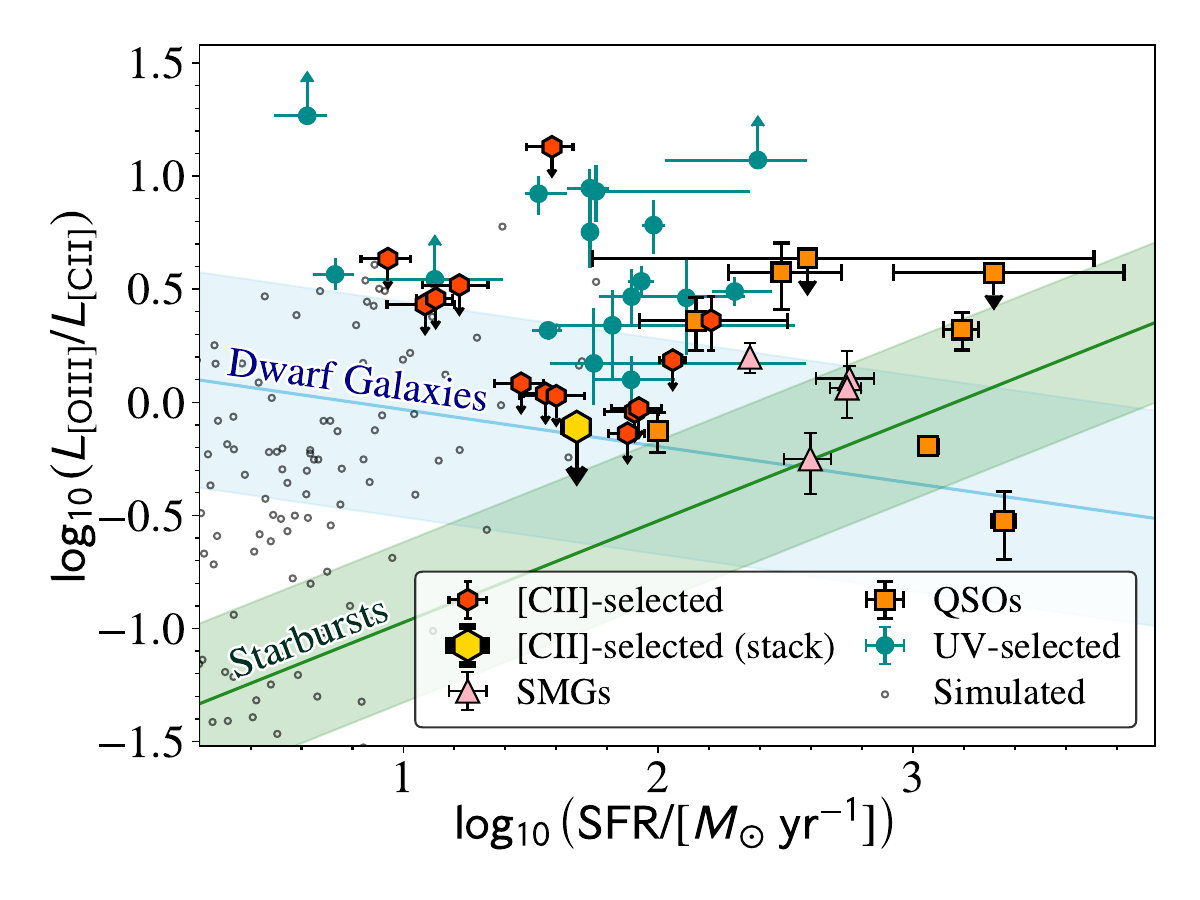}
    \caption{The \oiii{} / \cii{} emission line ratios of our \cii{}-selected sample (red hexagons), as a function of their \cii{}-based star formation rates. We compare to $z > 6$ galaxies in the literature, comprising UV-selected sources (cyan circles; \citealt{Harikane2019,Carniani2020,Akins2022,Witstok2022,Algera2023b}), sub-millimeter galaxies (pink triangles; \citealt{Marrone2018,Walter2018,Zavala2018,Tadaki2022}) and high-redshift quasars (orange squares; \citealt{Walter2018,Hashimoto2019,Novak2019,Decarli2023}; Algera \& Bakx et al.\ in prep). In addition, we overplot the relations for local starbursts and dwarf galaxies from \citet{delooze14}, and further compare to $z=7.7$ galaxies from the {\sc SERRA} simulations \citep{Pallottini2022}. Compared to UV-selected galaxies, the \cii{}-selected sample shows lower typical \oiii{} / \cii{} emission line ratios, indicating selection biases may play a role in setting the high \oiiitocii{} observed so far in $z > 6$ galaxies. In particular, the stack (yellow hexagon) suggests a typical \oiiitocii{}$\lesssim 0.8$ for the \cii{}-selected galaxy population.}
    \label{fig:OiiiCii}
\end{figure*}

We investigate the \oiiitocii{} ratios of our 13 \cii{}-selected galaxies found in the same fields as quasars. Given the strong correlation between SFR and \cii{} luminosity \citep[e.g.,][]{Schaerer2020}, these sources are star-formation-selected targets independent of their level of dust obscuration. 
We show the line ratios as a function of star formation rate in Figure \ref{fig:OiiiCii}, and compare to a compilation of (predominantly UV-selected) $z > 6$ galaxies in the literature.

For our single \oiii{}-detected source, we infer a line ratio of \oiiitocii{}$=2.3 \pm 0.6$, while for the remaining sources we place an upper limit on the line ratio, ranging from \oiiitocii{} $= 0.7$ for our deepest data, to \oiiitocii{} $= 4.3$ for the shallowest observations of \cii{}-faint systems ($< 3\sigma$).\footnote{This excludes the upper limit of \oiiitocii{}$< 14.2$ inferred for companion source J0842C2 in quasar field J0842+1218, which was covered at the very edge of the Band 8 primary beam ($\mathrm{PB} \approx 0.08$).} Our median upper limit on the line ratio is \oiiitocii{}$\lesssim 1.2$, which is significantly lower than the typical line ratios observed for high-redshift UV-selected galaxies (e.g., \citealt{Carniani2020,Harikane2019,Witstok2022}) with SFR $= 5 - 200$~$M_{\odot}$/yr. As a sanity check, we investigate the possibilities of extended \cii{} \citep{Fujimoto2019,Carniani2020} and \oiii{} emission through curve-of-growth analyses, but find no changes to our estimates.

In addition, the combined stack in \cii{} and \oiii{} provides even more stringent constraints on the typical line ratio of star-formation-selected high-redshift galaxies, with a $3\sigma$ limit of \oiiitocii{}$< 0.8$ at $\mathrm{SFR}_\text{\cii{}} \approx 40\,M_\odot\,\mathrm{yr}^{-1}$ -- similar to the typical line ratio of local dwarf galaxies \citep[$\approx 2$;][]{delooze14,Cormier2015} and spiral galaxies \citep[$\approx 0.5$;][]{Brauher2008}. We discuss the implications of these low \oiiitocii{} ratios in Section \ref{sec:OiiiCiiDiscussion}.

\subsection{Dust properties and limits}
\label{sec:dustProperties}

\begin{figure}
    \centering
    \includegraphics[width=0.5\textwidth]{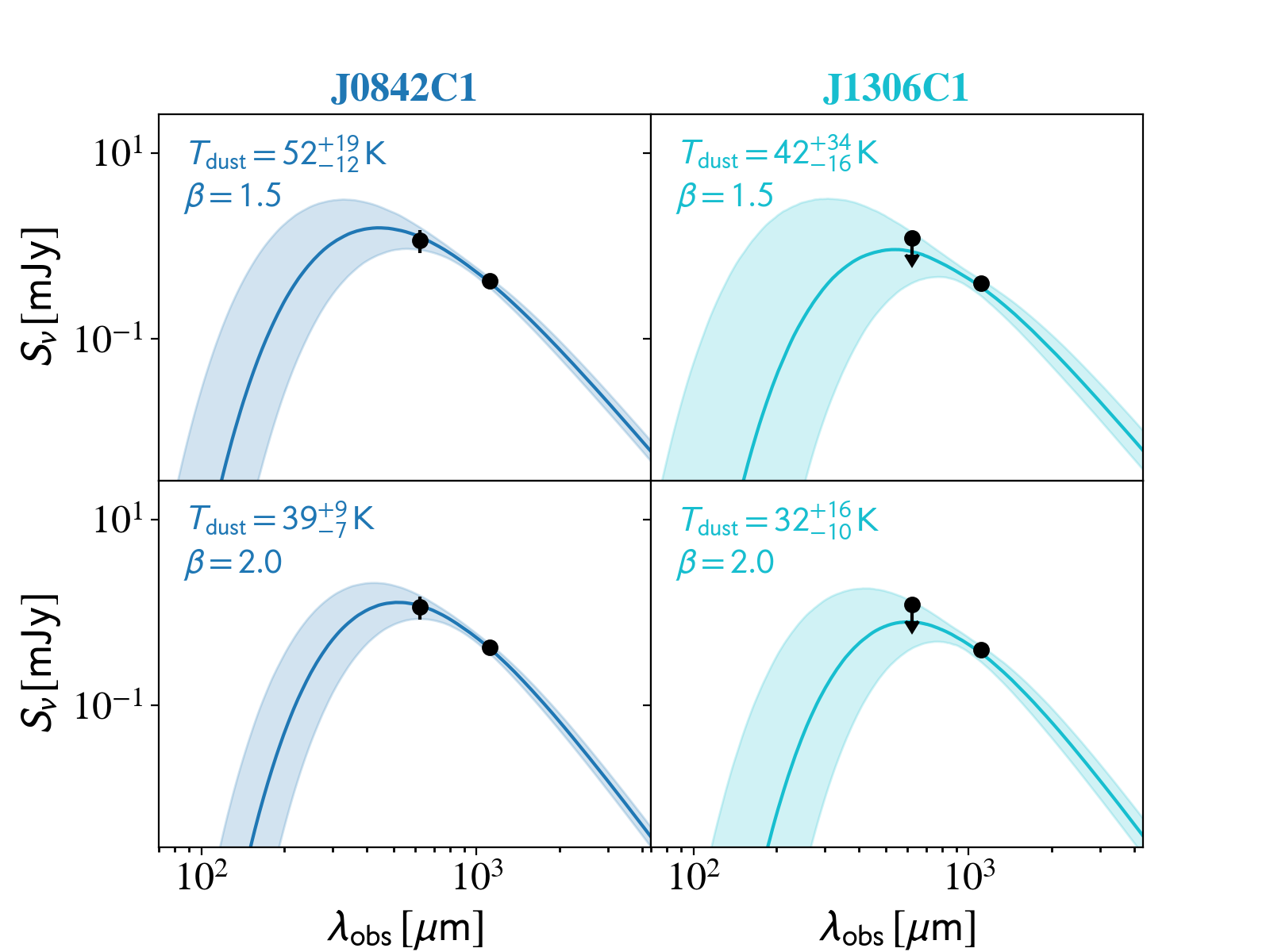}
    \caption{Fitted far-infrared SEDs of our two robustly continuum-detected targets. The two rows show the fit for an assumed $\beta = 1.5$ (top) and $\beta = 2.0$ (bottom), with the shaded region representing the confidence interval on the fit. These two bright \cii{}-emitters have dust temperatures that are typical for UV-selected $z\gtrsim5$ galaxies.}
    \label{fig:MBBfitting}
\end{figure}

We investigate the far-infrared spectral energy distributions (SEDs) of our two targets with robust continuum detections at rest-frame $160\,\mu$m (Section \ref{sec:dustContinuum}). We fit to their ALMA photometry following the framework of \citet{Algera2023b}, which assumes their dust SEDs can be characterized by an optically thin MBB similar to the models adopted in other high-$z$ works \citep[e.g.,][]{Bakx2020,Bakx2021,Witstok2022}. We correct for both heating by and contrast against the CMB following \citet{daCunha2013}. Given that we have measurements at only two distinct wavelengths, we adopt a fixed dust emissivity index $\beta$, while varying the dust temperature $T_\mathrm{dust}$ and dust mass $M_\mathrm{dust}$. We consider both $\beta = 1.5$ and $\beta = 2.0$, and show the MBB fits in Figure \ref{fig:MBBfitting}. The fitting outputs are tabulated in Table \ref{tab:mbb}.

\begin{table}
    \centering
    \caption{Results of optically thin modified blackbody fitting to the two continuum-detected quasar companion galaxies.}
    \label{tab:mbb}
    \begin{tabular}{llll}
    \hline 
& Parameter & J10842C1 & J1306C1 \\
\hline 
$\beta = 1.5$ & $\log_{10}(M_\mathrm{dust} / M_\odot)$ & $7.66 \pm 0.31$ & $7.84_{-0.61}^{+0.74}$ \\
& $T_\mathrm{dust}\,\mathrm{[K]}$ & $52_{-12}^{+19}$ & $42_{-16}^{+34}$ \\
& $\log_{10}(L_\mathrm{IR} / L_\odot)$ & $12.14_{-0.35}^{+0.45}$ & $11.81_{-0.48}^{+0.82}$ \\ 
\hline
$\beta = 2.0$ & $\log_{10}(M_\mathrm{dust} / M_\odot)$ & $7.96_{-0.27}^{+0.31}$ & $8.16_{-0.55}^{+0.76}$ \\
& $T_\mathrm{dust}\,\mathrm{[K]}$ & $39_{-7}^{+9}$ & $32_{-10}^{+16}$ \\
& $\log_{10}(L_\mathrm{IR} / L_\odot)$ & $11.96_{-0.27}^{+0.30} $ & $11.66_{-0.34}^{+0.55}$ \\ 
\hline 
\end{tabular}
    \raggedright \justify \vspace{-0.2cm}
\textbf{Notes:} 
The fits assume a fixed $\beta = 1.5$ (top three rows) or $\beta = 2.0$ (bottom three rows). Dust temperatures are corrected for the CMB, and IR luminosities are determined across rest-frame $8 - 1000\,\mu$m. 
\end{table}

Depending on the precise value adopted for $\beta$, we infer dust temperatures of $T_\mathrm{dust} \approx 30 - 50\,$K and dust masses of $\log(M_\mathrm{dust} / \mathrm{M}_\odot) \approx 7.7 - 8.1$ for both of our continuum-detected sources. While no stellar mass measurements for our targets are available, we can place a rough lower limit on $M_\star$ by appealing to models of dust production, which predict typical dust-to-stellar mass ratios of $M_\mathrm{dust} / M_\star \sim0.1 - 1\%$ \citep{Dayal2022,DiCesare2023}, in agreement with observations (e.g., \citealt{Witstok2023b,Algera2023b}). Assuming a conservative upper limit of $1\%$ on the dust-to-stellar mass ratio, our targets are likely to be relatively massive and hence chemically evolved galaxies ($M_\star \gtrsim 10^{10}\,\mathrm{M}_\odot$). Integrating our MBB fits across rest-frame $8 - 1000\,\mu$m, we find both targets to have infrared luminosities of $L_\mathrm{IR} \approx 10^{11.7 - 12.1}\,\mathrm{L}_\odot$, placing them into the class of (Ultra) Luminous Infrared Galaxies.

Adopting the conversion factor between $L_\mathrm{IR}$ and $\mathrm{SFR}_\mathrm{IR}$ from \citet{Kennicutt1998} and assuming a fiducial $\beta = 2$, we infer $\mathrm{SFR}_\mathrm{IR} = 158_{-73}^{+157}\,M_\odot\,\mathrm{yr}^{-1}$ and $\mathrm{SFR}_\mathrm{IR} = 79_{-43}^{+202}\,M_\odot\,\mathrm{yr}^{-1}$ for J10842C1 and J1306C1, respectively. These values are in agreement with their \cii{}-based SFRs of $\mathrm{SFR}_\text{\cii{}} = 161 \pm 12\,M_\odot\,\mathrm{yr}^{-1}$ and $\mathrm{SFR}_\text{\cii{}} = 114 \pm 13\,M_\odot\,\mathrm{yr}^{-1}$ (see Table \ref{tab:sample}), although observations at additional far-IR wavelengths are needed to improve the large uncertainties. Nevertheless, this consistency between IR and \cii{} emission suggests the bulk of the star formation in these two \cii{}-luminous companion galaxies is obscured. This is in agreement with the two \cii{}-selected, optically-dark galaxies at $z\sim7$ presented by \citet{fudamoto2021}, which have comparable \cii{} luminosities and little unobscured star formation.

The dust temperatures and masses of our two continuum-detected sources are in line with those of massive UV-selected galaxies at $z \approx 5 - 7.5$ (e.g., \citealt{Faisst2020,Bakx2021,Sommovigo2022,Algera2023b}), albeit on the massive end, which may suggest the dust properties of our \cii{}-selected sample to be similar to those of the wider high-redshift population. However, we note that the above analysis is limited to the two brightest continuum and \cii{} sources among our sample, which may therefore not be representative of the overall \cii{}-selected galaxy population.

\begin{figure}
    \centering
    \includegraphics[width=0.5\textwidth]{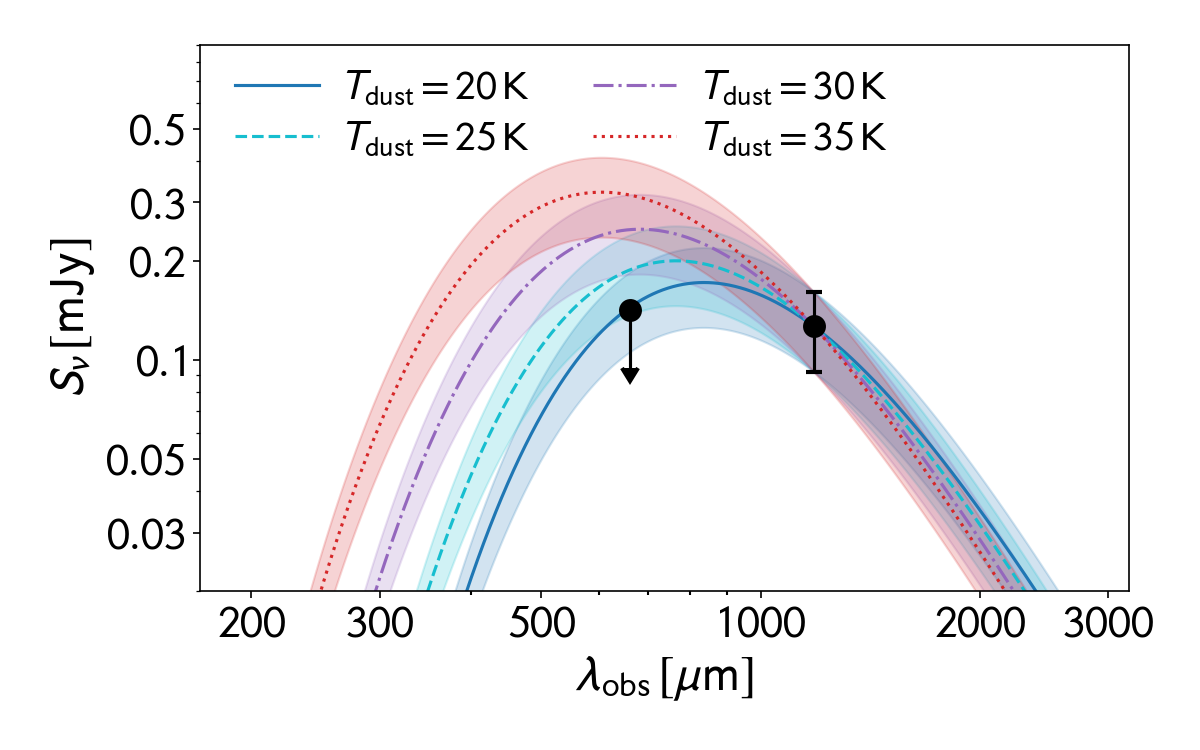}
    \caption{Optically thin modified blackbodies with $\beta = 2$ at different dust temperatures, anchored to the stacked continuum detection (and its measurement uncertainties) at rest-frame $160\,\mu$m. To match the stringent upper limit at rest-frame $90\,\mu$m, we require a MBB with cold dust ($T_\mathrm{dust} \lesssim 25 - 30\,$K).}
    \label{fig:stackedMBB}
\end{figure}

As such, we next turn to the dust continuum stacks (Section \ref{sec:stacking} and Fig.\ \ref{fig:stacks}), which treat the \cii{}-bright and -faint galaxies in an identical manner. The $160\,\mu$m continuum is clearly detected in the stack ($\mathrm{S/N}_\mathrm{peak} = 4.9\sigma$), while the continuum is undetected at rest-frame $90\,\mu$m ($< 3\sigma$). We show several MBBs with $\beta = 2.0$ anchored to the $160\,\mu$m detection in Fig. \ref{fig:stackedMBB}, which suggest the average dust SED of our \cii{}-selected sample to be characterized by a cold dust temperature of $T_\mathrm{dust} \lesssim 25\,$K. Even a shallower $\beta = 1.5$ would still imply cold dust of $T_\mathrm{dust} \lesssim 30\,$K for the stacked SED. 

Assuming $T_\mathrm{dust} = 25\,$K, we infer a massive average dust reservoir of $\log(M_\mathrm{dust} / M_\odot) \approx 8.0$, yet a rather modest infrared luminosity of $L_\mathrm{IR} \approx 10^{11}\,L_\odot$. This dust mass, similar to those of the individually-fitted sources, is on the massive end of the UV-selected sources studied in ALPINE \citep{Sommovigo2022} and REBELS \citep{Ferrara2022}, and on the low-end of sub-mm selected dusty populations \citep{daCunha2015}. In order to have built up this dust mass, the sample is likely to have previously undergone significant chemical evolution, while the low $L_\mathrm{IR}$ suggests it is currently in a state of lower star formation activity. This is qualitatively consistent the scenario whereby the dust temperature correlates with specific SFR (e.g., \citealt{Magnelli2014}), or inversely correlates with depletion time ($t_\mathrm{depl}$; \citealt{Sommovigo2021,Sommovigo2022,Vallini2024}). In particular, \citet{Sommovigo2021} predict that $T_\mathrm{dust} \propto t_\mathrm{depl}^{-1/(4+\beta)}$, which could suggest our sample to be characterized by long depletion times, and therefore large molecular gas masses. CO observations of two luminous $z\sim6.5$ quasar companion galaxies by \citet{Pensabene2021} have indeed revealed high molecular gas masses of $M_\mathrm{gas} \sim 10^9 - 10^{10}\,M_\odot$ in these sources, in qualitative agreement with the picture sketched above. 

While no direct gas mass measurements are available for our companion galaxy sample, some studies have suggested \cii{} can be utilized as a molecular gas mass tracer (e.g., \citealt{Zanella2018,Madden2020,Dessauges-Zavadsky2020,Vizgan2022,Aravena2024}). We caution, however, that due to the tight scaling between $L_\text{\cii{}}$ and SFR, a simultaneous relation between \cii{} and molecular gas mass may simply correspond to the implicit assumption of a linear Schmidt-Kennicutt relation \citep{Bethermin2023}. Nevertheless, adopting the conversion factor of $\alpha_\text{\cii{}} = 30\,M_\odot\,L_\odot^{-1}$ suggested by \citet{Zanella2018}, we infer a gas mass estimate of $M_\mathrm{gas} \sim (1 - 2) \times 10^{10}\,M_\odot$ from the \cii{} emission line stack (Section \ref{sec:stacking}). Combining this with the $\mathrm{SFR}_\mathrm{IR} \sim 20\,M_\odot\,\mathrm{yr}^{-1}$ inferred from the MBB fit to the continuum stacks suggests a depletion time of $t_\mathrm{depl} \lesssim 800\,$Myr, where we adopt this value as an upper limit given that any unobscured star formation is not accounted for. The inferred gas-to-dust ratio is approximately $\delta_\mathrm{GDR} \sim 100$, similar to that of local galaxies (e.g., \citealt{Genzel2015}), although with large and potentially systematic uncertainties that are difficult to quantify.

Qualitatively, this analysis suggests the average quasar companion galaxy to be relatively dust- and gas-rich, with a long depletion timescale. We do caveat, however, that the relatively low-S/N detection of the stacked Band 6 flux density prevents us from making any definite claims regarding the typical mass and temperature of the dust reservoir of our sample. As shown in Appendix \ref{app:dustSEDs}, a formal MBB fit leaves the possibility of a higher dust temperature, as the combination of a strict $90\,\mu$m upper limit and a low-S/N $160\,\mu$m measurement favours a fit that stays below the Band 6 detection (see also the discussion in \citealt{Algera2023b}). In addition, if the dust is optically thick -- as may be expected from the large inferred dust mass from optically thin models and the non-detection at $90\,\mu$m -- the true dust temperature could be larger, while the infrared luminosity is not significantly affected (\citealt{Algera2023b}; see also Algera et al.\ in prep). However, multi-band ALMA observations of the quasar companion galaxies are crucial to better constrain their typical dust SEDs.

\subsection{Sample completeness study}
\label{sec:sampleCompleteness}
Our companion galaxy sample was selected through its serendipitously discovered line emission seen in the fields of known quasars. 
These sources are robust line emitters ($> 5.5 \sigma$), although close to this detection limit, false-positives are not excluded as was shown in shallower observations \citep[e.g.,][]{Hayatsu2017,Hayatsu2019}, and as we will discuss in the subsequent section. {\color{referee} For four sources, additional deep JWST imaging allows us to look for their unobscured emission. Beyond direct} fidelity estimates, we evaluate the other assumption underlying the creation of this star-formation rate selected sample: \textit{is the line emission \cii{} emission?}

{\color{referee}
\subsubsection{Multi-wavelength counterparts of emission-line selected sources}
Several quasar fields have existing \textit{JWST} observations, namely J0100 (GTO 1243; P.I. Simon Lilly), J1342 (GO 1764; P.I. Xiaohui Fan), P036 (GO 2028; P.I. Feige Wang and GO 3990; P.I. Takahiro Morishita), and P231 (GO 2028; P.I. Feige Wang). At star formation rates in excess of $> 5$~M$_{\odot}$/yr, these galaxies are either strongly dust-extincted (see Section~\ref{sec:dustProperties}) or should be visible in \textit{JWST} imaging. The existing observations offer a chance to detect the \cii{} line emitters at near-infrared wavelengths using NIRCam. Moreover, MIRI imaging is available for the field containing the quasar J1342+0928. In this comparison, we use the product level 3 calibrated data from the Mikulski Archive for Space Telscopes\footnote{\url{https://mast.stsci.edu/}}, and correct for small astrometric offsets ($< 0.5$ arcsec) between ALMA and James Webb using the central quasar source, which agrees with the typical ALMA and James Webb astrometric uncertainties (ALMA Technical Handbook; see also \citealt{Killi2024} for a more thorough discussion on the astrometric uncertainties in ALMA). 

Figures~\ref{fig:JWST_2} and \ref{fig:JWST_1} show the NIRCam and MIRI imaging centered on the companion galaxies. {\color{referee2}A bright source is seen close to the central position ($< 0.5$~arcsec) for two of the fields (J0100+28C1; $f = 0.91$ \& $\Delta V_{\rm QSO} = 110$~km/s, and P23120C2; $f = 1.0$ \& $\Delta V_{\rm QSO} = 19000$~km/s), and for a source is seen at $< 1$ arcsec for J1342+0928C1 ($f = 1.0$ \& $\Delta V_{\rm QSO} = 240$~km/s), with a different object seen in MIRI, although this object lies close to the noise limits of the image. No obvious emitter is seen close to P036+03C1 ($f = 1.0$ \& $\Delta V_{\rm QSO} = 3200$~km/s). Importantly, the optical counterparts seen in the JWST/NIRCam imaging do not appear to evolve strongly with increasing wavelength beyond the larger PSFs \citep{Bowler2022}.
Consequently, the JWST imaging provide little evidence for rest-frame optical and near-infrared counterparts.
}

Additional observations with the HST and the infrared {\it Spitzer} telescope exist for the J1342+0928 field, as reported in \cite{Rojas-Ruiz2024}. They analyse the HST-imaging from the Advanced Camera for Surveys (ACS)/F814W, Wide Field Camera 3/F105W/F125W bands, and Spitzer/Infrared Array Camera at 3.6 and 4.5 $\mu$m. Although this imaging shows the emission seen in JWST imaging, they conclude there is no association between this emission and the line emitting galaxy J1342+0928C1 since the offset of the emission is {\color{referee2} $\sim 0.7$~arcsec. This offset could possibly be due to strong dust-obscuration with the caveat that no morphological change is seen with increasing wavelengths \citep[e.g.,][]{Wang2019,Bowler2022}. Similarly}, since this source is detected in NIRCam/F090W and ACS/F814W implies that the nearby object is likely a low-redshift interloper, particularly since the bright quasar J1342+0928 is completely extincted at these wavelengths due to the absorption by neutral gas at $z > 7$. {\color{referee2} As mentioned, since} some emission is seen in the MIRI image offset from the NIRCam emission, it could be that the companion galaxy is unrelated to this foreground object with $z_{\rm Ly break} < 6.4$, in line with previous studies and our analysis of the equivalent width of the ALMA-detected line (Sec.~\ref{sec:nonciiinterlopers}). 
Similarly, the non-detection of rest-frame optical emission at P036+03C1 is surprising, and could point to excessive dust obscuration or a systematic issue with the method for identifying line emitters. Consequently, subsequent analysis carefully accounts for the potential of false positives across the sample in order to ensure that the conclusions about a varying line luminosity ratio \oiiitocii{} are robust.

Beyond photometric analysis through broad- and medium band imaging, the J0100+28 system has been studied spectroscopically as part of the EIGER (Emission-line galaxies and Intergalactic Gas in the Epoch of Reionization) Wide Field Slitless Spectroscopy study reported in \citep{Kashino2023}. This study did not reveal any $z = 5.3 - 7$ counterpart from the \oiii{}$\lambda 5008$ line, with an average unobscured star-formation rate down to $\sim 5$~$M_{\odot}$/yr based on a stack \citep{Matthee2023}, although there is a preference towards low dust attenuation E($B-V$) $\approx 0.14$. Note however that the selection of this sample is by the \oiii{}$\lambda 5008$ line emission line, which is not a robust tracer of star-formation rate and particularly dust-obscuration of the \hii{} regions can impact the source selection. The spectroscopic redshift of J0100+2802C1 $z_{\rm CII} = 6.324$, and with an estimated SFR of 75.8~$M_{\odot}$/yr, the dust obscuration fraction needs to be at least $\sim 94$~per cent.

In an effort to estimate the effect of dust-obscuration in these sources, we use equation~11 from \cite{Ferrara2022}. As demonstrated in Section~\ref{sec:dustProperties}, the typical dust mass of these objects is $\approx 10^8$~$M_{\odot}$. The predicted UV dust attenuation at 1500~\AA{} depends modestly on the dust extinction cross-section, intermediately on the dust morphology, and strongly on the source size. For our sources, the resulting optical depth scales roughly as $\tau_{1500} \approx (0.5...2)/r^2_d$. For dusty galaxies, typical sizes are $\approx 0.1 - 0.5$~kpc \citep{Ferrara2022}, and could cause the UV to be predominantly absorbed with obscuration fractions above 95~per cent. As a consequence, these galaxies could be strongly dust-obscured \citep[e.g.,]{Wang2019}, even within the JWST wavelengths (rest-frame 1500 to 5000~\AA), and/or have strong dust gradients \citep{Bowler2022}.

\begin{figure*}
    \centering
    \includegraphics[width=\linewidth]{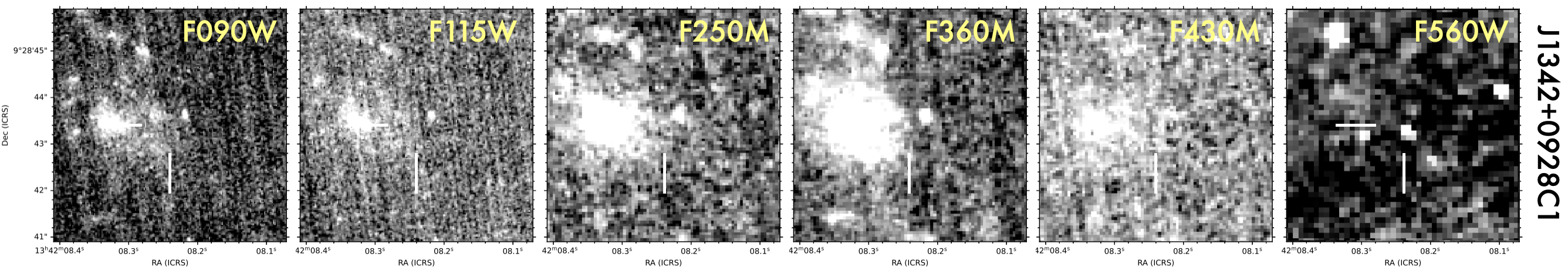}
    \caption{JWST imaging for J1342+0928C1 across the publically-available NIRCam (F090W, F115W, F250M, F360M, F420M) and MIRI (F560W) imaging show a bright object close ($< 0.5$~arcsec) to the central ALMA location. The poststamps are $5\farcs0 \times 5\farcs 0$.}
    \label{fig:JWST_2}
\end{figure*}
\begin{figure}
    \centering
    \includegraphics[width=\linewidth]{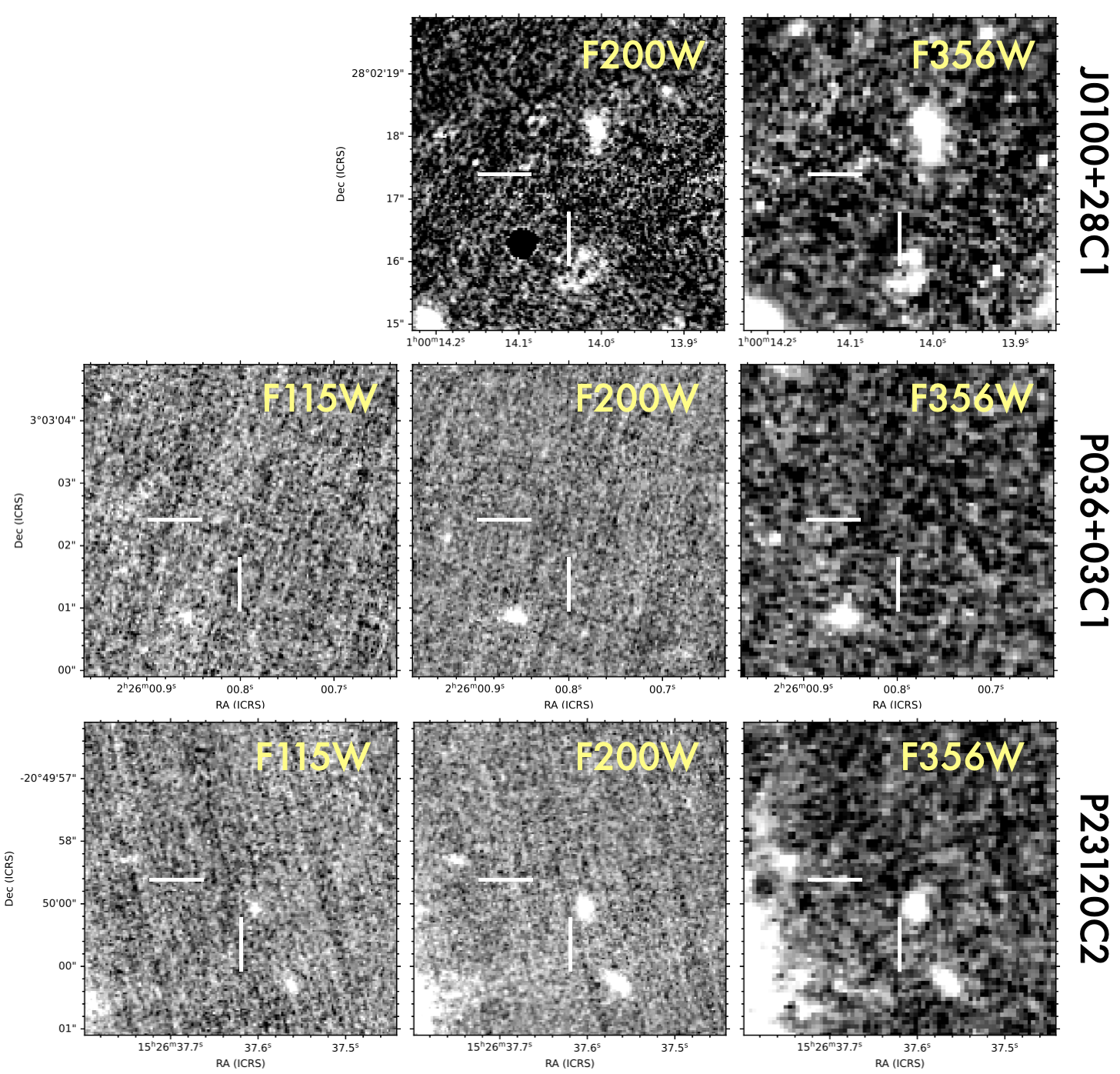}
    \caption{JWST imaging for J0100+28, P036+03C1, and P23120C2 across the publicly-available NIRCam (F200W, F356W, and F115W for the latter two sources) imaging show a bright object close {\color{referee2}($\sim 0.7$ and $\sim 0.5$~arcsec, resp.)} to the central ALMA location for two of the sources. The poststamps are $5\farcs0 \times 5\farcs 0$. No object is visible for P036+03C1.}
    \label{fig:JWST_1}
\end{figure}
}

\subsubsection{Fidelity of companion selection}
\label{sec:fidelity}
Our targets were identified using the \textsc{FindClump} algorithm \citep{Decarli2014,Walter2016,GonzalezLopez2019ApJ...882..139G}, which uses a combination of spectral convolution together with a {\sc sextractor} back-end to identify sources. By applying the method on both the positive and negative data cubes (i.e., the RA, DEC and frequency flux density tables), it is possible to calculate the false positive fraction of identifications, called the fidelity, 
\begin{equation}
    f = 1 - \frac{n_{\rm negative}}{n_{\rm positive}},
\end{equation}
where $n_{\rm negative}$ and $n_{\rm positive}$ are the number of negative and positive emission features found at a given signal-to-noise ratio. In this fidelity analysis, all negative features are taken to be spurious, and hence reflect the noise properties of the ALMA data cubes. 

Every source in this study has a fidelity of at least $f \ge 0.91$ (see Table~\ref{tab:sample}; \citealt{Venemans2020}). {\color{referee} The probability of no false-positive sources with these fidelities is equal to $\Pi_{n = 1 - 13} f = 72$~per cent. The probability of a single false-positive is equal to 25~per cent, leaving three per cent probability of more than one false-positive in the sample. The non-detection of rest-frame optical emission at P036+03C1 is surprising, and could point to some systematic issue with the method for identifying line emitters \citep{GonzalezLopez2019ApJ...882..139G,Venemans2020}. However, based on the fidelity estimates, it is likely that no false-positive line signals exist within this sample. 

}

\subsubsection{Non-\cii{} line interlopers}
\label{sec:nonciiinterlopers}
The creation of this sample further assumes that the emission lines seen in the quasar fields correspond to the \cii{} line. As (one of) the most dominant cooling lines of the interstellar medium, \ciil{} is likely to be bright and detectable even for low-mass, $z > 5$ galaxies. That said, lower-redshift line emission such as carbon-monoxide (CO) and atomic carbon (\ci{}) could also be responsible for the observed emission, as seen in for example blind surveys such as ASPECS \citep{Decarli2020}. Although other spectral lines (e.g., \oiii{}) could cause the confusion, these would require 
unreasonably-high redshifts ($z_{\rm OIII 88} \geq 11$) to be considered likely contaminants.

\begin{figure}
    \centering
    \includegraphics[width=\linewidth]{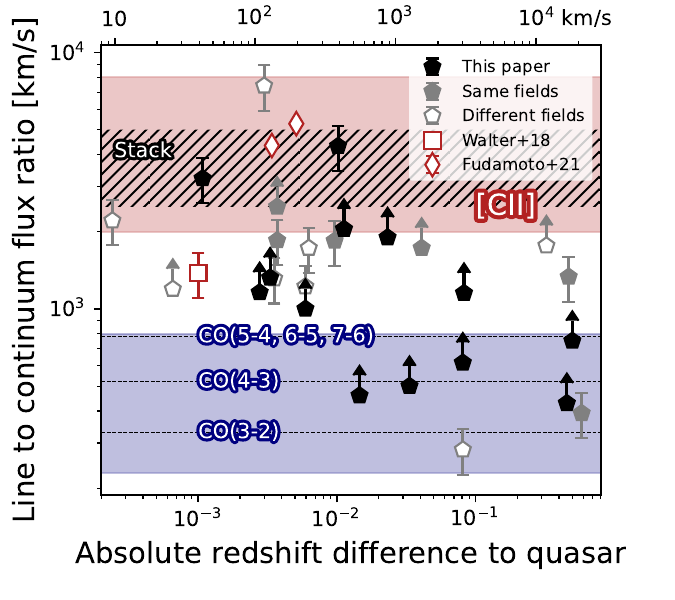}
    \caption{The equivalent width of the \cii{} line is shown against the redshift difference between the quasar and the companion galaxy. Since the \cii{} line is more luminous than the other contaminant lines (typically CO and \ci{}), the expected equivalent width of \cii{} emitters (\textit{red fill}; assuming $L_{\rm [CII]}/L_{\rm IR} = 0.05 - 0.2$\,per cent) is expected to be higher than those for the CO and \ci{} emission lines (\textit{blue fill}; \citealt{Hagimoto2023}). We show the equivalent widths of the sources observed in this paper ({\it black}), sources where we could not target the \oiii{} emission ({\it grey}; \citealt{Venemans2020}) where the sources with a grey fill indicate sources in fields that were targeted in this study, and the sources with a white fill indicate sources that were in fields that were not targeted. Three known companion sources are also shown in {\it red} (\citealt{Walter2018,fudamoto2021}). The stacked equivalent width of the sample (\textit{hatched region}) suggests the majority of sources in this sample are true \cii{} emitters.
    }
    \label{fig:dz_to_qso_vs_flux_ratio}
\end{figure}

In an effort to characterize the star-forming gas content across the Universe, deep pencil-beam observations with ALMA \citep{Decarli2016} and the VLA \citep{Riechers2019,Riechers2020} have worked to characterize the CO and \ci{} luminosity function (LF). We can now use these studies to estimate the chance of CO interlopers within the quasar fields.
The ASPECS survey \citep{Decarli2020} in particular is useful, as it uses Band~6 (and 3) to look for line emitting galaxies, which is similar to the selection band for most of the sources in our sample.
The expected number of CO and \ci{} line emitters in our observations is equal to the sum of the expected number of sources above $5.5 \sigma$ based on the \cite{Decarli2020} LFs for all possible CO and \ci{} emission lines. The luminosity functions of the spectral lines are converted to absolute numbers using the comoving cosmological volume probed by the observations that targeted the \cii{} of the initial quasar, carefully accounting for the sensitivity across the field-of-view and the associated redshift coverage for each spectral line. 
The companion sources are found in \cii{} observations that are typically centred on the quasar. The flux sensitivity of interferometric observations decrease as they move away further from the centre of the field, resulting in annuli with similar flux density sensitivity depths extending to a so-called primary-beam sensitivity of 0.2, roughly 17~arcseconds away from the central source. 
For each field, we split up the field-of-view into annuli that fall off using a Gaussian distribution. 
Using the $5.5 \sigma$ selection depth of companion galaxies, we are able to estimate the sensitivity limit for each potential CO and \ci{} emission line. Since each different potential emission line results in a different redshift of the companion source, we adjust the cosmological comoving volume that each annulus probes. 
We assume each observation that targets a quasar source covered a total of $\sim\pm8~\mathrm{GHz}$ of bandwidth centered around the observed frequency. 

In total, \cite{Venemans2020} identify 19 companion sources in the 10 quasar fields that are studied in this work. Out of these 19, we expect 4.2 to be CO or \ci{} emitters based on the ASPECS luminosity functions. In our study, we only target a subset of 13 out of these 19 companion galaxies with redshifts that facilitate observing \oiii{} from the ground. As such, we find that our sample of 13 galaxies may contain 2.8 CO or \ci{} emitters.
We check that the effect of the specific CO spectral line flux density ratios does not strongly affect the results, whether we use ordinary galaxies from blind fields \citep{Boogaard2020} or dusty star-forming galaxies \citep{Hagimoto2023}, although ratios that peak at lower-$J$ -- such as the Milky Way \citep{Fixsen1999} -- can reduce the total number of interlopers. Furthermore, we compare these results to the phenomenological CO luminosity model from \cite{Bisigello2022}, which provides similar results as the main approach using the CO LFs from \cite{Decarli2020}. The observed estimates of the CO LFs rely on a single field, which could affect the number of predicted interlopers (i.e., an overdensity of CO line emitters in the HUDF field), although the ASPECS study \citep{Decarli2020} finds that the effect of cosmic varience should be modest.

The analytical calculation shows that the central region has a higher probability of finding CO emitters, while at larger separations the area increases rapidly but the sensitivity drops faster. The observations that identified these companion sources are thus sensitive to CO line luminosities on the order of $\log L'$ from 10$^8$ to $10^{10}$~K~km/s~pc$^2$, before or on the predicted knee ($L'_*$) in the LF \citep{Decarli2020}, where its evolution with line luminosity is relatively mild. The shallow slope means that Eddington bias \citep{Eddington1913,Serjeant2023} is likely not an important effect in the study of CO interlopers as apparent quasar companion galaxies, although this may become an issue for larger-area studies such as ASPIRE.

In an effort to test which sources are at risk of being interlopers,  
Figure~\ref{fig:dz_to_qso_vs_flux_ratio} compares the equivalent width of the line detection of each source to the redshift difference between the companion galaxy and the central quasar.\footnote{As ALMA develops towards 2030 with the Wideband Sensitivity Upgrade \citep{Carpenter2022}, the diagnostic capabilities of the equivalent width (Fig.~\ref{fig:dz_to_qso_vs_flux_ratio}) will improve as more sources will be detected in continuum.}
The \cii{} line is typically responsible for a substantial fraction of the overall far-IR luminosity \citep{Gullberg2015}, and we indicate the sensitive region between the typical range of $L_\text{\cii{}} / L_\mathrm{IR} = 0.05 - 0.2$~per cent. For the underlying continuum shape we assume a modified blackbody (MBB) at a temperature of $T_\mathrm{dust} = 45\,$K, although we note that the expected equivalent widths reduce when a colder dust temperature is adopted (Section \ref{sec:dustProperties}). We compare to the line-to-continuum ratios expected for CO lines, based on the scaling relations derived from a stack by \cite{Hagimoto2023} based on dusty star-forming galaxies. As shown in \cite{Hagimoto2023}, these equivalent widths are in line with blind sources and models \citep[e.g.,][]{Boogaard2020,Bisigello2022}, seen by line luminosity relations that are (near-)linear for over four orders of magnitude. For consistency, in the scenario where the lines are due to CO-emitting galaxies at lower redshift, we adopt a typical underlying SED of $T_\mathrm{dust} = 35\,$K, consistent with the sources studied by \citet{Hagimoto2023} (see also \citealt{Bendo2023}).

The two continuum-detected sources lie firmly above the expected region for CO lines, and are consistent with being \cii{} emission. For the remaining 11 galaxies, the lower limits also hint at a relatively large line-to-continuum ratio indicative of \cii{} emission. Indeed, the stacked line-to-continuum ratio is consistent with a typical $L_\text{\cii{}} / L_\mathrm{IR} \approx 0.1$~per cent, indicative of a low contamination of line interlopers. 
Three companion sources with confirmed \cii{} emission \citep{Walter2018,fudamoto2021} provide a relative comparison, all lying above the CO equivalent widths, although the source from \cite{Walter2018} indicates a relatively low equivalent width (EW\,$\approx 1300$~km/s).
There appears to be a gentle preference towards a higher line-to-continuum ratio for sources with redshifts closer to the central source, while line-emitting galaxies in the other side-band tend to have lower effective equivalent widths (placed at $\pm \sim8$~GHz or $\pm 10^4$~km/s). 

All sources discussed in this paper that lie in the CO equivalent width region are only based on $3 \sigma$ upper limits, and thus do not provide conclusive evidence for interlopers. These sources are characterized by low signal-to-noise ratios and small line velocity widths, as the equivalent width scales with $\propto SNR \times{} \sqrt{\Delta V}$. {\color{referee}The J1342+0928C1 source is detected at $\sim 240$~km/s offset from the quasar, with an equivalent width of at least $> 1000$~km/s, placing it above the region typically expected for CO emission. Since a nearby foreground emitter ($z_{\rm LBG} < 6$) was seen at these wavelengths, this emission is likely unrelated to the ALMA line emission, in agreement with previous studies \citep{Rojas-Ruiz2024}.}
The two sources with low limits on their equivalent width above $\Delta z > 0.1$ are P183+05C2 and P23120C2, and could pose a risk for interlopers. The three other sources with equivalent line limits in the CO regime are J2318\_3113C2, J2318\_3113C1 and P036+03C1, from lowest to highest equivalent width estimates.
In order to estimate the effect of these two to five potential interlopers, we re-run the stacking studies on a sample excluding these two or five potential interlopers. The $3 \sigma$ upper limit of the \oiiitocii{} ratio does not appear to change between these three different tests. Although it is not possible to exclude interlopers with our current data, they do not significantly affect our estimate of \oiiitocii{} or the average dust temperature in our stack.

\subsection{Biases in a \cii{}-selected quasar companion sample}
\label{sec:biases}
The selection of galaxies from fields close to quasars by their line emission could affect the results of this study. Since the sources are selected from lines expected to be \cii{}, Eddington bias would imply we probe sources that happen to be bright in \cii{}, either through noise fluctuations or through their galaxy properties. 
The effect of noise fluctuations is linked to the slope of the \cii{} LF. Observationally, the estimates for \cii{} LFs in the $z > 5$ Universe are limited to targeted surveys \citep{Capak2015,Yamaguchi2017,Yan2020,Loiacono2021} or low number statistics from blind surveys \citep{Decarli2020,Fujimoto2022}. Simulations by \cite{Lagache2018} and \cite{Bisigello2022} predict a shallow evolution in the $L_{\rm [CII]} \approx 10^9$~L$_{\odot}$ regime, which diminishes the effect of Eddington bias (c.f., 
\citealt{Popping2016}).

As seen in Figure~\ref{fig:dz_to_qso_vs_flux_ratio}, 80~per cent of sources lie in the same sideband as the quasar (i.e., either the lower- or upper sideband, and thus within four GHz or roughly 5000~km/s from the \cii{} emission of the quasar), which is likely due to the ability of quasars to trace overdense regions \citep{Overzier2016}, even out to high redshifts \citep{Chiang2013,Chiang2017}.\footnote{These \cii{} observations were executed in Band~6, where the atmospheric transmission is relatively flat, which means there is no or little variation in the noise in the upper- and lower-side bands.} The majority of sources reside at most $\Delta z = 0.03$ from the central quasar, equivalent to roughly 1000~km/s at $z = 6$. This is in line with the expected velocity distribution of members galaxies of a cluster environment under construction \citep{Kurk2004,Koyama2013,Koyama2021}. 
Cosmic overdensities left over from the Big Bang collapsed to form the centres of future clusters, and evolution of galaxies in these overdense regions is accelerated in the $z > 2$ Universe \citep{Shimakawa2018,Smail2024}. As a consequence, simulations predict an excess of bright sources in the vicinity of quasars \citep{Chiang2013,Chiang2017}, and this has also been seen in observations \citep[e.g.,][]{Arrigoni2023,Bakx2024}. {\color{referee} The early onset of cosmic star-formation in these regions likely results in a more chemically-enriched environment \citep{Shimakawa2015,Biffi2018}. As a consequence, these metal-enhanced regions could be an important reason for the large dust masses and low \oiiitocii{} line luminosity ratio found in this study.} Although, since 30 per cent of all star-formation is predicted to occur in overdense regions \citep{Chiang2017} at $z = 6$, untargeted observations are also affected by the additional boost in star-formation due to these proto-cluster environments.

Beyond the immediate effect of earlier star-formation in overdense regions, the quasars themselves can affect the galaxies in their surroundings through outflows \citep{Zana2023_more_galaxies_around_highz_quasars}. The star-formation in more massive galaxies appears boosted, although the effect diminishes towards longer distances. A numerical test by \cite{Ferrara2023} predicts a higher star-formation efficiency ($\sim 25$~per cent). A comparison to targeted studies such as ALPINE \citep{Fevre2019} suggests an increase in star-formation by three-fold, although the effect of interlopers could affect this interpretation.

\section{The \oiiitocii{} ratio of High-redshift Galaxies}
\label{sec:OiiiCiiDiscussion}
Our observations provide deep estimates of the \oiiitocii{} line luminosity ratio (Section~\ref{sec:oiiitociiGraphSection}) of a robust (Section~\ref{sec:fidelity}) star-formation rate selected sample with reservoirs of cold dust (Section~\ref{sec:dustProperties}) in the distant Universe. Although there is the potential for CO-line interlopers (Section~\ref{sec:nonciiinterlopers}), {\color{referee} and the evolution in fields surrounding quasars could be accelerated resulting in more metal-rich galaxies} (Section~\ref{sec:biases}), {\color{referee2} this large sample offers the opportunity to investigate the wealth of explanations for the high \oiiitocii{} line ratio at $z > 6$.} It is good to acknowledge that the cause of the elevated line ratio could be through a combination of multiple factors, particularly since individual explanations are often inter-linked through the physics of galaxy evolution. 

The selection of objects in the $z > 6$ Universe has relied heavily on bright UV-luminous systems identified by {\it Hubble}. These systems are often characterized by recent bursts of star-formation \citep{Hashimoto2018,Tamura2019}, which could bias observations towards young systems with a low PDR covering fraction similar to local dwarf galaxies \citep{Madden2013,Madden2018,Cormier2019}. 
The resulting ISM conditions could therefore only be representative of a bursty phase of dust-free UV-luminous star-formation \citep{Sun_2023} with low gas-depletion timescales \citep{Vallini2024}, as \textit{Hubble} and \textit{JWST} observations tend to select sources with high surface star-formation rates \citep{Whitney2020}. 

Our sample of thirteen $z > 6$ galaxies selected through their \cii{} emission provides a complementary perspective, with twelve upper limits and a single detection in \oiii{} implying a break with the high-$z$ \oiii{} luminous perspective. These systems may represent a more metal-enriched, and less bursty population of galaxies found in the epoch of reionization, when compared to other $z > 6$ galaxies with similar star-formation rates{\color{referee}, potentially due to their selection close to nearby quasars.} As a result, they might have less strong radiation fields and/or a lower gas density \citep[e.g.,][]{Vallini2024}. 
Similar to previous studies \citep{Harikane2019,Katz2022}, we find it unlikely that inclination effects\footnote{Galaxy inclination can cause differences in line velocities between the \cii{} and \oiii{}, where \oiii{} would be typically emitted by more compact regions with smaller velocity widths, and thus would be detected at a higher signal-to-noise ratio \citep{Kohandel2019}.} or extended \cii{} can explain the enhanced emission seen for distant galaxies, since in those scenarios we would expect roughly half of our sample to be detected with bright \oiii{}.
Instead, only a single source is seen with high-enough \oiii{} emission to be detected, while none of the other sources show signs of enhanced \oiiitocii{} emission. Similarly, the fact that no \oiii{} was reported for the \cii{}-identified sources makes the extended emission less likely as an explanation. 

These $z > 6$ sources are detected by \cii{}, and thus already have evidence of Carbon-rich environments caused by nucleosynthesis in the last billion years. The rapid build-up of metals in the early Universe is thus likely not restricted to just Oxygen-producing supernovae for these sources \citep{Witstok2023}. We do not have a guarantee that these sources are a complete sample of galaxies within the quasar environment \citep{Zana2023}, and we might thus be prone to select the more star-forming and chemically-enriched systems 
\citep{Zana2022_enhanched_SF}. The atomic abundance of Carbon and Oxygen atoms might still run out of step for a portion of the distant galaxies compared to the local Universe \citep{Maiolino2019}. However, if this were the dominant effect, the identification of galaxies by their star-formation through \cii{} emission would likely also result in an enhanced \oiiitocii{} line ratio. This was not observed in this experiment. Meanwhile, the low dust temperature of the stacked spectrum indicates large volumes of dust in the early Universe. Dust reservoirs are known to be efficient at removing Carbon atoms from the gas-phase metals \citep{Konstantopoulou2022,Witstok2023}, although this does not exclude Oxygen removal \citep{Savage1996}. For these $z > 6$ galaxies, no such effect was observed. We can thus not exclude the metallicity as a cause for a high \oiiitocii{} ratio for UV-selected sources.

\begin{figure*}
    \centering
    \includegraphics[width=0.9\textwidth]{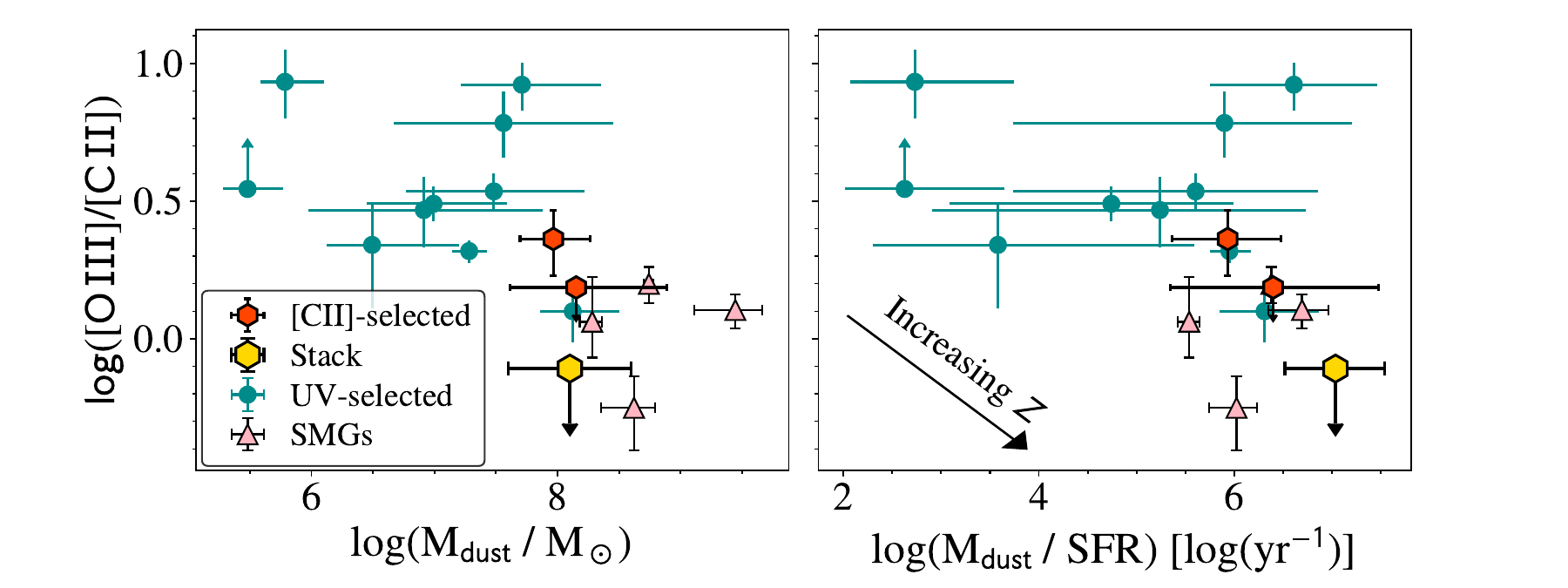}
    \caption{The \oiii{} to \cii{} ratio as a function of dust mass (left) and dust mass divided by IR (or IR + UV where available) star formation rate (right). The two \cii{}-selected sources with robust constraints on their dust SED are shown as red hexagons, while the stack is shown via the yellow diamond. Several UV-selected literature sources with good constraints on their IR SEDs are shown \citep[][and references therein]{Harikane2019,Witstok2022,Algera2023b}, as are the SMGs from \citet{Walter2018}, \citet{Marrone2018} and \citet{Zavala2018,Tadaki2022}. Galaxies with large dust masses, or larger dust mass-to-SFRs, appear to show a lower \oiiitocii{}. {\color{referee}This could be driven by metallicity effects, whereby more metal-enriched galaxies are dustier, and show stronger \cii{} emission relative to the \oiii{} line.}}
    \label{fig:lineratioVsMdust}
\end{figure*}

Although these sources do not have excessive \oiii{} emission, the nature of these \cii{} emitting galaxies remains still unknown. Future \textit{JWST} observations of their rest-UV and optical properties would provide accurate stellar masses (i.e., including potential older populations) of these sources to enable a study of their star-burstiness {\color{referee}across the entire sample, and particularly at the longer wavelengths} \citep{Vallini2021,Vallini2024}. Similarly, rest-frame optical spectroscopic observations can provide estimates of their metal enrichment and star-formation history in order to place these sources on a projected evolutionary pathway. However, as a proxy for metallicity we can investigate their \oiiitocii{} ratios in the context of their dust masses, given that dust consists fully of metals. 

We show the \oiiitocii{} ratio of the quasar companion sample as a function of dust mass (left) and dust mass divided by IR (or UV + IR, where available) star formation rate (right) in Figure \ref{fig:lineratioVsMdust}. Compared to the typical UV-selected population, our two \cii{}-selected galaxies with robust constraints on their IR SEDs and the full-sample stack -- taken to have a dust temperature in the range $T_\mathrm{dust} = 25 - 30\,$K -- cover a nearly totally different region of the parameter space: the \cii{}-selected sample, besides showing lower \oiiitocii{} ratios, is dustier and shows lower dust-to-SFR values. {\color{referee}This is likely indicative of the metal-enrichment of these emission-line selected galaxies, as low-metallicity systems may be fainter in \cii{} relative to their SFRs }\citep{Vallini2015,Vallini2020,Arata2020}. One potential consequence of a larger metal-content could be the rapid build-up of dust reservoirs through grain growth in the interstellar medium, which may constitute a significant fraction of total dust production (e.g., \citealt{Asano2013,DiCesare2023}; Algera et al.\ in prep), although accurate stellar masses are necessary to place the measured dust masses into further context. 

Moreover, a high $M_\mathrm{dust} / \mathrm{SFR}$ may represent a galaxy evolutionary phase beyond a likely UV-luminous early starburst. Such a burst will usher in rapid dust build-up, increasing $M_\mathrm{dust}$, while a subsequent drop in $\mathrm{SFR}$ after the burst will be followed by a similar drop in $L_\text{\cii{}}$, and hence a further increase in $M_\mathrm{dust} / \mathrm{SFR}$. As such, a high dust-to-SFR ratio could be used as an approximate means to pinpoint relatively evolved galaxy systems when only submillimeter observations are available. Moving beyond these estimates, direct metallicity measurements can be an important probe for characterizing their evolutionary phase. However, investigating the origins of this evolutionary phase would require high-resolution spectroscopic observations at rest-frame UV/optical and far-infrared wavelengths. 
These observations would be able to differentiate whether these galaxies are further evolved because they lie in cosmic overdensities, because of feedback from an AGN, or because they represent a population as of yet unprobed by existing blank-field studies in the infrared. 

In summary, a UV-based selection of high-redshift galaxies is expected to lead {\color{referee2}the selection of young, bursty systems,} characterized by a high \oiiitocii{}. Once the metaphorical dust has settled, and early metal-enrichment has taken place, these galaxies will move out of their UV-luminous starburst phase, and may no longer be selected in great numbers in UV-based surveys. Galaxies such as those picked up by studies such as REBELS, which are UV-luminous but simultaneously host large dust reservoirs (\citealt{Sommovigo2022,Inami2022,Algera2023b}; see also \citealt{Harikane2019,Mitsuhashi2023a}), may therefore fall into a transitionary phase between this early UV-bright bursty phase, and a more SMG-like, dusty phase. Only through dedicated IR-based selections, can this key evolutionary phase be mapped and understood in detail.

\section{Conclusions}
\label{sec:conclusions}
We report deep \oiii{} observations of thirteen SFR-selected galaxies with ALMA, {\color{referee2}in an effort to understand the elevated \oiiitocii{} ratios seen in the $z > 6$ Universe}. These galaxies were serendipitously detected as \cii{}-emitting quasar companion galaxies by \citet{Venemans2020}, and span \cii{}-based star formation rates from $10 - 160\,\mathrm{M}_\odot\,\mathrm{yr}^{-1}$. Our main conclusions are as follows: 

\begin{itemize}
\renewcommand\labelitemi{\tiny \textbf{$\blacksquare{}$}}

\item We detect only a single of the companion galaxies in \oiii{} emission, finding \oiiitocii{} $=2.3 \pm 0.6$. The remaining twelve galaxies are not detected at a typical $3\sigma$ depth of \oiiitocii{} $< 1.2$, and through a deep stack we place a stringent upper limit of \oiiitocii{} $< 0.8$ on their average line ratio. This is consistent with the \oiiitocii{} of galaxies in the local Universe, and an order of magnitude lower than the typical line ratios of UV-selected galaxies at $z > 6$.

\item Through detailed modelling, we assess that, at most, three galaxies among our sample are low-redshift CO- or \ci{} emitters that were misclassified as a \cii{}-emitting source. {\color{referee2} Similarly, we find no obvious rest-frame optical counterpart for the four sources with available JWST imaging, which could be due to strong dust obscuration or issues with the sample selection.} However, removing the most likely two to five interlopers (based on their low equivalent width and/or high redshift) from our study does not change the conclusions of this study{\color{referee2}, which provides confidence in the overall claims of this paper}.

\item For the two galaxies in our sample with robust continuum detections at rest-frame $160\,\mu$m --- one of which is also detected at rest-frame $90\,\mu$m -- we find typical dust temperatures of $T_\mathrm{dust} \approx 30 - 50\,$K and dust masses of $M_\mathrm{dust} \approx 10^8\,M_\odot$; massive relative to UV-selected studies at similar redshifts (\citealt{Sommovigo2022,Ferrara2022}) but with lower masses than sub-mm selected sources \citep{daCunha2015,Walter2018}. We do not detect rest-frame $90\,\mu$m continuum emission even in a deep stack of the full \cii{}-selected sample, implying a low average dust temperature of $T_\mathrm{dust} \lesssim 25 -30\,$K and a similarly high dust mass of $M_\mathrm{dust} \approx 10^8\,M_\odot$. 

\item The low \oiiitocii{} line ratio indicates that prior findings of enhanced \oiii{} emission were likely due to the UV-bright selection of bright, blue Lyman-break galaxies. This UV selection resulted in observational biases towards high surface star-formation rates or young stellar populations, with associated bright \oiii{} emission. As a consequence, future studies should {\color{referee2} also include sources selected through other means than UV- or sub-mm continuum fluxes.}

\end{itemize}

This study has characterized thirteen quasar companion sources identified through their \cii{} emission. Studies, including with the \textit{JWST}, should focus on a better characterization of the properties of these galaxies. Additional information on the nature of these sources are still needed, as there are currently no estimates to the stellar emission, any source-to-source variation in the dust attenuation, nor how representative these systems are compared to typical (i.e., $L_*$) galaxies in the $z > 6$ Universe. Direct observations of the circum-galactic medium of these sources could further assess the extent of potential interactions with the nearby quasars, and is particularly important given the advent of the new ASPIRE ALMA Large Program, which has the study of companion galaxies as a particular science goal.

\section*{Acknowledgements}
{\color{referee} The authors kindly thank the referee, Prof. Roberto Maiolino, for his insightful comments and suggestions to improve this manuscript.}
This paper made use of the following software packages: {\tt{spectral-cube}} \citep{Ginsburg2019}, {\tt{radio-beam}} \citep{Koch2021} and {\tt{interferopy}} \citep{Boogaard2021}. 

This work was supported by NAOJ ALMA Scientific Research Grant Nos. 2018-09B (TJLCB) 2021-19A (HSBA) and JSPS KAKENHI No.~17H06130, 22H04939, 22J21948, and 22KJ1598.
This paper makes use of the following ALMA data: ADS/JAO.ALMA\#2021.1.00668.S.
T.H. was supported by Leading Initiative for Excellent Young Researchers, MEXT, Japan (HJH02007) and by JSPS KAKENHI grant No. 22H01258he.

\section*{Data Availability}
The reduced, calibrated and science-ready
ALMA data are available from the ALMA Science Archive
at {\tt https://almascience.eso.org/asax/}. The data will be made available under any reasonable request to the author.



\bibliographystyle{mnras}
\bibliography{example} 

\begin{thebibliography}{}
\makeatletter
\relax
\def\mn@urlcharsother{\let\do\@makeother \do\$\do\&\do\#\do\^\do\_\do\%\do\~}
\def\mn@doi{\begingroup\mn@urlcharsother \@ifnextchar [ {\mn@doi@}
  {\mn@doi@[]}}
\def\mn@doi@[#1]#2{\def\@tempa{#1}\ifx\@tempa\@empty \href
  {http://dx.doi.org/#2} {doi:#2}\else \href {http://dx.doi.org/#2} {#1}\fi
  \endgroup}
\def\mn@eprint#1#2{\mn@eprint@#1:#2::\@nil}
\def\mn@eprint@arXiv#1{\href {http://arxiv.org/abs/#1} {{\tt arXiv:#1}}}
\def\mn@eprint@dblp#1{\href {http://dblp.uni-trier.de/rec/bibtex/#1.xml}
  {dblp:#1}}
\def\mn@eprint@#1:#2:#3:#4\@nil{\def\@tempa {#1}\def\@tempb {#2}\def\@tempc
  {#3}\ifx \@tempc \@empty \let \@tempc \@tempb \let \@tempb \@tempa \fi \ifx
  \@tempb \@empty \def\@tempb {arXiv}\fi \@ifundefined
  {mn@eprint@\@tempb}{\@tempb:\@tempc}{\expandafter \expandafter \csname
  mn@eprint@\@tempb\endcsname \expandafter{\@tempc}}}

\bibitem[\protect\citeauthoryear{{Adams} et~al.,}{{Adams}
  et~al.}{2022}]{Adams2022a}
{Adams} N.~J.,  et~al., 2022, arXiv e-prints, \href
  {https://ui.adsabs.harvard.edu/abs/2022arXiv220711217A} {p. arXiv:2207.11217}

\bibitem[\protect\citeauthoryear{{Akins} et~al.,}{{Akins}
  et~al.}{2022}]{Akins2022}
{Akins} H.~B.,  et~al., 2022, \mn@doi [\apj] {10.3847/1538-4357/ac795b}, \href
  {https://ui.adsabs.harvard.edu/abs/2022ApJ...934...64A} {934, 64}

\bibitem[\protect\citeauthoryear{{Algera} et~al.,}{{Algera}
  et~al.}{2023}]{Algera2023}
{Algera} H. S.~B.,  et~al., 2023, \mn@doi [\mnras] {10.1093/mnras/stac3195},
  \href {https://ui.adsabs.harvard.edu/abs/2023MNRAS.518.6142A} {518, 6142}

\bibitem[\protect\citeauthoryear{{Algera} et~al.,}{{Algera}
  et~al.}{2024}]{Algera2023b}
{Algera} H. S.~B.,  et~al., 2024, \mn@doi [\mnras] {10.1093/mnras/stad3111},
  \href {https://ui.adsabs.harvard.edu/abs/2024MNRAS.527.6867A} {527, 6867}

\bibitem[\protect\citeauthoryear{{Appleton} et~al.,}{{Appleton}
  et~al.}{2013}]{Appleton2013}
{Appleton} P.~N.,  et~al., 2013, \mn@doi [\apj] {10.1088/0004-637X/777/1/66},
  \href {https://ui.adsabs.harvard.edu/abs/2013ApJ...777...66A} {777, 66}

\bibitem[\protect\citeauthoryear{{Arata}, {Yajima}, {Nagamine}, {Li}  \&
  {Khochfar}}{{Arata} et~al.}{2019}]{Arata2019}
{Arata} S.,  {Yajima} H.,  {Nagamine} K.,  {Li} Y.,   {Khochfar} S.,  2019,
  \mn@doi [\mnras] {10.1093/mnras/stz1887}, \href
  {https://ui.adsabs.harvard.edu/abs/2019MNRAS.tmp.1825A} {p.~1825}

\bibitem[\protect\citeauthoryear{Arata, Yajima, Nagamine, Abe  \&
  Khochfar}{Arata et~al.}{2020}]{Arata2020}
Arata S.,  Yajima H.,  Nagamine K.,  Abe M.,   Khochfar S.,  2020, arXiv
  e-prints

\bibitem[\protect\citeauthoryear{{Aravena} et~al.,}{{Aravena}
  et~al.}{2024}]{Aravena2024}
{Aravena} M.,  et~al., 2024, \mn@doi [\aap] {10.1051/0004-6361/202347281},
  \href {https://ui.adsabs.harvard.edu/abs/2024A&A...682A..24A} {682, A24}

\bibitem[\protect\citeauthoryear{{Arrigoni Battaia} et~al.,}{{Arrigoni Battaia}
  et~al.}{2023}]{Arrigoni2023}
{Arrigoni Battaia} F.,  et~al., 2023, \mn@doi [\aap]
  {10.1051/0004-6361/202245520}, \href
  {https://ui.adsabs.harvard.edu/abs/2023A&A...676A..51A} {676, A51}

\bibitem[\protect\citeauthoryear{{Asano}, {Takeuchi}, {Hirashita}  \&
  {Inoue}}{{Asano} et~al.}{2013}]{Asano2013}
{Asano} R.~S.,  {Takeuchi} T.~T.,  {Hirashita} H.,   {Inoue} A.~K.,  2013,
  \mn@doi [Earth, Planets, and Space] {10.5047/eps.2012.04.014}, \href
  {https://ui.adsabs.harvard.edu/abs/2013EP&S...65..213A} {65, 213}

\bibitem[\protect\citeauthoryear{{Atek}, {Shuntov}, {Furtak}, {Richard},
  {Kneib}, {Mahler Adi Zitrin}  \& {McCracken Clotilde Laigle St{\'e}phane
  Charlot}}{{Atek} et~al.}{2022}]{Atek2022}
{Atek} H.,  {Shuntov} M.,  {Furtak} L.~J.,  {Richard} J.,  {Kneib} J.-P.,
  {Mahler Adi Zitrin} G.,   {McCracken Clotilde Laigle St{\'e}phane Charlot}
  H.~J.,  2022, arXiv e-prints, \href
  {https://ui.adsabs.harvard.edu/abs/2022arXiv220712338A} {p. arXiv:2207.12338}

\bibitem[\protect\citeauthoryear{{Bakx} et~al.,}{{Bakx}
  et~al.}{2020}]{Bakx2020}
{Bakx} T. J.~L.~C.,  et~al., 2020, \mn@doi [\mnras] {10.1093/mnras/staa509},
  \href {https://ui.adsabs.harvard.edu/abs/2020MNRAS.493.4294B} {493, 4294}

\bibitem[\protect\citeauthoryear{{Bakx} et~al.,}{{Bakx}
  et~al.}{2021}]{Bakx2021}
{Bakx} T. J.~L.~C.,  et~al., 2021, \mn@doi [\mnras] {10.1093/mnrasl/slab104},
  \href {https://ui.adsabs.harvard.edu/abs/2021MNRAS.508L..58B} {508, L58}

\bibitem[\protect\citeauthoryear{{Bakx} et~al.,}{{Bakx}
  et~al.}{2024}]{Bakx2024}
{Bakx} T. J.~L.~C.,  et~al., 2024, \mn@doi [\mnras] {10.1093/mnras/stae1155},
  \href {https://ui.adsabs.harvard.edu/abs/2024MNRAS.530.4578B} {530, 4578}

\bibitem[\protect\citeauthoryear{{Barrufet} et~al.,}{{Barrufet}
  et~al.}{2023}]{Barrufet2023}
{Barrufet} L.,  et~al., 2023, \mn@doi [\mnras] {10.1093/mnras/stad1259}, \href
  {https://ui.adsabs.harvard.edu/abs/2023MNRAS.522.3926B} {522, 3926}

\bibitem[\protect\citeauthoryear{{Behrens}, {Pallottini}, {Ferrara},
  {Gallerani}  \& {Vallini}}{{Behrens} et~al.}{2018}]{Behrens2018}
{Behrens} C.,  {Pallottini} A.,  {Ferrara} A.,  {Gallerani} S.,   {Vallini} L.,
   2018, \mn@doi [\mnras] {10.1093/mnras/sty552}, \href
  {https://ui.adsabs.harvard.edu/abs/2018MNRAS.477..552B} {477, 552}

\bibitem[\protect\citeauthoryear{{Bendo} et~al.,}{{Bendo}
  et~al.}{2023}]{Bendo2023}
{Bendo} G.~J.,  et~al., 2023, \mn@doi [\mnras] {10.1093/mnras/stac3771}, \href
  {https://ui.adsabs.harvard.edu/abs/2023MNRAS.522.2995B} {522, 2995}

\bibitem[\protect\citeauthoryear{{B{\'e}thermin} et~al.,}{{B{\'e}thermin}
  et~al.}{2023}]{Bethermin2023}
{B{\'e}thermin} M.,  et~al., 2023, \mn@doi [\aap]
  {10.1051/0004-6361/202348115}, \href
  {https://ui.adsabs.harvard.edu/abs/2023A&A...680L...8B} {680, L8}

\bibitem[\protect\citeauthoryear{{Biffi}, {Planelles}, {Borgani}, {Rasia},
  {Murante}, {Fabjan}  \& {Gaspari}}{{Biffi} et~al.}{2018}]{Biffi2018}
{Biffi} V.,  {Planelles} S.,  {Borgani} S.,  {Rasia} E.,  {Murante} G.,
  {Fabjan} D.,   {Gaspari} M.,  2018, \mn@doi [\mnras] {10.1093/mnras/sty363},
  \href {https://ui.adsabs.harvard.edu/abs/2018MNRAS.476.2689B} {476, 2689}

\bibitem[\protect\citeauthoryear{{Bisbas} et~al.,}{{Bisbas}
  et~al.}{2022}]{Bisbas2022}
{Bisbas} T.~G.,  et~al., 2022, \mn@doi [\apj] {10.3847/1538-4357/ac7960}, \href
  {https://ui.adsabs.harvard.edu/abs/2022ApJ...934..115B} {934, 115}

\bibitem[\protect\citeauthoryear{{Bisigello} et~al.,}{{Bisigello}
  et~al.}{2022}]{Bisigello2022}
{Bisigello} L.,  et~al., 2022, \mn@doi [\aap] {10.1051/0004-6361/202244019},
  \href {https://ui.adsabs.harvard.edu/abs/2022A&A...666A.193B} {666, A193}

\bibitem[\protect\citeauthoryear{{Boogaard} et~al.,}{{Boogaard}
  et~al.}{2020}]{Boogaard2020}
{Boogaard} L.~A.,  et~al., 2020, \mn@doi [\apj] {10.3847/1538-4357/abb82f},
  \href {https://ui.adsabs.harvard.edu/abs/2020ApJ...902..109B} {902, 109}

\bibitem[\protect\citeauthoryear{{Boogaard}, {Meyer}  \& {Novak}}{{Boogaard}
  et~al.}{2021}]{Boogaard2021}
{Boogaard} L.,  {Meyer} R.~A.,   {Novak} M.,  2021, {Interferopy: analysing
  datacubes from radio-to-submm observations}, Zenodo,
  \mn@doi{10.5281/zenodo.5775604}

\bibitem[\protect\citeauthoryear{{Bouwens} et~al.,}{{Bouwens}
  et~al.}{2015}]{Bouwens2015}
{Bouwens} R.~J.,  et~al., 2015, \mn@doi [\apj] {10.1088/0004-637X/803/1/34},
  \href {https://ui.adsabs.harvard.edu/abs/2015ApJ...803...34B} {803, 34}

\bibitem[\protect\citeauthoryear{{Bouwens} et~al.,}{{Bouwens}
  et~al.}{2021}]{Bouwens2021Rebels}
{Bouwens} R.~J.,  et~al., 2021, arXiv e-prints, \href
  {https://ui.adsabs.harvard.edu/abs/2021arXiv210613719B} {p. arXiv:2106.13719}

\bibitem[\protect\citeauthoryear{{Bowler}, {Cullen}, {McLure}, {Dunlop}  \&
  {Avison}}{{Bowler} et~al.}{2022}]{Bowler2022}
{Bowler} R.~A.~A.,  {Cullen} F.,  {McLure} R.~J.,  {Dunlop} J.~S.,   {Avison}
  A.,  2022, \mn@doi [\mnras] {10.1093/mnras/stab3744}, \href
  {https://ui.adsabs.harvard.edu/abs/2022MNRAS.510.5088B} {510, 5088}

\bibitem[\protect\citeauthoryear{{Boylan-Kolchin}}{{Boylan-Kolchin}}{2022}]{Boylan-Kolchin2022}
{Boylan-Kolchin} M.,  2022, arXiv e-prints, \href
  {https://ui.adsabs.harvard.edu/abs/2022arXiv220801611B} {p. arXiv:2208.01611}

\bibitem[\protect\citeauthoryear{{Brauher}, {Dale}  \& {Helou}}{{Brauher}
  et~al.}{2008}]{Brauher2008}
{Brauher} J.~R.,  {Dale} D.~A.,   {Helou} G.,  2008, \mn@doi [\apjs]
  {10.1086/590249}, \href
  {https://ui.adsabs.harvard.edu/abs/2008ApJS..178..280B} {178, 280}

\bibitem[\protect\citeauthoryear{{Bunker} et~al.,}{{Bunker}
  et~al.}{2023}]{Bunker2023}
{Bunker} A.~J.,  et~al., 2023, \mn@doi [\aap] {10.1051/0004-6361/202346159},
  \href {https://ui.adsabs.harvard.edu/abs/2023A&A...677A..88B} {677, A88}

\bibitem[\protect\citeauthoryear{{CASA Team} et~al.,}{{CASA Team}
  et~al.}{2022}]{CASATEAM2022}
{CASA Team} et~al., 2022, \mn@doi [\pasp] {10.1088/1538-3873/ac9642}, \href
  {https://ui.adsabs.harvard.edu/abs/2022PASP..134k4501C} {134, 114501}

\bibitem[\protect\citeauthoryear{{Capak} et~al.,}{{Capak}
  et~al.}{2015}]{Capak2015}
{Capak} P.~L.,  et~al., 2015, \mn@doi [\nat] {10.1038/nature14500}, \href
  {https://ui.adsabs.harvard.edu/abs/2015Natur.522..455C} {522, 455}

\bibitem[\protect\citeauthoryear{{Carniani} et~al.,}{{Carniani}
  et~al.}{2017}]{carniani:2017oiii}
{Carniani} S.,  et~al., 2017, \mn@doi [\aap] {10.1051/0004-6361/201630366},
  \href {https://ui.adsabs.harvard.edu/abs/2017A&A...605A..42C} {605, A42}

\bibitem[\protect\citeauthoryear{{Carniani} et~al.,}{{Carniani}
  et~al.}{2020}]{Carniani2020}
{Carniani} S.,  et~al., 2020, \mn@doi [\mnras] {10.1093/mnras/staa3178}, \href
  {https://ui.adsabs.harvard.edu/abs/2020MNRAS.499.5136C} {499, 5136}

\bibitem[\protect\citeauthoryear{{Carniani} et~al.,}{{Carniani}
  et~al.}{2023a}]{Carniani2023}
{Carniani} S.,  et~al., 2023a, \mn@doi [arXiv e-prints]
  {10.48550/arXiv.2306.11801}, \href
  {https://ui.adsabs.harvard.edu/abs/2023arXiv230611801C} {p. arXiv:2306.11801}

\bibitem[\protect\citeauthoryear{{Carniani} et~al.,}{{Carniani}
  et~al.}{2023b}]{Xu2023}
{Carniani} S.,  et~al., 2023b, \mn@doi [arXiv e-prints]
  {10.48550/arXiv.2306.11801}, \href
  {https://ui.adsabs.harvard.edu/abs/2023arXiv230611801C} {p. arXiv:2306.11801}

\bibitem[\protect\citeauthoryear{{Carpenter}, {Brogan}, {Iono}  \&
  {Mroczkowski}}{{Carpenter} et~al.}{2022}]{Carpenter2022}
{Carpenter} J.,  {Brogan} C.~L.,  {Iono} D.,   {Mroczkowski} T.,  2022, arXiv
  e-prints, \href {https://ui.adsabs.harvard.edu/abs/2022arXiv221100195C} {p.
  arXiv:2211.00195}

\bibitem[\protect\citeauthoryear{{Casey}, {Narayanan}  \& {Cooray}}{{Casey}
  et~al.}{2014}]{Casey2014}
{Casey} C.~M.,  {Narayanan} D.,   {Cooray} A.,  2014, \mn@doi [\physrep]
  {10.1016/j.physrep.2014.02.009}, \href
  {https://ui.adsabs.harvard.edu/abs/2014PhR...541...45C} {541, 45}

\bibitem[\protect\citeauthoryear{{Castellano} et~al.,}{{Castellano}
  et~al.}{2022}]{Castellano2022}
{Castellano} M.,  et~al., 2022, arXiv e-prints, \href
  {https://ui.adsabs.harvard.edu/abs/2022arXiv220709436C} {p. arXiv:2207.09436}

\bibitem[\protect\citeauthoryear{{Chiang}, {Overzier}  \& {Gebhardt}}{{Chiang}
  et~al.}{2013}]{Chiang2013}
{Chiang} Y.-K.,  {Overzier} R.,   {Gebhardt} K.,  2013, \mn@doi [\apj]
  {10.1088/0004-637X/779/2/127}, \href
  {https://ui.adsabs.harvard.edu/abs/2013ApJ...779..127C} {779, 127}

\bibitem[\protect\citeauthoryear{{Chiang}, {Overzier}, {Gebhardt}  \&
  {Henriques}}{{Chiang} et~al.}{2017}]{Chiang2017}
{Chiang} Y.-K.,  {Overzier} R.~A.,  {Gebhardt} K.,   {Henriques} B.,  2017,
  \mn@doi [\apjl] {10.3847/2041-8213/aa7e7b}, \href
  {https://ui.adsabs.harvard.edu/abs/2017ApJ...844L..23C} {844, L23}

\bibitem[\protect\citeauthoryear{{Cormier} et~al.,}{{Cormier}
  et~al.}{2015}]{Cormier2015}
{Cormier} D.,  et~al., 2015, \mn@doi [\aap] {10.1051/0004-6361/201425207},
  \href {https://ui.adsabs.harvard.edu/abs/2015A&A...578A..53C} {578, A53}

\bibitem[\protect\citeauthoryear{{Cormier} et~al.,}{{Cormier}
  et~al.}{2019}]{Cormier2019}
{Cormier} D.,  et~al., 2019, \mn@doi [\aap] {10.1051/0004-6361/201834457},
  \href {https://ui.adsabs.harvard.edu/abs/2019A&A...626A..23C} {626, A23}

\bibitem[\protect\citeauthoryear{{Curti} et~al.,}{{Curti}
  et~al.}{2024}]{Curti2024}
{Curti} M.,  et~al., 2024, \mn@doi [\aap] {10.1051/0004-6361/202346698}, \href
  {https://ui.adsabs.harvard.edu/abs/2024A&A...684A..75C} {684, A75}

\bibitem[\protect\citeauthoryear{{Dayal} et~al.,}{{Dayal}
  et~al.}{2022}]{Dayal2022}
{Dayal} P.,  et~al., 2022, \mn@doi [\mnras] {10.1093/mnras/stac537}, \href
  {https://ui.adsabs.harvard.edu/abs/2022MNRAS.512..989D} {512, 989}

\bibitem[\protect\citeauthoryear{{De Cia}, {Ledoux}, {Savaglio}, {Schady}  \&
  {Vreeswijk}}{{De Cia} et~al.}{2013}]{DeCia2013}
{De Cia} A.,  {Ledoux} C.,  {Savaglio} S.,  {Schady} P.,   {Vreeswijk} P.~M.,
  2013, \mn@doi [\aap] {10.1051/0004-6361/201321834}, \href
  {https://ui.adsabs.harvard.edu/abs/2013A&A...560A..88D} {560, A88}

\bibitem[\protect\citeauthoryear{{De Looze} et~al.,}{{De Looze}
  et~al.}{2014}]{delooze14}
{De Looze} I.,  et~al., 2014, \mn@doi [\aap] {10.1051/0004-6361/201322489},
  \href {https://ui.adsabs.harvard.edu/abs/2014A&A...568A..62D} {568, A62}

\bibitem[\protect\citeauthoryear{{Decarli} et~al.,}{{Decarli}
  et~al.}{2014}]{Decarli2014}
{Decarli} R.,  et~al., 2014, \mn@doi [\apj] {10.1088/0004-637X/782/2/78}, \href
  {https://ui.adsabs.harvard.edu/abs/2014ApJ...782...78D} {782, 78}

\bibitem[\protect\citeauthoryear{{Decarli} et~al.,}{{Decarli}
  et~al.}{2016}]{Decarli2016}
{Decarli} R.,  et~al., 2016, \mn@doi [\apj] {10.3847/1538-4357/833/1/69}, \href
  {https://ui.adsabs.harvard.edu/abs/2016ApJ...833...69D} {833, 69}

\bibitem[\protect\citeauthoryear{{Decarli} et~al.,}{{Decarli}
  et~al.}{2020}]{Decarli2020}
{Decarli} R.,  et~al., 2020, \mn@doi [\apj] {10.3847/1538-4357/abaa3b}, \href
  {https://ui.adsabs.harvard.edu/abs/2020ApJ...902..110D} {902, 110}

\bibitem[\protect\citeauthoryear{{Decarli} et~al.,}{{Decarli}
  et~al.}{2023}]{Decarli2023}
{Decarli} R.,  et~al., 2023, \mn@doi [\aap] {10.1051/0004-6361/202245674},
  \href {https://ui.adsabs.harvard.edu/abs/2023A&A...673A.157D} {673, A157}

\bibitem[\protect\citeauthoryear{{Dessauges-Zavadsky}
  et~al.,}{{Dessauges-Zavadsky} et~al.}{2020}]{Dessauges-Zavadsky2020}
{Dessauges-Zavadsky} M.,  et~al., 2020, \mn@doi [\aap]
  {10.1051/0004-6361/202038231}, \href
  {https://ui.adsabs.harvard.edu/abs/2020A&A...643A...5D} {643, A5}

\bibitem[\protect\citeauthoryear{{Di Cesare}, {Graziani}, {Schneider},
  {Ginolfi}, {Venditti}, {Santini}  \& {Hunt}}{{Di Cesare}
  et~al.}{2023}]{DiCesare2023}
{Di Cesare} C.,  {Graziani} L.,  {Schneider} R.,  {Ginolfi} M.,  {Venditti} A.,
   {Santini} P.,   {Hunt} L.~K.,  2023, \mn@doi [\mnras]
  {10.1093/mnras/stac3702}, \href
  {https://ui.adsabs.harvard.edu/abs/2023MNRAS.519.4632D} {519, 4632}

\bibitem[\protect\citeauthoryear{{Donnan} et~al.,}{{Donnan}
  et~al.}{2022}]{Donnan2022}
{Donnan} C.~T.,  et~al., 2022, arXiv e-prints, \href
  {https://ui.adsabs.harvard.edu/abs/2022arXiv220712356D} {p. arXiv:2207.12356}

\bibitem[\protect\citeauthoryear{{Eddington}}{{Eddington}}{1913}]{Eddington1913}
{Eddington} A.~S.,  1913, \mn@doi [\mnras] {10.1093/mnras/73.5.359}, \href
  {https://ui.adsabs.harvard.edu/abs/1913MNRAS..73..359E} {73, 359}

\bibitem[\protect\citeauthoryear{{Faisst}, {Fudamoto}, {Oesch}, {Scoville},
  {Riechers}, {Pavesi}  \& {Capak}}{{Faisst} et~al.}{2020}]{Faisst2020}
{Faisst} A.~L.,  {Fudamoto} Y.,  {Oesch} P.~A.,  {Scoville} N.,  {Riechers}
  D.~A.,  {Pavesi} R.,   {Capak} P.,  2020, \mn@doi [\mnras]
  {10.1093/mnras/staa2545}, \href
  {https://ui.adsabs.harvard.edu/abs/2020MNRAS.498.4192F} {498, 4192}

\bibitem[\protect\citeauthoryear{{Faisst} et~al.,}{{Faisst}
  et~al.}{2022}]{Faisst2022}
{Faisst} A.~L.,  et~al., 2022, \mn@doi [Universe] {10.3390/universe8060314},
  \href {https://ui.adsabs.harvard.edu/abs/2022Univ....8..314F} {8, 314}

\bibitem[\protect\citeauthoryear{{Ferland} et~al.,}{{Ferland}
  et~al.}{2017}]{Ferland2017}
{Ferland} G.~J.,  et~al., 2017, \rmxaa, \href
  {https://ui.adsabs.harvard.edu/abs/2017RMxAA..53..385F} {53, 385}

\bibitem[\protect\citeauthoryear{{Ferrara}, {Hirashita}, {Ouchi}  \&
  {Fujimoto}}{{Ferrara} et~al.}{2017}]{Ferrara2017}
{Ferrara} A.,  {Hirashita} H.,  {Ouchi} M.,   {Fujimoto} S.,  2017, \mn@doi
  [\mnras] {10.1093/mnras/stx1898}, \href
  {https://ui.adsabs.harvard.edu/abs/2017MNRAS.471.5018F} {471, 5018}

\bibitem[\protect\citeauthoryear{{Ferrara}, {Vallini}, {Pallottini},
  {Gallerani}, {Carniani}, {Kohandel}, {Decataldo}  \& {Behrens}}{{Ferrara}
  et~al.}{2019}]{Ferrara2019}
{Ferrara} A.,  {Vallini} L.,  {Pallottini} A.,  {Gallerani} S.,  {Carniani} S.,
   {Kohandel} M.,  {Decataldo} D.,   {Behrens} C.,  2019, \mn@doi [\mnras]
  {10.1093/mnras/stz2031}, \href
  {https://ui.adsabs.harvard.edu/abs/2019MNRAS.489....1F} {489, 1}

\bibitem[\protect\citeauthoryear{{Ferrara}, {Pallottini}  \& {Dayal}}{{Ferrara}
  et~al.}{2022}]{Ferrara2022}
{Ferrara} A.,  {Pallottini} A.,   {Dayal} P.,  2022, arXiv e-prints, \href
  {https://ui.adsabs.harvard.edu/abs/2022arXiv220800720F} {p. arXiv:2208.00720}

\bibitem[\protect\citeauthoryear{{Ferrara}, {Zana}, {Gallerani}  \&
  {Sommovigo}}{{Ferrara} et~al.}{2023}]{Ferrara2023}
{Ferrara} A.,  {Zana} T.,  {Gallerani} S.,   {Sommovigo} L.,  2023, \mn@doi
  [\mnras] {10.1093/mnras/stad299}, \href
  {https://ui.adsabs.harvard.edu/abs/2023MNRAS.520.3089F} {520, 3089}

\bibitem[\protect\citeauthoryear{{Finkelstein} et~al.,}{{Finkelstein}
  et~al.}{2022}]{Finkelstein2022}
{Finkelstein} S.~L.,  et~al., 2022, arXiv e-prints, \href
  {https://ui.adsabs.harvard.edu/abs/2022arXiv220712474F} {p. arXiv:2207.12474}

\bibitem[\protect\citeauthoryear{{Fixsen}, {Bennett}  \& {Mather}}{{Fixsen}
  et~al.}{1999}]{Fixsen1999}
{Fixsen} D.~J.,  {Bennett} C.~L.,   {Mather} J.~C.,  1999, \mn@doi [\apj]
  {10.1086/307962}, \href
  {https://ui.adsabs.harvard.edu/abs/1999ApJ...526..207F} {526, 207}

\bibitem[\protect\citeauthoryear{{Fudamoto} et~al.,}{{Fudamoto}
  et~al.}{2021}]{fudamoto2021}
{Fudamoto} Y.,  et~al., 2021, \mn@doi [\nat] {10.1038/s41586-021-03846-z},
  \href {https://ui.adsabs.harvard.edu/abs/2021Natur.597..489F} {597, 489}

\bibitem[\protect\citeauthoryear{{Fudamoto}, {Inoue}  \& {Sugahara}}{{Fudamoto}
  et~al.}{2022}]{Fudamoto2022}
{Fudamoto} Y.,  {Inoue} A.~K.,   {Sugahara} Y.,  2022, arXiv e-prints, \href
  {https://ui.adsabs.harvard.edu/abs/2022arXiv220601879F} {p. arXiv:2206.01879}

\bibitem[\protect\citeauthoryear{{Fujimoto} et~al.,}{{Fujimoto}
  et~al.}{2019}]{Fujimoto2019}
{Fujimoto} S.,  et~al., 2019, arXiv e-prints, \href
  {https://ui.adsabs.harvard.edu/abs/2019arXiv190206760F} {p. arXiv:1902.06760}

\bibitem[\protect\citeauthoryear{{Fujimoto} et~al.,}{{Fujimoto}
  et~al.}{2021}]{Fujimoto2021}
{Fujimoto} S.,  et~al., 2021, \mn@doi [\apj] {10.3847/1538-4357/abd7ec}, \href
  {https://ui.adsabs.harvard.edu/abs/2021ApJ...911...99F} {911, 99}

\bibitem[\protect\citeauthoryear{{Fujimoto} et~al.,}{{Fujimoto}
  et~al.}{2022}]{Fujimoto2022}
{Fujimoto} S.,  et~al., 2022, arXiv e-prints, \href
  {https://ui.adsabs.harvard.edu/abs/2022arXiv221103896F} {p. arXiv:2211.03896}

\bibitem[\protect\citeauthoryear{{Fujimoto} et~al.,}{{Fujimoto}
  et~al.}{2023}]{Fujimoto2023}
{Fujimoto} S.,  et~al., 2023, \mn@doi [arXiv e-prints]
  {10.48550/arXiv.2303.01658}, \href
  {https://ui.adsabs.harvard.edu/abs/2023arXiv230301658F} {p. arXiv:2303.01658}

\bibitem[\protect\citeauthoryear{{Gallerani}, {Pallottini}, {Feruglio},
  {Ferrara}, {Maiolino}, {Vallini}, {Riechers}  \& {Pavesi}}{{Gallerani}
  et~al.}{2018}]{Gallerani2018}
{Gallerani} S.,  {Pallottini} A.,  {Feruglio} C.,  {Ferrara} A.,  {Maiolino}
  R.,  {Vallini} L.,  {Riechers} D.~A.,   {Pavesi} R.,  2018, \mn@doi [\mnras]
  {10.1093/mnras/stx2458}, \href
  {https://ui.adsabs.harvard.edu/abs/2018MNRAS.473.1909G} {473, 1909}

\bibitem[\protect\citeauthoryear{{Genzel} et~al.,}{{Genzel}
  et~al.}{2015}]{Genzel2015}
{Genzel} R.,  et~al., 2015, \mn@doi [\apj] {10.1088/0004-637X/800/1/20}, \href
  {https://ui.adsabs.harvard.edu/abs/2015ApJ...800...20G} {800, 20}

\bibitem[\protect\citeauthoryear{{Ginolfi} et~al.,}{{Ginolfi}
  et~al.}{2019}]{Ginolfi2019}
{Ginolfi} M.,  et~al., 2019, arXiv e-prints, \href
  {https://ui.adsabs.harvard.edu/abs/2019arXiv191004770G} {p. arXiv:1910.04770}

\bibitem[\protect\citeauthoryear{{Ginsburg} et~al.,}{{Ginsburg}
  et~al.}{2019}]{Ginsburg2019}
{Ginsburg} A.,  et~al., 2019, {radio-astro-tools/spectral-cube: Release
  v0.4.5}, Zenodo, \mn@doi{10.5281/zenodo.3558614}

\bibitem[\protect\citeauthoryear{{Gonz{\'a}lez-L{\'o}pez}
  et~al.,}{{Gonz{\'a}lez-L{\'o}pez}
  et~al.}{2019}]{GonzalezLopez2019ApJ...882..139G}
{Gonz{\'a}lez-L{\'o}pez} J.,  et~al., 2019, \mn@doi [\apj]
  {10.3847/1538-4357/ab3105}, \href
  {https://ui.adsabs.harvard.edu/abs/2019ApJ...882..139G} {882, 139}

\bibitem[\protect\citeauthoryear{{Gullberg} et~al.,}{{Gullberg}
  et~al.}{2015}]{Gullberg2015}
{Gullberg} B.,  et~al., 2015, \mn@doi [\mnras] {10.1093/mnras/stv372}, \href
  {https://ui.adsabs.harvard.edu/abs/2015MNRAS.449.2883G} {449, 2883}

\bibitem[\protect\citeauthoryear{{Hagimoto} et~al.,}{{Hagimoto}
  et~al.}{2023}]{Hagimoto2023}
{Hagimoto} M.,  et~al., 2023, \mn@doi [\mnras] {10.1093/mnras/stad784}, \href
  {https://ui.adsabs.harvard.edu/abs/2023MNRAS.521.5508H} {521, 5508}

\bibitem[\protect\citeauthoryear{{Harikane} et~al.,}{{Harikane}
  et~al.}{2020}]{Harikane2019}
{Harikane} Y.,  et~al., 2020, \mn@doi [\apj] {10.3847/1538-4357/ab94bd}, \href
  {https://ui.adsabs.harvard.edu/abs/2020ApJ...896...93H} {896, 93}

\bibitem[\protect\citeauthoryear{{Harikane}, {Nakajima}, {Ouchi}, {Umeda},
  {Isobe}, {Ono}, {Xu}  \& {Zhang}}{{Harikane} et~al.}{2023a}]{Harikane2023}
{Harikane} Y.,  {Nakajima} K.,  {Ouchi} M.,  {Umeda} H.,  {Isobe} Y.,  {Ono}
  Y.,  {Xu} Y.,   {Zhang} Y.,  2023a, \mn@doi [arXiv e-prints]
  {10.48550/arXiv.2304.06658}, \href
  {https://ui.adsabs.harvard.edu/abs/2023arXiv230406658H} {p. arXiv:2304.06658}

\bibitem[\protect\citeauthoryear{{Harikane} et~al.,}{{Harikane}
  et~al.}{2023b}]{Harikane2022}
{Harikane} Y.,  et~al., 2023b, \mn@doi [\apjs] {10.3847/1538-4365/acaaa9},
  \href {https://ui.adsabs.harvard.edu/abs/2023ApJS..265....5H} {265, 5}

\bibitem[\protect\citeauthoryear{{Hashimoto} et~al.,}{{Hashimoto}
  et~al.}{2018}]{Hashimoto2018}
{Hashimoto} T.,  et~al., 2018, \mn@doi [\nat] {10.1038/s41586-018-0117-z},
  \href {https://ui.adsabs.harvard.edu/abs/2018Natur.557..392H} {557, 392}

\bibitem[\protect\citeauthoryear{{Hashimoto} et~al.,}{{Hashimoto}
  et~al.}{2019}]{Hashimoto2019}
{Hashimoto} T.,  et~al., 2019, \mn@doi [\pasj] {10.1093/pasj/psz049}, \href
  {https://ui.adsabs.harvard.edu/abs/2019PASJ..tmp...70H} {p.~70}

\bibitem[\protect\citeauthoryear{{Hayatsu} et~al.,}{{Hayatsu}
  et~al.}{2017}]{Hayatsu2017}
{Hayatsu} N.~H.,  et~al., 2017, \mn@doi [\pasj] {10.1093/pasj/psx018}, \href
  {https://ui.adsabs.harvard.edu/abs/2017PASJ...69...45H} {69, 45}

\bibitem[\protect\citeauthoryear{{Hayatsu} et~al.,}{{Hayatsu}
  et~al.}{2019}]{Hayatsu2019}
{Hayatsu} N.~H.,  et~al., 2019, \mn@doi [Research Notes of the American
  Astronomical Society] {10.3847/2515-5172/ab3228}, \href
  {https://ui.adsabs.harvard.edu/abs/2019RNAAS...3...97H} {3, 97}

\bibitem[\protect\citeauthoryear{{Inami} et~al.,}{{Inami}
  et~al.}{2022}]{Inami2022}
{Inami} H.,  et~al., 2022, \mn@doi [\mnras] {10.1093/mnras/stac1779}, \href
  {https://ui.adsabs.harvard.edu/abs/2022MNRAS.515.3126I} {515, 3126}

\bibitem[\protect\citeauthoryear{{Inoue} et~al.,}{{Inoue}
  et~al.}{2016}]{Inoue2016}
{Inoue} A.~K.,  et~al., 2016, \mn@doi [Science] {10.1126/science.aaf0714},
  \href {https://ui.adsabs.harvard.edu/abs/2016Sci...352.1559I} {352, 1559}

\bibitem[\protect\citeauthoryear{{Kashino}, {Lilly}, {Matthee}, {Eilers},
  {Mackenzie}, {Bordoloi}  \& {Simcoe}}{{Kashino} et~al.}{2023}]{Kashino2023}
{Kashino} D.,  {Lilly} S.~J.,  {Matthee} J.,  {Eilers} A.-C.,  {Mackenzie} R.,
  {Bordoloi} R.,   {Simcoe} R.~A.,  2023, \mn@doi [\apj]
  {10.3847/1538-4357/acc588}, \href
  {https://ui.adsabs.harvard.edu/abs/2023ApJ...950...66K} {950, 66}

\bibitem[\protect\citeauthoryear{{Katz} et~al.,}{{Katz}
  et~al.}{2019}]{Katz2019}
{Katz} H.,  et~al., 2019, \mn@doi [\mnras] {10.1093/mnras/stz1672}, \href
  {https://ui.adsabs.harvard.edu/abs/2019MNRAS.487.5902K} {487, 5902}

\bibitem[\protect\citeauthoryear{{Katz} et~al.,}{{Katz}
  et~al.}{2021}]{Katz2021}
{Katz} H.,  et~al., 2021, \mn@doi [\mnras] {10.1093/mnras/stab2148}, \href
  {https://ui.adsabs.harvard.edu/abs/2021MNRAS.507.1254K} {507, 1254}

\bibitem[\protect\citeauthoryear{{Katz} et~al.,}{{Katz}
  et~al.}{2022}]{Katz2022}
{Katz} H.,  et~al., 2022, \mn@doi [\mnras] {10.1093/mnras/stac028}, \href
  {https://ui.adsabs.harvard.edu/abs/2022MNRAS.510.5603K} {510, 5603}

\bibitem[\protect\citeauthoryear{{Kennicutt}}{{Kennicutt}}{1998}]{Kennicutt1998}
{Kennicutt} Robert~C. J.,  1998, \mn@doi [\apj] {10.1086/305588}, \href
  {https://ui.adsabs.harvard.edu/abs/1998ApJ...498..541K} {498, 541}

\bibitem[\protect\citeauthoryear{{Killi} et~al.,}{{Killi}
  et~al.}{2024}]{Killi2024}
{Killi} M.,  et~al., 2024, \mn@doi [arXiv e-prints]
  {10.48550/arXiv.2402.07982}, \href
  {https://ui.adsabs.harvard.edu/abs/2024arXiv240207982K} {p. arXiv:2402.07982}

\bibitem[\protect\citeauthoryear{{Koch} et~al.,}{{Koch}
  et~al.}{2021}]{Koch2021}
{Koch} E.,  et~al., 2021, {radio-astro-tools/radio-beam: v0.3.3}, Zenodo,
  \mn@doi{10.5281/zenodo.4623788}

\bibitem[\protect\citeauthoryear{{Kohandel}, {Pallottini}, {Ferrara},
  {Zanella}, {Behrens}, {Carniani}, {Gallerani}  \& {Vallini}}{{Kohandel}
  et~al.}{2019}]{Kohandel2019}
{Kohandel} M.,  {Pallottini} A.,  {Ferrara} A.,  {Zanella} A.,  {Behrens} C.,
  {Carniani} S.,  {Gallerani} S.,   {Vallini} L.,  2019, \mn@doi [\mnras]
  {10.1093/mnras/stz1486}, \href
  {https://ui.adsabs.harvard.edu/abs/2019MNRAS.487.3007K} {487, 3007}

\bibitem[\protect\citeauthoryear{{Konstantopoulou} et~al.,}{{Konstantopoulou}
  et~al.}{2022}]{Konstantopoulou2022}
{Konstantopoulou} C.,  et~al., 2022, \mn@doi [\aap]
  {10.1051/0004-6361/202243994}, \href
  {https://ui.adsabs.harvard.edu/abs/2022A&A...666A..12K} {666, A12}

\bibitem[\protect\citeauthoryear{{Koyama} et~al.,}{{Koyama}
  et~al.}{2013}]{Koyama2013}
{Koyama} Y.,  et~al., 2013, \mn@doi [\mnras] {10.1093/mnras/stt1035}, \href
  {https://ui.adsabs.harvard.edu/abs/2013MNRAS.434..423K} {434, 423}

\bibitem[\protect\citeauthoryear{{Koyama} et~al.,}{{Koyama}
  et~al.}{2021}]{Koyama2021}
{Koyama} Y.,  et~al., 2021, \mn@doi [\mnras] {10.1093/mnrasl/slab013}, \href
  {https://ui.adsabs.harvard.edu/abs/2021MNRAS.503L...1K} {503, L1}

\bibitem[\protect\citeauthoryear{{Kurk}, {Pentericci}, {R{\"o}ttgering}  \&
  {Miley}}{{Kurk} et~al.}{2004}]{Kurk2004}
{Kurk} J.~D.,  {Pentericci} L.,  {R{\"o}ttgering} H.~J.~A.,   {Miley} G.~K.,
  2004, \mn@doi [\aap] {10.1051/0004-6361:20040075}, \href
  {https://ui.adsabs.harvard.edu/abs/2004A&A...428..793K} {428, 793}

\bibitem[\protect\citeauthoryear{{Lagache}, {Cousin}  \& {Chatzikos}}{{Lagache}
  et~al.}{2018}]{Lagache2018}
{Lagache} G.,  {Cousin} M.,   {Chatzikos} M.,  2018, \mn@doi [\aap]
  {10.1051/0004-6361/201732019}, \href
  {https://ui.adsabs.harvard.edu/abs/2018A&A...609A.130L} {609, A130}

\bibitem[\protect\citeauthoryear{{Laporte} et~al.,}{{Laporte}
  et~al.}{2019}]{Laporte2019}
{Laporte} N.,  et~al., 2019, \mn@doi [\mnras] {10.1093/mnrasl/slz094}, \href
  {https://ui.adsabs.harvard.edu/abs/2019MNRAS.487L..81L} {487, L81}

\bibitem[\protect\citeauthoryear{{Le F{\`e}vre}, {B{\'e}thermin}, {Faisst},
  {Capak}, {Cassata}, {Silverman}, {Schaerer}  \& {Yan}}{{Le F{\`e}vre}
  et~al.}{2019}]{Fevre2019}
{Le F{\`e}vre} O.,  {B{\'e}thermin} M.,  {Faisst} A.,  {Capak} P.,  {Cassata}
  P.,  {Silverman} J.~D.,  {Schaerer} D.,   {Yan} L.,  2019, arXiv e-prints,
  \href {https://ui.adsabs.harvard.edu/abs/2019arXiv191009517L} {p.
  arXiv:1910.09517}

\bibitem[\protect\citeauthoryear{{Le F{\`e}vre} et~al.,}{{Le F{\`e}vre}
  et~al.}{2020}]{LeFevre2020}
{Le F{\`e}vre} O.,  et~al., 2020, \mn@doi [\aap] {10.1051/0004-6361/201936965},
  \href {https://ui.adsabs.harvard.edu/abs/2020A&A...643A...1L} {643, A1}

\bibitem[\protect\citeauthoryear{{Le{\'s}niewska} \&
  {Micha{\l}owski}}{{Le{\'s}niewska} \&
  {Micha{\l}owski}}{2019}]{Lesniewska2019}
{Le{\'s}niewska} A.,  {Micha{\l}owski} M.~J.,  2019, \mn@doi [\aap]
  {10.1051/0004-6361/201935149}, \href
  {https://ui.adsabs.harvard.edu/abs/2019A&A...624L..13L} {624, L13}

\bibitem[\protect\citeauthoryear{{Loiacono} et~al.,}{{Loiacono}
  et~al.}{2021}]{Loiacono2021}
{Loiacono} F.,  et~al., 2021, \mn@doi [\aap] {10.1051/0004-6361/202038607},
  \href {https://ui.adsabs.harvard.edu/abs/2021A&A...646A..76L} {646, A76}

\bibitem[\protect\citeauthoryear{{Madden} \& {Cormier}}{{Madden} \&
  {Cormier}}{2018}]{Madden2018}
{Madden} S.~C.,  {Cormier} D.,  2018, arXiv e-prints, \href
  {https://ui.adsabs.harvard.edu/abs/2018arXiv181009953M} {p. arXiv:1810.09953}

\bibitem[\protect\citeauthoryear{{Madden}, {Poglitsch}, {Geis}, {Stacey}  \&
  {Townes}}{{Madden} et~al.}{1997}]{Madden1997}
{Madden} S.~C.,  {Poglitsch} A.,  {Geis} N.,  {Stacey} G.~J.,   {Townes} C.~H.,
   1997, \mn@doi [\apj] {10.1086/304247}, \href
  {https://ui.adsabs.harvard.edu/abs/1997ApJ...483..200M} {483, 200}

\bibitem[\protect\citeauthoryear{{Madden} et~al.,}{{Madden}
  et~al.}{2013}]{Madden2013}
{Madden} S.~C.,  et~al., 2013, \mn@doi [\pasp] {10.1086/671138}, \href
  {https://ui.adsabs.harvard.edu/abs/2013PASP..125..600M} {125, 600}

\bibitem[\protect\citeauthoryear{{Madden} et~al.,}{{Madden}
  et~al.}{2020}]{Madden2020}
{Madden} S.~C.,  et~al., 2020, \mn@doi [\aap] {10.1051/0004-6361/202038860},
  \href {https://ui.adsabs.harvard.edu/abs/2020A&A...643A.141M} {643, A141}

\bibitem[\protect\citeauthoryear{{Magnelli} et~al.,}{{Magnelli}
  et~al.}{2014}]{Magnelli2014}
{Magnelli} B.,  et~al., 2014, \mn@doi [\aap] {10.1051/0004-6361/201322217},
  \href {https://ui.adsabs.harvard.edu/abs/2014A&A...561A..86M} {561, A86}

\bibitem[\protect\citeauthoryear{{Maiolino} \& {Mannucci}}{{Maiolino} \&
  {Mannucci}}{2019}]{Maiolino2019}
{Maiolino} R.,  {Mannucci} F.,  2019, \mn@doi [\aapr]
  {10.1007/s00159-018-0112-2}, \href
  {https://ui.adsabs.harvard.edu/abs/2019A&ARv..27....3M} {27, 3}

\bibitem[\protect\citeauthoryear{{Marrone} et~al.,}{{Marrone}
  et~al.}{2018}]{Marrone2018}
{Marrone} D.~P.,  et~al., 2018, \mn@doi [\nat] {10.1038/nature24629}, \href
  {https://ui.adsabs.harvard.edu/abs/2018Natur.553...51M} {553, 51}

\bibitem[\protect\citeauthoryear{{Matthee}, {Mackenzie}, {Simcoe}, {Kashino},
  {Lilly}, {Bordoloi}  \& {Eilers}}{{Matthee} et~al.}{2023}]{Matthee2023}
{Matthee} J.,  {Mackenzie} R.,  {Simcoe} R.~A.,  {Kashino} D.,  {Lilly} S.~J.,
  {Bordoloi} R.,   {Eilers} A.-C.,  2023, \mn@doi [\apj]
  {10.3847/1538-4357/acc846}, \href
  {https://ui.adsabs.harvard.edu/abs/2023ApJ...950...67M} {950, 67}

\bibitem[\protect\citeauthoryear{{Meijerink}, {Spaans}  \&
  {Israel}}{{Meijerink} et~al.}{2007}]{Meijerink2007}
{Meijerink} R.,  {Spaans} M.,   {Israel} F.~P.,  2007, \mn@doi [\aap]
  {10.1051/0004-6361:20066130}, \href
  {https://ui.adsabs.harvard.edu/abs/2007A&A...461..793M} {461, 793}

\bibitem[\protect\citeauthoryear{{Mitsuhashi} et~al.,}{{Mitsuhashi}
  et~al.}{2023}]{Mitsuhashi2023a}
{Mitsuhashi} I.,  et~al., 2023, \mn@doi [arXiv e-prints]
  {10.48550/arXiv.2311.16857}, \href
  {https://ui.adsabs.harvard.edu/abs/2023arXiv231116857M} {p. arXiv:2311.16857}

\bibitem[\protect\citeauthoryear{{Morishita} \& {Stiavelli}}{{Morishita} \&
  {Stiavelli}}{2022}]{morishita22}
{Morishita} T.,  {Stiavelli} M.,  2022, arXiv e-prints, \href
  {https://ui.adsabs.harvard.edu/abs/2022arXiv220711671M} {p. arXiv:2207.11671}

\bibitem[\protect\citeauthoryear{{Naidu} et~al.,}{{Naidu}
  et~al.}{2022}]{Naidu2022}
{Naidu} R.~P.,  et~al., 2022, arXiv e-prints, \href
  {https://ui.adsabs.harvard.edu/abs/2022arXiv220709434N} {p. arXiv:2207.09434}

\bibitem[\protect\citeauthoryear{{Nakajima}, {Ouchi}, {Isobe}, {Harikane},
  {Zhang}, {Ono}, {Umeda}  \& {Oguri}}{{Nakajima} et~al.}{2023}]{Nakajima2023}
{Nakajima} K.,  {Ouchi} M.,  {Isobe} Y.,  {Harikane} Y.,  {Zhang} Y.,  {Ono}
  Y.,  {Umeda} H.,   {Oguri} M.,  2023, \mn@doi [\apjs]
  {10.3847/1538-4365/acd556}, \href
  {https://ui.adsabs.harvard.edu/abs/2023ApJS..269...33N} {269, 33}

\bibitem[\protect\citeauthoryear{{Novak} et~al.,}{{Novak}
  et~al.}{2019}]{Novak2019}
{Novak} M.,  et~al., 2019, \mn@doi [\apj] {10.3847/1538-4357/ab2beb}, \href
  {https://ui.adsabs.harvard.edu/abs/2019ApJ...881...63N} {881, 63}

\bibitem[\protect\citeauthoryear{{Oesch}, {Bouwens}, {Illingworth}, {Labb{\'e}}
   \& {Stefanon}}{{Oesch} et~al.}{2018}]{Oesch2018}
{Oesch} P.~A.,  {Bouwens} R.~J.,  {Illingworth} G.~D.,  {Labb{\'e}} I.,
  {Stefanon} M.,  2018, \mn@doi [\apj] {10.3847/1538-4357/aab03f}, \href
  {https://ui.adsabs.harvard.edu/abs/2018ApJ...855..105O} {855, 105}

\bibitem[\protect\citeauthoryear{{Overzier}}{{Overzier}}{2016}]{Overzier2016}
{Overzier} R.~A.,  2016, \mn@doi [\aapr] {10.1007/s00159-016-0100-3}, \href
  {https://ui.adsabs.harvard.edu/abs/2016A&ARv..24...14O} {24, 14}

\bibitem[\protect\citeauthoryear{{Pallottini}, {Ferrara}, {Bovino}, {Vallini},
  {Gallerani}, {Maiolino}  \& {Salvadori}}{{Pallottini}
  et~al.}{2017}]{Pallottini2017_Chemistry}
{Pallottini} A.,  {Ferrara} A.,  {Bovino} S.,  {Vallini} L.,  {Gallerani} S.,
  {Maiolino} R.,   {Salvadori} S.,  2017, \mn@doi [\mnras]
  {10.1093/mnras/stx1792}, \href
  {https://ui.adsabs.harvard.edu/abs/2017MNRAS.471.4128P} {471, 4128}

\bibitem[\protect\citeauthoryear{{Pallottini} et~al.,}{{Pallottini}
  et~al.}{2022}]{Pallottini2022}
{Pallottini} A.,  et~al., 2022, \mn@doi [\mnras] {10.1093/mnras/stac1281},
  \href {https://ui.adsabs.harvard.edu/abs/2022MNRAS.513.5621P} {513, 5621}

\bibitem[\protect\citeauthoryear{{Pensabene} et~al.,}{{Pensabene}
  et~al.}{2021}]{Pensabene2021}
{Pensabene} A.,  et~al., 2021, \mn@doi [\aap] {10.1051/0004-6361/202039696},
  \href {https://ui.adsabs.harvard.edu/abs/2021A&A...652A..66P} {652, A66}

\bibitem[\protect\citeauthoryear{{Pizzati}, {Ferrara}, {Pallottini},
  {Gallerani}, {Vallini}, {Decataldo}  \& {Fujimoto}}{{Pizzati}
  et~al.}{2020}]{Pizzati2020}
{Pizzati} E.,  {Ferrara} A.,  {Pallottini} A.,  {Gallerani} S.,  {Vallini} L.,
  {Decataldo} D.,   {Fujimoto} S.,  2020, \mn@doi [\mnras]
  {10.1093/mnras/staa1163}, \href
  {https://ui.adsabs.harvard.edu/abs/2020MNRAS.495..160P} {495, 160}

\bibitem[\protect\citeauthoryear{{Pizzati}, {Ferrara}, {Pallottini},
  {Sommovigo}, {Kohandel}  \& {Carniani}}{{Pizzati} et~al.}{2023}]{Pizzati2023}
{Pizzati} E.,  {Ferrara} A.,  {Pallottini} A.,  {Sommovigo} L.,  {Kohandel} M.,
    {Carniani} S.,  2023, \mn@doi [\mnras] {10.1093/mnras/stac3816}, \href
  {https://ui.adsabs.harvard.edu/abs/2023MNRAS.519.4608P} {519, 4608}

\bibitem[\protect\citeauthoryear{{Planck Collaboration} et~al.,}{{Planck
  Collaboration} et~al.}{2020}]{Planck2020}
{Planck Collaboration} et~al., 2020, \mn@doi [\aap]
  {10.1051/0004-6361/201833910}, \href
  {https://ui.adsabs.harvard.edu/abs/2020A&A...641A...6P} {641, A6}

\bibitem[\protect\citeauthoryear{{Planck Collaboration} et~al.,}{{Planck
  Collaboration} et~al.}{2021}]{Planck2021}
{Planck Collaboration} et~al., 2021, \mn@doi [\aap]
  {10.1051/0004-6361/201833910e}, \href
  {https://ui.adsabs.harvard.edu/abs/2021A&A...652C...4P} {652, C4}

\bibitem[\protect\citeauthoryear{{Popping}, {van Kampen}, {Decarli}, {Spaans},
  {Somerville}  \& {Trager}}{{Popping} et~al.}{2016}]{Popping2016}
{Popping} G.,  {van Kampen} E.,  {Decarli} R.,  {Spaans} M.,  {Somerville}
  R.~S.,   {Trager} S.~C.,  2016, \mn@doi [\mnras] {10.1093/mnras/stw1323},
  \href {https://ui.adsabs.harvard.edu/abs/2016MNRAS.461...93P} {461, 93}

\bibitem[\protect\citeauthoryear{{Riechers} et~al.,}{{Riechers}
  et~al.}{2019}]{Riechers2019}
{Riechers} D.~A.,  et~al., 2019, \mn@doi [\apj] {10.3847/1538-4357/aafc27},
  \href {https://ui.adsabs.harvard.edu/abs/2019ApJ...872....7R} {872, 7}

\bibitem[\protect\citeauthoryear{{Riechers} et~al.,}{{Riechers}
  et~al.}{2020}]{Riechers2020}
{Riechers} D.~A.,  et~al., 2020, \mn@doi [\apjl] {10.3847/2041-8213/ab9595},
  \href {https://ui.adsabs.harvard.edu/abs/2020ApJ...896L..21R} {896, L21}

\bibitem[\protect\citeauthoryear{{Rojas-Ruiz} et~al.,}{{Rojas-Ruiz}
  et~al.}{2024}]{Rojas-Ruiz2024}
{Rojas-Ruiz} S.,  et~al., 2024, \mn@doi [arXiv e-prints]
  {10.48550/arXiv.2404.02960}, \href
  {https://ui.adsabs.harvard.edu/abs/2024arXiv240402960R} {p. arXiv:2404.02960}

\bibitem[\protect\citeauthoryear{{Rybak} et~al.,}{{Rybak}
  et~al.}{2019}]{Rybak2019}
{Rybak} M.,  et~al., 2019, \mn@doi [\apj] {10.3847/1538-4357/ab0e0f}, \href
  {https://ui.adsabs.harvard.edu/abs/2019ApJ...876..112R} {876, 112}

\bibitem[\protect\citeauthoryear{{Savage} \& {Sembach}}{{Savage} \&
  {Sembach}}{1996}]{Savage1996}
{Savage} B.~D.,  {Sembach} K.~R.,  1996, \mn@doi [\araa]
  {10.1146/annurev.astro.34.1.279}, \href
  {https://ui.adsabs.harvard.edu/abs/1996ARA&A..34..279S} {34, 279}

\bibitem[\protect\citeauthoryear{{Schaerer} et~al.,}{{Schaerer}
  et~al.}{2020}]{Schaerer2020}
{Schaerer} D.,  et~al., 2020, \mn@doi [\aap] {10.1051/0004-6361/202037617},
  \href {https://ui.adsabs.harvard.edu/abs/2020A&A...643A...3S} {643, A3}

\bibitem[\protect\citeauthoryear{{Schneider} \& {Maiolino}}{{Schneider} \&
  {Maiolino}}{2024}]{Schneider2024}
{Schneider} R.,  {Maiolino} R.,  2024, \mn@doi [\aapr]
  {10.1007/s00159-024-00151-2}, \href
  {https://ui.adsabs.harvard.edu/abs/2024A&ARv..32....2S} {32, 2}

\bibitem[\protect\citeauthoryear{{Serjeant} \& {Bakx}}{{Serjeant} \&
  {Bakx}}{2023}]{Serjeant2023}
{Serjeant} S.,  {Bakx} T. J.~L.~C.,  2023, \mn@doi [Nature Astronomy]
  {10.1038/s41550-023-02093-8}, \href
  {https://ui.adsabs.harvard.edu/abs/2023NatAs.tmp..206S} {}

\bibitem[\protect\citeauthoryear{{Shimakawa}, {Kodama}, {Tadaki}, {Hayashi},
  {Koyama}  \& {Tanaka}}{{Shimakawa} et~al.}{2015}]{Shimakawa2015}
{Shimakawa} R.,  {Kodama} T.,  {Tadaki} K.-i.,  {Hayashi} M.,  {Koyama} Y.,
  {Tanaka} I.,  2015, \mn@doi [\mnras] {10.1093/mnras/stv051}, \href
  {https://ui.adsabs.harvard.edu/abs/2015MNRAS.448..666S} {448, 666}

\bibitem[\protect\citeauthoryear{{Shimakawa} et~al.,}{{Shimakawa}
  et~al.}{2018}]{Shimakawa2018}
{Shimakawa} R.,  et~al., 2018, \mn@doi [\mnras] {10.1093/mnras/stx2494}, \href
  {https://ui.adsabs.harvard.edu/abs/2018MNRAS.473.1977S} {473, 1977}

\bibitem[\protect\citeauthoryear{{Smail}}{{Smail}}{2024}]{Smail2024}
{Smail} I.,  2024, \mn@doi [arXiv e-prints] {10.48550/arXiv.2401.08761}, \href
  {https://ui.adsabs.harvard.edu/abs/2024arXiv240108761S} {p. arXiv:2401.08761}

\bibitem[\protect\citeauthoryear{{Sommovigo}, {Ferrara}, {Carniani}, {Zanella},
  {Pallottini}, {Gallerani}  \& {Vallini}}{{Sommovigo}
  et~al.}{2021}]{Sommovigo2021}
{Sommovigo} L.,  {Ferrara} A.,  {Carniani} S.,  {Zanella} A.,  {Pallottini} A.,
   {Gallerani} S.,   {Vallini} L.,  2021, \mn@doi [\mnras]
  {10.1093/mnras/stab720}, \href
  {https://ui.adsabs.harvard.edu/abs/2021MNRAS.503.4878S} {503, 4878}

\bibitem[\protect\citeauthoryear{{Sommovigo} et~al.,}{{Sommovigo}
  et~al.}{2022}]{Sommovigo2022}
{Sommovigo} L.,  et~al., 2022, \mn@doi [\mnras] {10.1093/mnras/stac302}, \href
  {https://ui.adsabs.harvard.edu/abs/2022MNRAS.513.3122S} {513, 3122}

\bibitem[\protect\citeauthoryear{{Spilker} et~al.,}{{Spilker}
  et~al.}{2020}]{Spilker2020}
{Spilker} J.~S.,  et~al., 2020, \mn@doi [\apj] {10.3847/1538-4357/abc4e6},
  \href {https://ui.adsabs.harvard.edu/abs/2020ApJ...905...86S} {905, 86}

\bibitem[\protect\citeauthoryear{{Stacey}, {Hailey-Dunsheath}, {Ferkinhoff},
  {Nikola}, {Parshley}, {Benford}, {Staguhn}  \& {Fiolet}}{{Stacey}
  et~al.}{2010}]{Stacey2010}
{Stacey} G.~J.,  {Hailey-Dunsheath} S.,  {Ferkinhoff} C.,  {Nikola} T.,
  {Parshley} S.~C.,  {Benford} D.~J.,  {Staguhn} J.~G.,   {Fiolet} N.,  2010,
  \mn@doi [\apj] {10.1088/0004-637X/724/2/957}, \href
  {https://ui.adsabs.harvard.edu/abs/2010ApJ...724..957S} {724, 957}

\bibitem[\protect\citeauthoryear{{Sugahara}, {Ouchi}, {Harikane}, {Bouch{\'e}},
  {Mitchell}  \& {Blaizot}}{{Sugahara} et~al.}{2019}]{Sugahara2019}
{Sugahara} Y.,  {Ouchi} M.,  {Harikane} Y.,  {Bouch{\'e}} N.,  {Mitchell}
  P.~D.,   {Blaizot} J.,  2019, \mn@doi [\apj] {10.3847/1538-4357/ab49fe},
  \href {https://ui.adsabs.harvard.edu/abs/2019ApJ...886...29S} {886, 29}

\bibitem[\protect\citeauthoryear{{Sugahara} et~al.,}{{Sugahara}
  et~al.}{2021}]{Sugahara2021}
{Sugahara} Y.,  et~al., 2021, arXiv e-prints, \href
  {https://ui.adsabs.harvard.edu/abs/2021arXiv210402201S} {p. arXiv:2104.02201}

\bibitem[\protect\citeauthoryear{Sun, Faucher-Giguère, Hayward, Shen, Wetzel
  \& Cochrane}{Sun et~al.}{2023}]{Sun_2023}
Sun G.,  Faucher-Giguère C.-A.,  Hayward C.~C.,  Shen X.,  Wetzel A.,
  Cochrane R.~K.,  2023, \mn@doi [The Astrophysical Journal Letters]
  {10.3847/2041-8213/acf85a}, 955, L35

\bibitem[\protect\citeauthoryear{{Tadaki} et~al.,}{{Tadaki}
  et~al.}{2022}]{Tadaki2022}
{Tadaki} K.-i.,  et~al., 2022, \mn@doi [\pasj] {10.1093/pasj/psac018}, \href
  {https://ui.adsabs.harvard.edu/abs/2022PASJ...74L...9T} {74, L9}

\bibitem[\protect\citeauthoryear{{Tamura} et~al.,}{{Tamura}
  et~al.}{2019}]{Tamura2019}
{Tamura} Y.,  et~al., 2019, \mn@doi [\apj] {10.3847/1538-4357/ab0374}, \href
  {https://ui.adsabs.harvard.edu/abs/2019ApJ...874...27T} {874, 27}

\bibitem[\protect\citeauthoryear{{Tamura} et~al.,}{{Tamura}
  et~al.}{2023}]{Tamura2023}
{Tamura} Y.,  et~al., 2023, \mn@doi [\apj] {10.3847/1538-4357/acd637}, \href
  {https://ui.adsabs.harvard.edu/abs/2023ApJ...952....9T} {952, 9}

\bibitem[\protect\citeauthoryear{{Ura} et~al.,}{{Ura} et~al.}{2023}]{Ura2023}
{Ura} R.,  et~al., 2023, \mn@doi [\apj] {10.3847/1538-4357/acc530}, \href
  {https://ui.adsabs.harvard.edu/abs/2023ApJ...948....3U} {948, 3}

\bibitem[\protect\citeauthoryear{{Vallini}, {Gallerani}, {Ferrara},
  {Pallottini}  \& {Yue}}{{Vallini} et~al.}{2015}]{Vallini2015}
{Vallini} L.,  {Gallerani} S.,  {Ferrara} A.,  {Pallottini} A.,   {Yue} B.,
  2015, \mn@doi [\apj] {10.1088/0004-637X/813/1/36}, \href
  {https://ui.adsabs.harvard.edu/abs/2015ApJ...813...36V} {813, 36}

\bibitem[\protect\citeauthoryear{{Vallini}, {Ferrara}, {Pallottini}, {Carniani}
   \& {Gallerani}}{{Vallini} et~al.}{2020}]{Vallini2020}
{Vallini} L.,  {Ferrara} A.,  {Pallottini} A.,  {Carniani} S.,   {Gallerani}
  S.,  2020, \mn@doi [\mnras] {10.1093/mnrasl/slaa047}, \href
  {https://ui.adsabs.harvard.edu/abs/2020MNRAS.tmpL..44V} {}

\bibitem[\protect\citeauthoryear{{Vallini}, {Ferrara}, {Pallottini}, {Carniani}
   \& {Gallerani}}{{Vallini} et~al.}{2021}]{Vallini2021}
{Vallini} L.,  {Ferrara} A.,  {Pallottini} A.,  {Carniani} S.,   {Gallerani}
  S.,  2021, \mn@doi [\mnras] {10.1093/mnras/stab1674}, \href
  {https://ui.adsabs.harvard.edu/abs/2021MNRAS.505.5543V} {505, 5543}

\bibitem[\protect\citeauthoryear{{Vallini} et~al.,}{{Vallini}
  et~al.}{2024}]{Vallini2024}
{Vallini} L.,  et~al., 2024, \mn@doi [\mnras] {10.1093/mnras/stad3150}, \href
  {https://ui.adsabs.harvard.edu/abs/2024MNRAS.527...10V} {527, 10}

\bibitem[\protect\citeauthoryear{{Venemans} et~al.,}{{Venemans}
  et~al.}{2020}]{Venemans2020}
{Venemans} B.~P.,  et~al., 2020, \mn@doi [\apj] {10.3847/1538-4357/abc563},
  \href {https://ui.adsabs.harvard.edu/abs/2020ApJ...904..130V} {904, 130}

\bibitem[\protect\citeauthoryear{{Vizgan} et~al.,}{{Vizgan}
  et~al.}{2022}]{Vizgan2022}
{Vizgan} D.,  et~al., 2022, \mn@doi [\apj] {10.3847/1538-4357/ac5cba}, \href
  {https://ui.adsabs.harvard.edu/abs/2022ApJ...929...92V} {929, 92}

\bibitem[\protect\citeauthoryear{{Walter} et~al.,}{{Walter}
  et~al.}{2016}]{Walter2016}
{Walter} F.,  et~al., 2016, \mn@doi [\apj] {10.3847/1538-4357/833/1/67}, \href
  {https://ui.adsabs.harvard.edu/abs/2016ApJ...833...67W} {833, 67}

\bibitem[\protect\citeauthoryear{{Walter} et~al.,}{{Walter}
  et~al.}{2018}]{Walter2018}
{Walter} F.,  et~al., 2018, \mn@doi [\apjl] {10.3847/2041-8213/aaf4fa}, \href
  {https://ui.adsabs.harvard.edu/abs/2018ApJ...869L..22W} {869, L22}

\bibitem[\protect\citeauthoryear{{Wang} et~al.,}{{Wang}
  et~al.}{2019}]{Wang2019}
{Wang} T.,  et~al., 2019, \mn@doi [\nat] {10.1038/s41586-019-1452-4}, \href
  {https://ui.adsabs.harvard.edu/abs/2019Natur.572..211W} {572, 211}

\bibitem[\protect\citeauthoryear{{Watson}, {Christensen}, {Knudsen}, {Richard},
  {Gallazzi}  \& {Micha{\l}owski}}{{Watson} et~al.}{2015}]{Watson2015}
{Watson} D.,  {Christensen} L.,  {Knudsen} K.~K.,  {Richard} J.,  {Gallazzi}
  A.,   {Micha{\l}owski} M.~J.,  2015, \mn@doi [\nat] {10.1038/nature14164},
  \href {https://ui.adsabs.harvard.edu/abs/2015Natur.519..327W} {519, 327}

\bibitem[\protect\citeauthoryear{{Whitney}, {Conselice}, {Duncan}  \&
  {Spitler}}{{Whitney} et~al.}{2020}]{Whitney2020}
{Whitney} A.,  {Conselice} C.~J.,  {Duncan} K.,   {Spitler} L.~R.,  2020,
  \mn@doi [\apj] {10.3847/1538-4357/abb824}, \href
  {https://ui.adsabs.harvard.edu/abs/2020ApJ...903...14W} {903, 14}

\bibitem[\protect\citeauthoryear{{Witstok} et~al.,}{{Witstok}
  et~al.}{2022}]{Witstok2022}
{Witstok} J.,  et~al., 2022, \mn@doi [\mnras] {10.1093/mnras/stac1905}, \href
  {https://ui.adsabs.harvard.edu/abs/2022MNRAS.515.1751W} {515, 1751}

\bibitem[\protect\citeauthoryear{{Witstok}, {Jones}, {Maiolino}, {Smit}  \&
  {Schneider}}{{Witstok} et~al.}{2023a}]{Witstok2023b}
{Witstok} J.,  {Jones} G.~C.,  {Maiolino} R.,  {Smit} R.,   {Schneider} R.,
  2023a, \mn@doi [\mnras] {10.1093/mnras/stad1470}, \href
  {https://ui.adsabs.harvard.edu/abs/2023MNRAS.523.3119W} {523, 3119}

\bibitem[\protect\citeauthoryear{{Witstok} et~al.,}{{Witstok}
  et~al.}{2023b}]{Witstok2023}
{Witstok} J.,  et~al., 2023b, \mn@doi [\nat] {10.1038/s41586-023-06413-w},
  \href {https://ui.adsabs.harvard.edu/abs/2023Natur.621..267W} {621, 267}

\bibitem[\protect\citeauthoryear{{Wolfire}, {Vallini}  \& {Chevance}}{{Wolfire}
  et~al.}{2022}]{Wolfire2022}
{Wolfire} M.~G.,  {Vallini} L.,   {Chevance} M.,  2022, \mn@doi [\araa]
  {10.1146/annurev-astro-052920-010254}, \href
  {https://ui.adsabs.harvard.edu/abs/2022ARA&A..60..247W} {60, 247}

\bibitem[\protect\citeauthoryear{{Yajima}, {Nagamine}, {Zhu}, {Khochfar}  \&
  {Dalla Vecchia}}{{Yajima} et~al.}{2017}]{Yajima2017}
{Yajima} H.,  {Nagamine} K.,  {Zhu} Q.,  {Khochfar} S.,   {Dalla Vecchia} C.,
  2017, \mn@doi [\apj] {10.3847/1538-4357/aa82b5}, \href
  {https://ui.adsabs.harvard.edu/abs/2017ApJ...846...30Y} {846, 30}

\bibitem[\protect\citeauthoryear{{Yamaguchi} et~al.,}{{Yamaguchi}
  et~al.}{2017}]{Yamaguchi2017}
{Yamaguchi} Y.,  et~al., 2017, \mn@doi [\apj] {10.3847/1538-4357/aa80e0}, \href
  {https://ui.adsabs.harvard.edu/abs/2017ApJ...845..108Y} {845, 108}

\bibitem[\protect\citeauthoryear{{Yan} et~al.,}{{Yan} et~al.}{2020}]{Yan2020}
{Yan} L.,  et~al., 2020, \mn@doi [\apj] {10.3847/1538-4357/abc41c}, \href
  {https://ui.adsabs.harvard.edu/abs/2020ApJ...905..147Y} {905, 147}

\bibitem[\protect\citeauthoryear{{Yan}, {Ma}, {Ling}, {Cheng}, {Huang}  \&
  {Zitrin}}{{Yan} et~al.}{2022}]{Yan2022}
{Yan} H.,  {Ma} Z.,  {Ling} C.,  {Cheng} C.,  {Huang} J.-s.,   {Zitrin} A.,
  2022, arXiv e-prints, \href
  {https://ui.adsabs.harvard.edu/abs/2022arXiv220711558Y} {p. arXiv:2207.11558}

\bibitem[\protect\citeauthoryear{{Zana}, {Gallerani}, {Carniani}, {Vito},
  {Ferrara}, {Lupi}, {Di Mascia}  \& {Barai}}{{Zana}
  et~al.}{2022}]{Zana2022_enhanched_SF}
{Zana} T.,  {Gallerani} S.,  {Carniani} S.,  {Vito} F.,  {Ferrara} A.,  {Lupi}
  A.,  {Di Mascia} F.,   {Barai} P.,  2022, \mn@doi [\mnras]
  {10.1093/mnras/stac978}, \href
  {https://ui.adsabs.harvard.edu/abs/2022MNRAS.513.2118Z} {513, 2118}

\bibitem[\protect\citeauthoryear{{Zana}, {Carniani}, {Prelogovi{\'c}}, {Vito},
  {Allevato}, {Ferrara}, {Gallerani}  \& {Parlanti}}{{Zana}
  et~al.}{2023a}]{Zana2023_more_galaxies_around_highz_quasars}
{Zana} T.,  {Carniani} S.,  {Prelogovi{\'c}} D.,  {Vito} F.,  {Allevato} V.,
  {Ferrara} A.,  {Gallerani} S.,   {Parlanti} E.,  2023a, \mn@doi [arXiv
  e-prints] {10.48550/arXiv.2309.03940}, \href
  {https://ui.adsabs.harvard.edu/abs/2023arXiv230903940Z} {p. arXiv:2309.03940}

\bibitem[\protect\citeauthoryear{{Zana}, {Carniani}, {Prelogovi{\'c}}, {Vito},
  {Allevato}, {Ferrara}, {Gallerani}  \& {Parlanti}}{{Zana}
  et~al.}{2023b}]{Zana2023}
{Zana} T.,  {Carniani} S.,  {Prelogovi{\'c}} D.,  {Vito} F.,  {Allevato} V.,
  {Ferrara} A.,  {Gallerani} S.,   {Parlanti} E.,  2023b, \mn@doi [arXiv
  e-prints] {10.48550/arXiv.2309.03940}, \href
  {https://ui.adsabs.harvard.edu/abs/2023arXiv230903940Z} {p. arXiv:2309.03940}

\bibitem[\protect\citeauthoryear{{Zanella} et~al.,}{{Zanella}
  et~al.}{2018}]{Zanella2018}
{Zanella} A.,  et~al., 2018, \mn@doi [\mnras] {10.1093/mnras/sty2394}, \href
  {https://ui.adsabs.harvard.edu/abs/2018MNRAS.481.1976Z} {481, 1976}

\bibitem[\protect\citeauthoryear{{Zavala}, {Casey}, {da Cunha}, {Spilker},
  {Staguhn}, {Hodge}  \& {Drew}}{{Zavala} et~al.}{2018}]{Zavala2018}
{Zavala} J.~A.,  {Casey} C.~M.,  {da Cunha} E.,  {Spilker} J.,  {Staguhn} J.,
  {Hodge} J.,   {Drew} P.~M.,  2018, \mn@doi [\apj] {10.3847/1538-4357/aaecd2},
  \href {https://ui.adsabs.harvard.edu/abs/2018ApJ...869...71Z} {869, 71}

\bibitem[\protect\citeauthoryear{{Zavala} et~al.,}{{Zavala}
  et~al.}{2021}]{Zavala2021}
{Zavala} J.~A.,  et~al., 2021, \mn@doi [\apj] {10.3847/1538-4357/abdb27}, \href
  {https://ui.adsabs.harvard.edu/abs/2021ApJ...909..165Z} {909, 165}

\bibitem[\protect\citeauthoryear{{da Cunha} et~al.,}{{da Cunha}
  et~al.}{2013}]{daCunha2013}
{da Cunha} E.,  et~al., 2013, \mn@doi [\apj] {10.1088/0004-637X/766/1/13},
  \href {https://ui.adsabs.harvard.edu/abs/2013ApJ...766...13D} {766, 13}

\bibitem[\protect\citeauthoryear{{da Cunha} et~al.,}{{da Cunha}
  et~al.}{2015}]{daCunha2015}
{da Cunha} E.,  et~al., 2015, \mn@doi [\apj] {10.1088/0004-637X/806/1/110},
  \href {https://ui.adsabs.harvard.edu/abs/2015ApJ...806..110D} {806, 110}

\makeatother
\end{thebibliography}



\appendix

\section{Dust continuum emission at 88 and 158 micron}
\begin{figure*}
    \centering
    \includegraphics[width=0.195\textwidth]{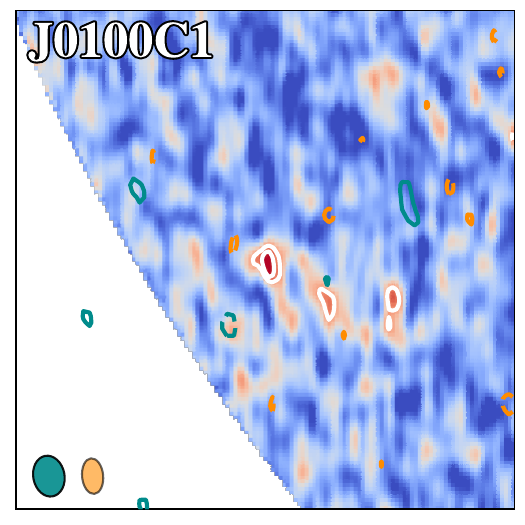}
    \includegraphics[width=0.195\textwidth]{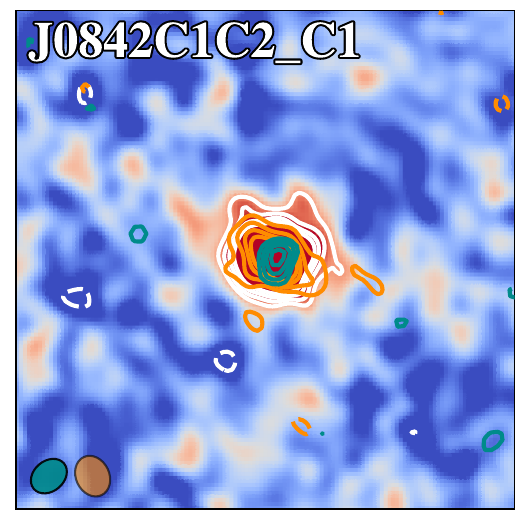}
    \includegraphics[width=0.195\textwidth]{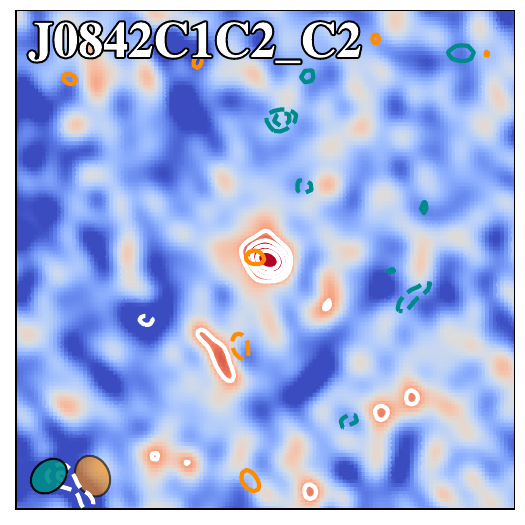}
    \includegraphics[width=0.195\textwidth]{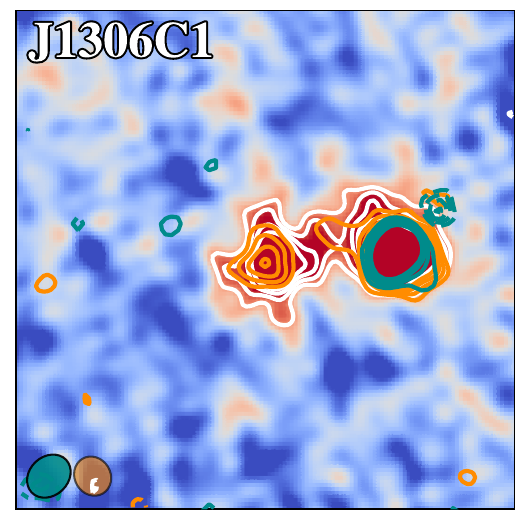}
    \includegraphics[width=0.195\textwidth]{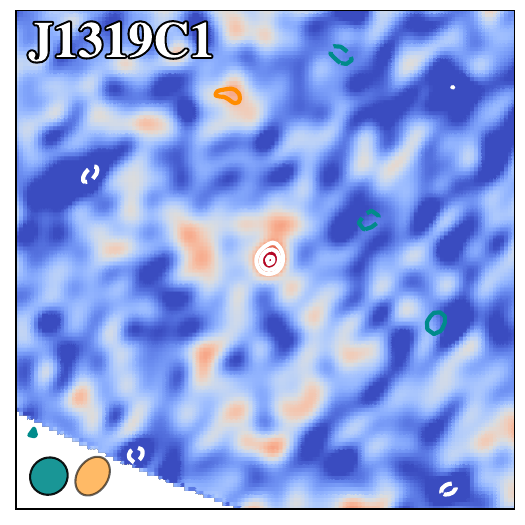}
    \includegraphics[width=0.195\textwidth]{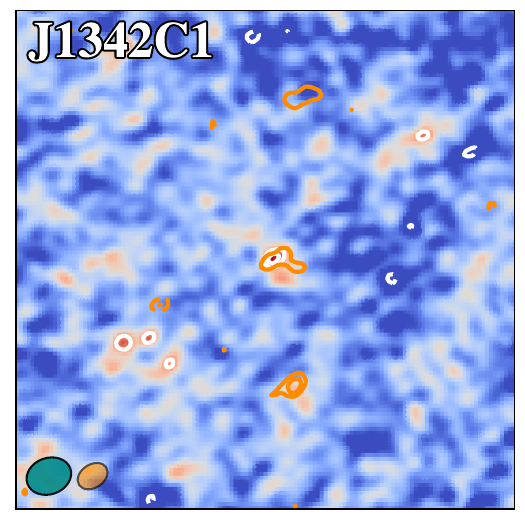}
    \includegraphics[width=0.195\textwidth]{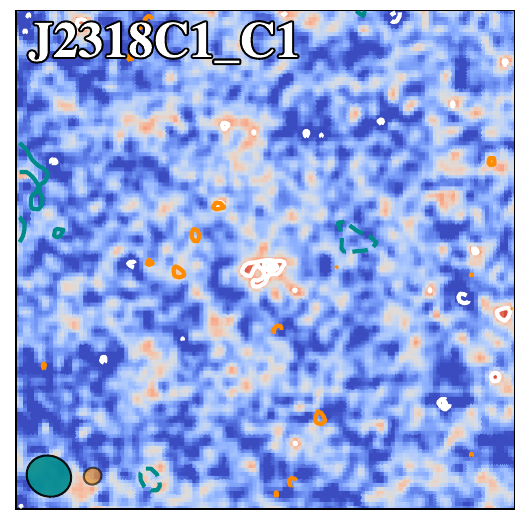}
    \includegraphics[width=0.195\textwidth]{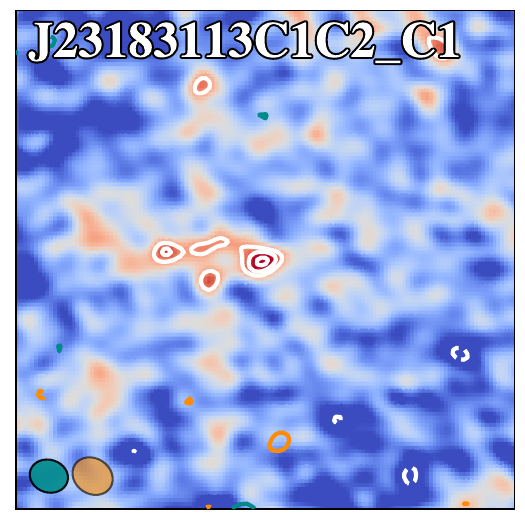}
    \includegraphics[width=0.195\textwidth]{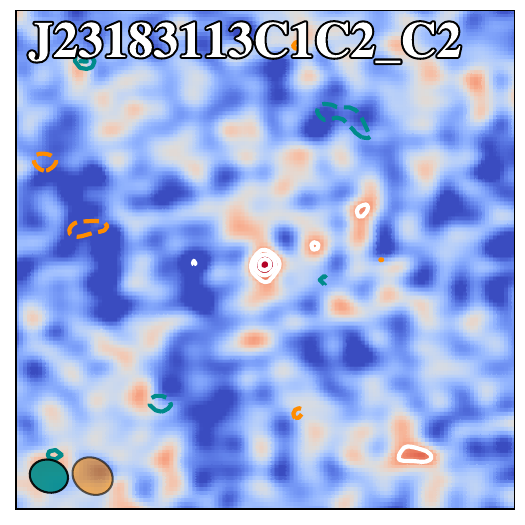}
    \includegraphics[width=0.195\textwidth]{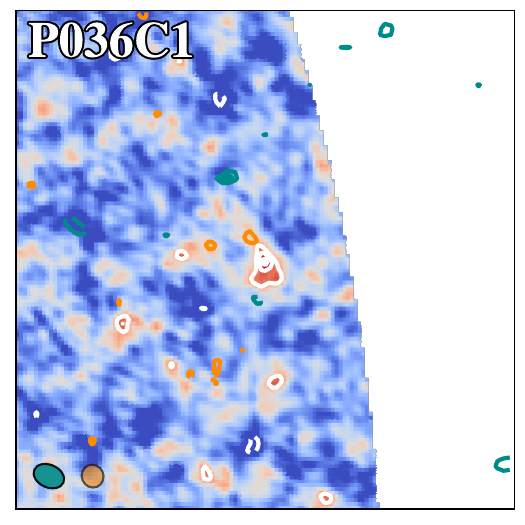}
    \includegraphics[width=0.195\textwidth]{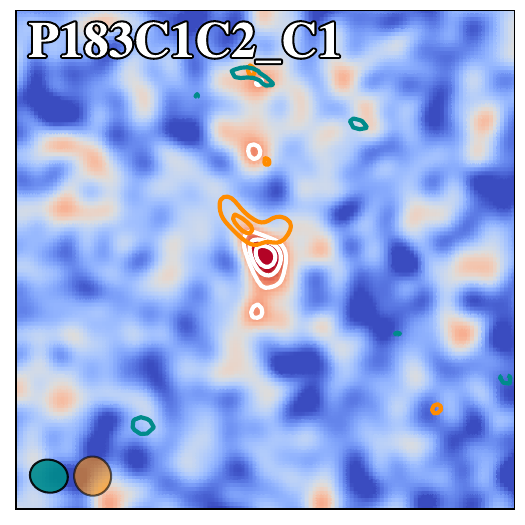}
    \includegraphics[width=0.195\textwidth]{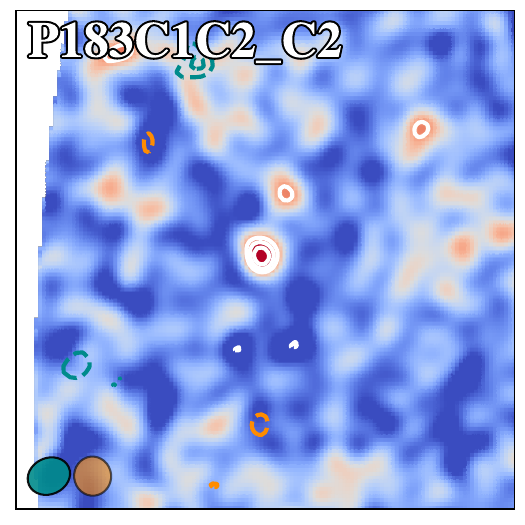}
    \includegraphics[width=0.195\textwidth]{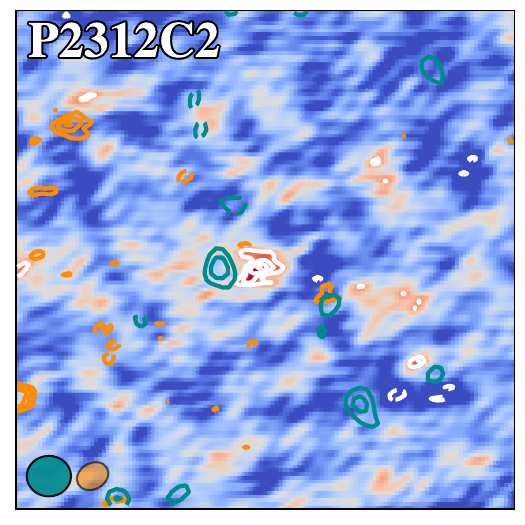}
    \caption{$4''\times4''$ cutouts of the \cii{} moment-0 maps of our 13 targets. The \cii{} emission is further indicated through white contours, with rest-frame $160\,\mu$m and $90\,\mu$m continua overlaid via orange and cyan contours, respectively. All contours are $\pm3, 4, 5, 6\sigma$, and negative contours are dashed. {\color{referee}The colorscale runs from $-1.5\sigma$ to $+4.5\sigma$.} Only one of our targets, J0842C1, is detected at rest-frame $90\,\mu$m.}
    \label{fig:continuum}
\end{figure*}

In Figure \ref{fig:continuum}, we show $4''\times4''$ cutouts of the \cii{} moment-0 maps, and overplot contours of rest-frame $90$ and $160\,\mu$m continuum emission. Two of our targets -- J0842C1 and J1306C1 -- are robustly continuum-detected at $160\,\mu$m, with the former also being detected at $90\,\mu$m.

\section{Modified Blackbody fit to the Stacked Dust SED}
\label{app:dustSEDs}

\begin{figure}
    \centering
    \includegraphics[width=0.35\textwidth]{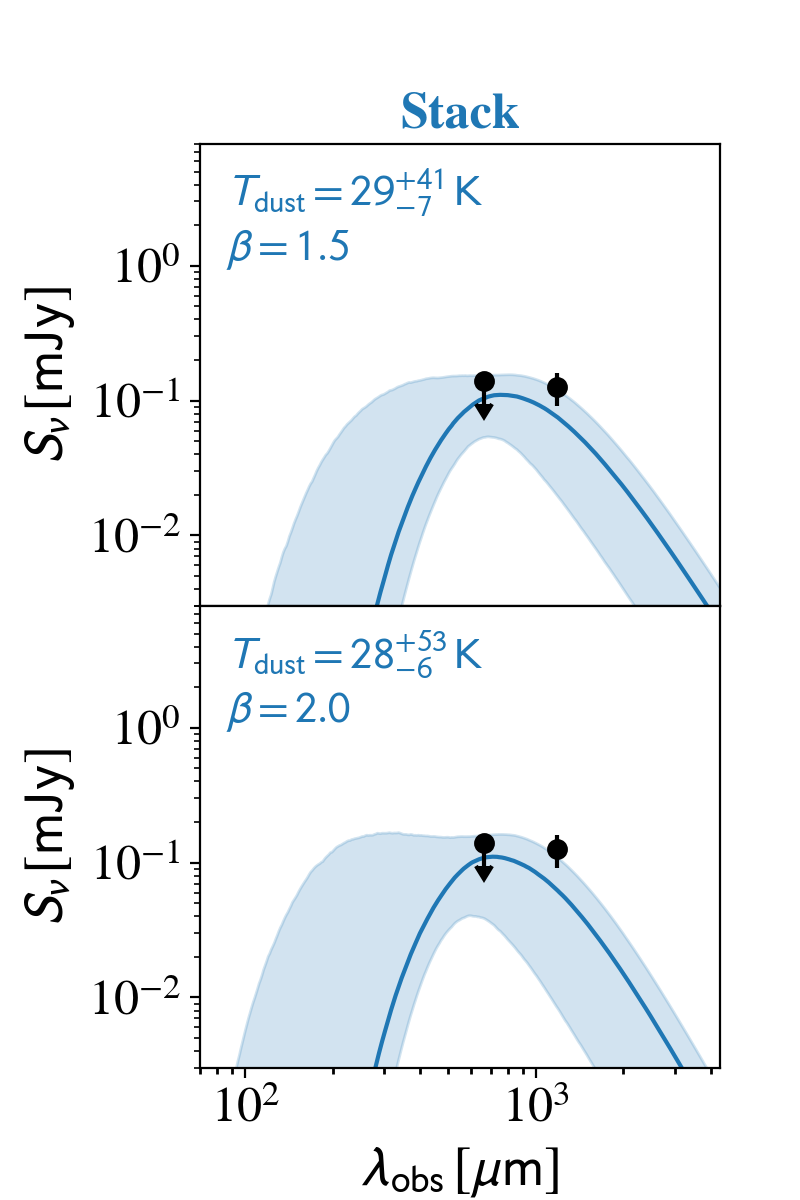}
    \caption{MBB fit to the stacked $90\,\mu$m and $160\,\mu$m continua. Due to the large uncertainty on the Band 6 flux density, the dust SED cannot accurately be constrained. However, the non-detection at rest-frame $90\,\mu$m suggests a cold average dust temperature, as discussed in the main text.}
    \label{fig:MBBfittingLowSNR}
\end{figure}

We here perform a formal MBB fit to the stacked continua in Fig.\ \ref{fig:MBBfittingLowSNR}. Due to the relatively large uncertainty on the $160\,\mu$m aperture flux density, our fit does not provide accurate constraints on the average dust temperature (and therefore average dust mass) of the \cii{}-selected companion sample, resulting in a massive spread ranging roughly from the CMB temperature to $\sim80\,$K. The inferred dust mass, assuming $\beta = 2$, is found to be $\log_{10}(M_\mathrm{dust} / M_\star) = 7.7_{-2.2}^{+0.9}$, and is thus similarly unconstrained. Deeper continuum observations at rest-frame $160\,\mu$m are needed to tighten these constraints.

\section{Stacking based on SFR weighting}
\label{app:stackingBasedOnSFR}
There is an argument towards stacking the \oiii{} and \cii{} emission normalized to the star-formation rate of each of the galaxies. However, this places the strongest weights on the brightest sources, and one of these sources has been detected in \oiii{} directly (J0842C1). As the main novelty of this work consists of the deep \oiii{} and $90\,\mu$m continuum observations of \textit{faint} \cii{}-emitters, we choose not to adopt an SFR-weighting in our fiducial stacks. Nevertheless, in the following we discuss the results of such an SFR-weighted stack.

\begin{figure}
    \centering
    \includegraphics[width=0.45 \textwidth]{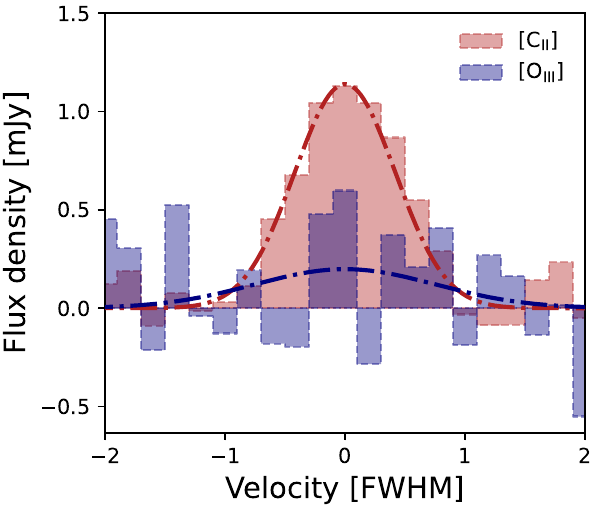}
    \includegraphics[width=0.23\textwidth]{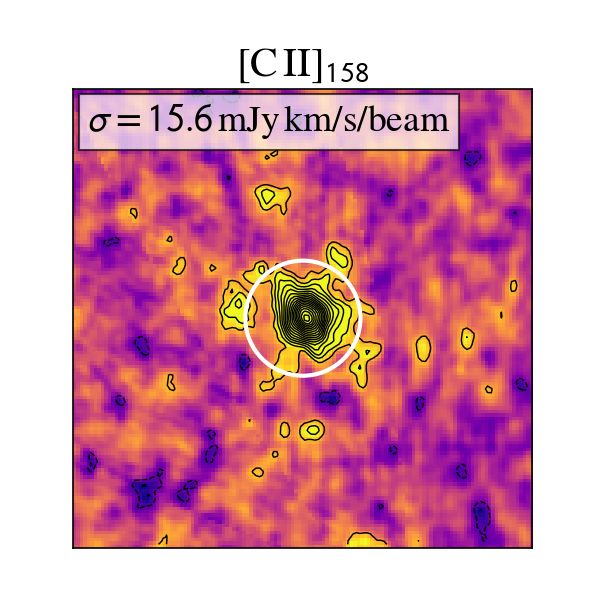}
    \includegraphics[width=0.23\textwidth]{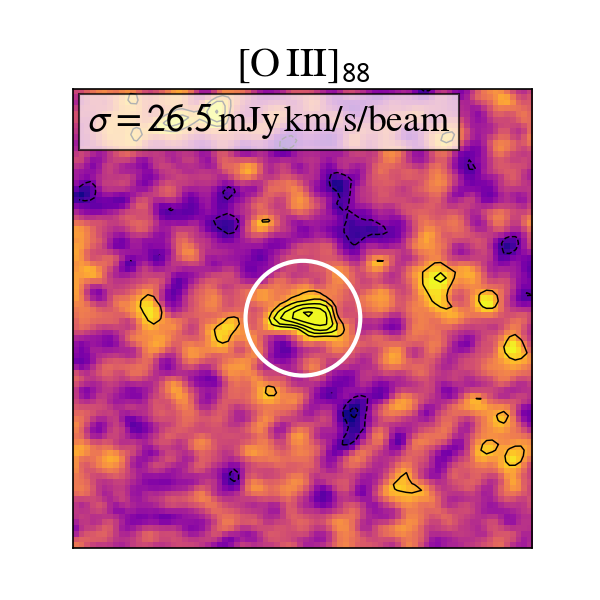}
    \caption{\textbf{Top:} The SFR-weighted, stacked spectrum of \oiii{} ({\it blue}) and \cii{} ({\it red}) normalized to a full-width at half-maximum per source prior to stacking. \textbf{Bottom:} 2D image stacks of the \cii{} (left) and \oiii{} moment-0 maps using the same SFR-weighting. In contrast to our fiducial analysis, this weighting results in the bright \oiii{}-detected source (J0842C1) dominating the stacked spectrum. As a result, a $\sim6\sigma$ \oiii{} signal is detected in the 1D and 2D stacks. However, upon excluding the single \oiii{}-detected source, this signal disappears completely.}
    \label{fig:SFR_based_stacking_result}
\end{figure}

In order to account for the fact that our galaxies span a range of \cii{}-based star formation rates, we normalize each source to a value of $\mathrm{SFR} = 50\,M_\odot\,\mathrm{yr}^{-1}$ prior to stacking, similar to the mean SFR across our sample. Given that the fiducial \cii{}-SFR relation from \citet{Schaerer2020} has a slope of unity, this comes down to a simple linear scaling in \cii{} line luminosity. We show the 2D \cii{} and \oiii{} stacks as well as the 1D spectra in Figure~\ref{fig:SFR_based_stacking_result}, which shows a modest detection in the stacked \oiii{}. This detection disappears completely when we remove the detected source from the sample, and as a result, we use the variance-weighted result in the main part of our paper.

\bsp	
\label{lastpage}
\end{document}